\documentclass[preprint,english,showpacs,preprintnumbers,amsmath,amssymb,floatfix]{revtex4}
\usepackage[T1]{fontenc}
\usepackage{adjustbox}

\usepackage[latin9,utf8]{inputenc}
\usepackage{color}
\usepackage{array}
\usepackage{amstext}
\usepackage{graphicx}
\usepackage{epstopdf}
\usepackage{esint}
\usepackage{float}
 \usepackage{tabularx}
\usepackage{dcolumn}
\usepackage{bm}
\usepackage{subfigure}
\usepackage{color}
\usepackage{float}
\usepackage{multirow}
\usepackage{xspace}
\usepackage{soul}
\usepackage{adjustbox}

\usepackage{booktabs}
\usepackage{amsmath}
\usepackage{hhline,multirow,tabularx}  
\DeclareGraphicsExtensions{.pdf}
\usepackage{latexsym,amssymb}
\usepackage[mathscr]{eucal}
\newcommand{\dzero}{D\O\xspace}
\newcommand{\ttbar}{\ensuremath{t\bar t}\xspace}
\newcommand{\x}{\ensuremath{x}\xspace}
\newcommand{\invfb}{fb$^{-1}$\xspace}
\makeatletter

\begin{document}
\title{ Impact of the top quark cross section data on parton distribution functions}

\newcommand{\SemnanUni}{Faculty of Physics, Semnan University, P.O. Box 35195-363, Semnan, Iran}
\newcommand{\IPM}{School of Particles and Accelerators, Institute for Research in Fundamental Sciences (IPM), P.O.Box 19395-5531, Tehran, Iran}
\newcommand{\Yazd}{Faculty of Physics, Yazd University, P.O. Box 89195-741, Yazd, Iran}

\author{M. Azizi} \email[]{majid.azizi650@gmail.com}  \affiliation{\SemnanUni}

\author{A. Khorramian}\email[]{khorramiana@semnan.ac.ir} \affiliation{\SemnanUni}
\author{H. Abdolmaleki} \email[]{Abdolmaleki@semnan.ac.ir} \affiliation{\SemnanUni}

\author{S. Paktinat Mehdiabadi} \email[]{paktinat@ipm.ir}  \affiliation{\Yazd} \affiliation{\IPM}

\begin{abstract}

Recent measurements of top quark pair production cross section, which is performed at the LHC and the Tevatron collider, are studied using Hessian profiling technique to obtain their impact on
the parton distribution functions (PDFs). The top quark production data covers different center-of-mass energies $\sqrt{s}$= 1.96, 5.02, 7, 8 and 13 TeV in either $pp$ or $p\bar{p}$ collisions.  It is explained how the Hessian profiling method may be used to assess the impact of these new data on PDFs and  consequently on their predictions. In this research, the impact of recent measurements of top quark pair cross sections on different CT14, MMHT2014, and NNPDF3.0 PDF sets  is investigated. 
 The analysis results show that the recent top quark production at the LHC and Tevatron data provide significant constraints in particular on the central value, relative uncertainties or both for the $s$-quark distribution and the gluon PDFs in both of CT14, MMHT2014 PDF sets and are insensitive to valence- quark PDFs.  A small constraint on the $\bar u$- sea quark distribution for  CT14 PDF is also observed. There is no impact on the NNPDF3.0 PDF set in presence of these data.
\end{abstract}

\maketitle

\section{Introduction}
\label{sec:introduction}
 
Parton distribution functions (PDFs) are a fundamental input into lepton-hadron and hadron-hadron collider physics for both the experimental and theoretical high energy particle physics. A recent review of the progress in determination of PDFs,
with the main emphasis on the usages for accurate phenomenology at the Large Hadron Collider (LHC), is reported in Ref.~\cite{Gao:2017yyd}. 
Since PDFs and their associated uncertainties play an important role in various LHC applications, so there are enough motivations to improve our understanding of the internal structure of the proton.

By utilizing experimental data, physical theories and proper mathematical methods implemented in computational tools, we are able to find better descriptions for PDFs inside the nucleons \cite{Dittmar:2009ii,Placakyte:2011az}.  As it is expected, released data by experimental groups in colliders have played a prominent role in increasing our understanding of hadrons. Almost all of the theoretical predictions in a hadron collider depend on the choice of PDFs. On the other hand, the measurements from the colliders can be used to improve the PDFs.
  
In the recent years, variety of  PDF sets \cite{Dulat:2015mca,Harland-Lang:2014zoa,Ball:2014uwa,Accardi:2016qay,Ball:2017nwa,Butterworth:2015oua} are extracted and published by various research groups. There are continuous efforts to improve PDFs, either by fitting new experimental data or by using new computational methods. 

Top quark is the heaviest elementary particle ($m_{t} >$ 170 GeV) \cite{Patrignani:2016xqp} which was discovered  in CDF \cite{Abe:1995hr} and \dzero \cite{D0:1995jca} experiments in 1995. 
It is expected that top quark plays an important role in the electroweak symmetry breaking because it is the only fermion which is close to its scale.

 Knowing top quark properties provides a unique opportunity to test the predictions of the standard model (SM) of the particles physics.
Top quark is considered as a window to physics beyond the standard model (BSM).  In hadron colliders, top quark is produced dominantly in pairs via QCD interactions, where quark-antiquark or gluon-gluon are fused to a high energy gluon that mediates the momentum to a top-antitop (\ttbar) pair. 
Top quark pair production can be considered as a motivation of many researches \cite{Thorne:2015rch,Pecjak:2016nee,Alekhin:2016jjz,Rojo:2015acz,Czakon:2016olj,Kieseler:2015jzh,Saha:2015lna,Wang:2017kyd}
seeking a deviation from the SM predictions as a signature of BSM. The total and differential  \ttbar production cross section are among the important observables at the hadron colliders, which have been measured with a high precision. The recent measurements of the total \ttbar pair production cross section are reported by ATLAS \cite{Aad:2014kva,ATLAS:2017fyu,Aaboud:2016pbd} and CMS \cite{CMS:2016pqu,Khachatryan:2016mqs,Khachatryan:2016yzq,
	Khachatryan:2016kzg,Sirunyan:2017uhy,CMS:2016rtp} collaborations at LHC, and \dzero \cite{Abazov:2016ekt} collaboration at Tevatron.
The  measurements of the differential \ttbar cross section is provided by ATLAS \cite{Aad:2014zka} and CMS \cite{Chatrchyan:2012saa} collaborations  at LHC.
Both experimental measurements are used in this analysis to constrain the PDF's.

In order to obtain the most comprehensive PDF constraints, different theoretical groups perform the global QCD fits of the experimental data.  To study the impact of new experimental measurements, one can perform a QCD global analysis with including the new data to the base data. 

To estimate the impact of new experimental measurements on the PDFs, we can use the approximate methods that can be used instead of a complete QCD fit, as an alternative approach. In this regard, one can use the Bayesian Monte Carlo reweighting and Hessian profiling techniques, as the approximate methods. As an example, the impact of the W-boson
charge asymmetry and of Z-boson production cross
sections data based on the Bayesian Monte Carlo reweighting   and  Hessian profiling techniques are reported in
 Refs. \cite{Ball:2011gg,Ball:2010gb,Camarda:2015zba,Watt:2012tq}.

	Comparing theoretical predictions and experimental measurements can be used to constrain the PDFs, strong coupling constant ($\alpha_s$) and top quark mass ($m_t$).   The central values and theoretical uncertainties for CT14, MMHT2014 and NNPDF3.0  are not all close to each other, except in the certain regions of $x$. The goal of this analysis is finding the impact of the new measurements of the production cross section of the top quark pair (\ttbar) \cite{Aad:2014kva,ATLAS:2017fyu,Aaboud:2016pbd,CMS:2016pqu,Khachatryan:2016mqs,Khachatryan:2016yzq,
		Khachatryan:2016kzg,Sirunyan:2017uhy,CMS:2016rtp,Abazov:2016ekt,Aad:2014zka,Chatrchyan:2012saa} on the modern  CT14 \cite{Dulat:2015mca}, NNPDF3.0 \cite{Ball:2014uwa} and MMHT2014 \cite{Harland-Lang:2014zoa} PDF sets using the Hessian profiling technique \cite{Paukkunen:2014zia}, without need of having a complete baseline global PDF fit procedure.

In this article, the QCD  analysis is performed based on xFitter open source framework \cite{xFitter,Alekhin:2014irh}.  The recent top quark production data which are not included in the main xFitter package are added. In Refs.~\cite{Vafaee:2017nze,Vafaee:2017jiv,Abdolmaleki:2017wlg,Vafaee:2017jnt,Rostami:2015iva,Salimi-Amiri:2018had}, we used xFitter for different QCD analyses, such as the study of different schemes in the QCD analysis and determination of the strong coupling constant.

The focus of this paper is to show how top quark new data can be used to constrain the PDFs (especially in the central value) and uncertainties or both for $s$-quark PDF and the gluon PDF at the large-$x$.   In Ref. \cite{Camarda:2015zba}, the impact of the Tevatron $W$ and $Z$ data on the MMHT2014 NLO set is reported  that shows a reduction of the PDF uncertainties in the  $d$-valence PDF. 

The outline of the paper is as follows. In Sec.~\ref{sec:data}, the data samples
are introduced and  in Sec.~\ref{sec:Hess}, a brief review of Hessian profiling method is described . In Sec.~\ref{sec:topquark}, the theoretical calculation and tools of the present analysis are explained. The impact of the new data on the central value and
uncertainty of the PDFs is shown in Sec.~\ref{sec:results}. Finally,  the results obtained in the paper are summarized in Sec.~\ref{sec:conclusion}.

\section{The top cross-section measurements and uncertainties}
\label{sec:data}

Before remarking the general impact of the \ttbar pair production cross section data, a brief explanation about these data is useful. The detailed discussion is given in Ref.~\cite{Czakon:2013tha,Czakon:2016olj} for the \ttbar pair production cross section data, but
for completeness we present a summary below with only focusing to recent \ttbar pair production cross section measurements.

The recent measurements of \ttbar pair production cross section by ATLAS \cite{Aad:2014kva,ATLAS:2017fyu,Aaboud:2016pbd,Aad:2014zka} and CMS \cite{CMS:2016pqu,Khachatryan:2016mqs,Khachatryan:2016yzq,
Khachatryan:2016kzg,Sirunyan:2017uhy,CMS:2016rtp,Chatrchyan:2012saa} collaborations at LHC and \dzero \cite{Abazov:2016ekt} collaboration at Tevatron, are considered in the present study. The ATLAS experiment \cite{Aad:2014kva} at LHC has measured the \ttbar production cross section in events containing an opposite-charge electron-$\mu$ ($e\mu$) pair. 
The measurement uses 4.6  (20.3) \invfb of data in $\sqrt{s}=$ 7 TeV (8 TeV). 
The corresponding measurement in $\sqrt{s}=$ 13 TeV  uses 3.2 \invfb of data \cite{Aaboud:2016pbd}. 
The same experiment , has also measured the cross section in 
lepton+jets final state in  $\sqrt{s}=$ 8 TeV with 20.2 \invfb of data \cite{ATLAS:2017fyu}. In another analysis, the ATLAS experiment has reported a measurement for the differential \ttbar cross section as a function of the top-quark transverse momentum \cite{Aad:2014zka}. The analysis uses 4.6 \invfb of data in $\sqrt{s}=$ 7 TeV  in lepton+jets final state.

The other main experiment at LHC, the CMS experiment, has also provided several results for the \ttbar cross section measurement. In a unique analysis, the cross-section is measured in $\sqrt{s}=$ 5.02 TeV, using 0.026 \invfb of data. The events are required to have an opposite-charge $e\mu$ pair and at least two jets. The same final state is used in $\sqrt{s}=$ 7 TeV (8 TeV) with 5 (19.7) \invfb of data to measure the cross section \cite{Khachatryan:2016mqs}. The  measurement in $\sqrt{s}=$ 13 TeV uses 2.2 \invfb of data collected in 2015 \cite{Khachatryan:2016kzg}.
The CMS experiment has measured the cross section also in lepton+jets final 
state \cite{Khachatryan:2016yzq}. The analysis uses 5  (19.6) \invfb of data in $\sqrt{s}=$ 7 TeV (8 TeV). The measurement in this final state in $\sqrt{s}=$ 13 TeV  using 3.2 \invfb of data is reported in Ref. \cite{Sirunyan:2017uhy}.
The CMS experiment has also published a result for the \ttbar cross section measurement in the fully hadronic final state based on 2.53 \invfb of data collected in $\sqrt{s}=$ 13 TeV \cite{CMS:2016rtp}. This experiment has combined the data from dilepton and  lepton+jets final states to measure the differential \ttbar cross section as a function of different kinematic variables \cite{Chatrchyan:2012saa}. The analysis uses 5 \invfb of data in $\sqrt{s}=$ 7 TeV.

The \dzero experiment at Tevatron has recently published a paper \cite{Abazov:2016ekt} on \ttbar cross section measurement in $p\bar{p}$ collisions in $\sqrt{s}=$ 1.96 TeV. The analysis uses 9.7 \invfb of data and each event is forced to have either one or two leptons.

In Tables \ref{tab:XsectionTable-ATLAS} and \ref{tab:XsectionTable-CMS},the  differential cross sections of top quark reported by  ATLAS \cite{Aad:2014zka} and CMS \cite{Chatrchyan:2012saa}  are summarized. 

\begin{table}[!htb]
    \begin{center}
		\begin{tabular}{c|c||c|c}

			\hline
			\hline          $p_T$ [GeV] & $1/\!\sigma\, d \sigma\!/\!d p_T$[GeV$^{-1}$] & Stat. [\%] & Sys. [\%] \\
         		\hline
   0 to   50 & $3.4 \cdot 10^{-3}$ & $\pm$     2.4 & $\pm$     5.1 \\ 
  50 to  100 &$ 6.7 \cdot 10^{-3}$ & $\pm$     1.2 & $\pm$     1.9 \\ 
 100 to  150 &$ 5.3 \cdot 10^{-3}$ & $\pm$     2.5 & $\pm$     2.6 \\ 
 150 to  200 &$ 2.6 \cdot 10^{-3}$ & $\pm$     2.0 & $\pm$     4.8 \\ 
 200 to  250 &$ 1.12 \cdot 10^{-3}$ & $\pm$     2.4 & $\pm$     4.8 \\ 
 250 to  350 &$ 0.32  \cdot 10^{-3}$ & $\pm$     3.5 & $\pm$     5.5 \\ 
 350 to  800 & $0.018  \cdot 10^{-3}$& $\pm$     6.1 & $\pm$    11 \\ 
\hline

\end{tabular}

\caption{Normalized differential cross-sections as a function of the transverse momentum measured by the ATLAS collaboration. The statistical and systematic uncertainties are also reported.}
\label{tab:XsectionTable-ATLAS}
          \end{center}
\end{table}

\begin{table}[h]
	\begin{center}
		\begin{tabular}{c|c||c|c}
			\hline
			\hline          $p_T$ [GeV] & $ 1/\!\sigma\, d \sigma\!/\!d p_T$[GeV$^{-1}$] & Stat. [\%] & Sys. [\%] \\
         		\hline
            $   0$ to $ 60$ & $4.54 \cdot 10^{-3}$ &  $\pm$  2.5 & $\pm$   3.6 \\
            $  60$ to $100$ & $6.66 \cdot 10^{-3}$ & $\pm$   2.4 & $\pm$   4.9 \\
            $ 100$ to $150$ & $4.74 \cdot 10^{-3}$ &  $\pm$  2.4 &  $\pm$  3.2 \\
            $ 150$ to $200$ & $2.50 \cdot 10^{-3}$ &  $\pm$   2.6 &  $\pm$   5.1 \\
            $ 200$ to $260$ & $1.04 \cdot 10^{-3}$ &  $\pm$   2.9 &  $\pm$   5.5 \\
            $ 260$ to $320$ & $0.38 \cdot 10^{-3}$ & $\pm$   3.7 &  $\pm$   8.2 \\
            $ 320$ to $400$ & $0.12 \cdot 10^{-3}$ & $\pm$   5.8 &  $\pm$   9.5\\
                      \hline
		\end{tabular}

\caption{ Normalized differential cross-sections as a function of the transverse momentum measured by the ATLAS collaboration. The statistical and systematic uncertainties are also reported.}
\label{tab:XsectionTable-CMS}
	\end{center}
\end{table}

In Table \ref{tab:Data},
\begin{table}[!htb]
    \begin{center}
        \begin{tabular}{ccc}
            \hline
            \hline
            $\sqrt{s}$ &   Ref.    & $\sigma_{Exp.}^{\rm tot}(t\bar{t})$ [pb]   \\
            \hline
            \hline
             \multicolumn{3}{c}{ATLAS Experiment at LHC}\\\hline
             7 TeV     &  \cite{Aad:2014kva} &  182.9 $\pm$ 3.1(stat.) $\pm$ 4.2(syst.) $\pm$ 3.6(lumi.) $\pm$ 3.3(beam) \\
             8 TeV &  \cite{Aad:2014kva}&  242.9 $\pm$ 1.7(stat.) $\pm$ 5.5(syst.) $\pm$ 5.1(lumi.) $\pm$ 4.2(beam)   \\
             8 TeV      & \cite{ATLAS:2017fyu}  & 248.3 $\pm$ 0.7(stat.) $\pm$ 13.4(syst.) $\pm$ 4.7(lumi.)   \\
             13 TeV      & \cite{Aaboud:2016pbd} &  818 $\pm$ 8(stat.) $\pm$ 27(syst.) $\pm$ 19(lumi.) $\pm$ 12(beam)  \\  
            \hline  
            \multicolumn{3}{c}{CMS Experiment at LHC}\\\hline
             5.02 TeV   &  \cite{CMS:2016pqu} &  82 $\pm$ 20(stat.)$\pm$ 5(syst.) $\pm$ 10(lumi.)  \\ 
             7 TeV    &  \cite{Khachatryan:2016mqs} & 173.6 $\pm$ 2.1(stat.) $^{+4.5}_{-4}$(syst.) $\pm$ 3.8(lumi.)  \\
             7 TeV  &  \cite{Khachatryan:2016yzq} & 161.7 $\pm$ 6(stat.) $\pm$ 12(syst.) $\pm$ 3.6(lumi.)  \\
             8 TeV &  \cite{Khachatryan:2016yzq} & 227.4 $\pm$ 3.8(stat.) $\pm$ 13.7(syst.) $\pm$ 6(lumi.) \\
             8 TeV   &  \cite{Khachatryan:2016mqs} & 244.9 $\pm$ 1.4(stat.) $^{+6.3}_{-5.5}$(syst.) $\pm$ 6.4(lumi.)  \\
             13 TeV   &  \cite{Khachatryan:2016kzg} & 815 $\pm$ 9(stat.) $\pm$ 38(syst.) $\pm$ 19(lumi.)  \\
             13 TeV & \cite{Sirunyan:2017uhy}  & 888 $\pm$ 2(stat.)  $^{+26}_{-28}$(syst.) $\pm$ 20(lumi.)  \\
             13 TeV  &  \cite{CMS:2016rtp}  & 834 $\pm$ 25 (stat.) $^{+118}_{-104}$(syst.) $\pm$ 23(lumi.)  \\
            \hline
            \multicolumn{3}{c}{\dzero Experiment at Tevatron}\\\hline
            1.96 TeV & \cite{Abazov:2016ekt} & 7.26 $\pm$ 0.13 (stat.) $^{+0.57}_{-0.50}$(syst.) \\
            \hline
            \hline
        \end{tabular}
    \end{center}
    \caption{\label{tab:Data}The recent measurements of top quark pair production total cross section in different center-of-mass energies with corresponding uncertainties and information.}
\end{table}
the specifications of the recent experimental measurements of the top quark pair production cross section, accompanied by their statistical, systematic, luminosity and beam (if reported) uncertainties are summarized.

\section{Hessian profiling technique}
\label{sec:Hess}
To study the impact of new experimental measurements on the PDFs,
 one can perform a QCD global fit analysis using the experimental data.
As an alternative approach, an approximate method can be used instead of a complete QCD fit. The profiling technique is the approximate method that can be applied for PDFs extracted by Hessian method \cite{xFitter}.

Generally,  there are two approximate methods such as Bayesian Monte Carlo reweighting \cite{Ball:2011gg,Ball:2010gb,Camarda:2015zba} and Hessian profiling \cite{Watt:2012tq} techniques.  The main benefit of using these two techniques is that they can be applied to find the impact of new experimental data on a preexisting   PDF.  It should be noted that these approximate methods have a number of limitations. For example, if the impact of new  measurements is very large, these methods can not be useful and in particular are not able to explain the effect on the input PDF parametrization, or in the theoretical calculations. Therefore, not only when using these  approximate methods some care should be taken but also we should care when interpreting their results.

 The Hessian profiling technique is based on the $\chi^2$ minimization method using  a comparison between  the theoretical predictions extracted with a given input Hessian PDF set  and the new experimental data. According to this method, the $\chi^2$ definition with taking into account the uncertainties of  experimental data and the effects from the variations of PDF  which  is encoded by the   Hessian eigenvectors, is as following  \cite{Gao:2017yyd,Camarda:2015zba}
\begin{eqnarray}
\chi^2 ({\beta_{exp}},{ \beta_{th}})
=\sum_{i=1}^{N_{data}}\frac{\left( [\sigma_i^{exp}
+\sum_j \Gamma_{ij}^{exp}\beta_{j, exp}]
 -[\sigma_i^{th}
 +\sum_k\Gamma_{ik}^{th}\,\beta_{k, th}] \right)^2}{\delta_i^2}
 +\sum_j \beta_{j, exp}^2+\sum_k \beta_{k, th}^2
 \label{eq:chi2hessian}
\end{eqnarray}
where $\delta_i$ is the total experimental
 uncorrelated uncertainty, $\beta_{j, exp}$  and  $\beta_{k, th}$ are the parameters corresponding
 to the set of fully correlated experimental systematic
 uncertainties and the PDF Hessian eigenvectors, respectively. 
Also in above equation, $N_{data}$ is the number of experimental data
 points which is being added into the fit,  and finally  the matrices $\Gamma_{ij}^{exp}$ and
 $\Gamma_{ik}^{th}$ encode the effects of the corresponding
$\beta_{j, exp}$  and  $\beta_{k, th}$ parameters on the experimental data and on the
 theory predictions, respectively.
 
After minimizing the $\chi^2$  in Eq.~(\ref{eq:chi2hessian}),
the corresponding values of the theoretical $\beta_{k, th}^{min}$ parameters can be interpreted as leading  to optimized PDFs (``profiled'') to explain the new specific measurement. In the next sections it will be seen how profiling method  modifies both central values and total PDF uncertainties.

\section{Theoretical predictions of top quark production}
\label{sec:topquark}

The cross-section of the \ttbar production is one of the most important measurements among different top quark measurements. The SM predictions for this measurement involve both the QCD calculations of the partonic processes and also PDF used to integrate the partonic cross section. The next-to-leading-order (NLO) production cross section of un-polarized and polarized top quark pair are calculated in Refs. \cite{Nason:1987xz,Beenakker:1988bq,Bernreuther:2004jv}. Beyond the NLO accuracy, the resummation of the soft gluon emission at next-to-leading-logarithmic (NNL) correction is investigated in Refs. \cite{Kidonakis:1997gm,Bonciani:1998vc}. 
At this time, the next-to-next-to-leading-order (NNLO)
corrections to inclusive production of \ttbar pair, accomplished in Refs. \cite{Czakon:2013goa,Czakon:2012pz,Czakon:2012zr,Baernreuther:2012ws}, are needed to improve the computational tools \cite{Czakon:2007wk,Czakon:2007ej,Mitov:2006xs,Ferroglia:2009ep,Ferroglia:2009ii,Czakon:2010td,Bierenbaum:2011gg,Baernreuther:2013caa}. 

There are many different computational tools to calculate the \ttbar production cross section such as HATHOR \cite{Aliev:2010zk}, Top++ \cite{Czakon:2011xx}, DiffTop \cite{Guzzi:2014wia} and MCFM \cite{Campbell:2015qma}. Although DiffTop is capable to calculate both total and differential cross section of \ttbar pair production.

As it is expected from the PDFs of proton and antiproton, gluon-gluon fusion is dominant in \ttbar production in proton-proton colliders like LHC. About 80\% of \ttbar pairs in LHC at the center-of-mass energy of ($ \sqrt{s} = $) 7 TeV are from  gluon-gluon fusion. The fraction grows with $ \sqrt{s} $ and can reach 90\% in $ \sqrt{s} = $ 14 TeV \cite{Patrignani:2016xqp}. So the \ttbar production cross section measurement from LHC can mainly constrain the gluon PDF.
Due to high mass of the top quark, the \ttbar production cross section receives the main contribution from high-\x gluon distribution which is affected by considerable uncertainty.

In this analysis, the HATHOR and DiffTop computational programs at NNLO which are implemented in xFitter \cite{xFitter}, the new version of HeraFitter \cite{Alekhin:2014irh}, is used to include the \ttbar cross section measurements, following the profiling method.

\section{Results and Discussion}
\label{sec:results}
To study the impact of top quark cross section measurements on a given PDF set,  the Hessian profiling method is used. This approximate method incorporates  the information contained in new measurements into an existing specific PDF sets without the need for refitting. 

The top cross section from LHC and Tevatron are used to update the proton PDFs using the profiling method, utilized by the Thorne-Roberts (TR) scheme \cite{Thorne:2012az} of General-Mass Variable Flavour Number (GM-VFN) scheme.  The values of top quark mass, $ m_t$, and strong coupling constant at Z boson mass, $\alpha_{s}(M_Z)$, are set to 173.3 GeV and 0.118, respectively.

The CT14, MMHT2014 and NNPDF3.0 parton distribution functions, in different confidence level are available in LHAPDF library \cite{Buckley:2014ana} which is interfaced to xFitter. The theoretical calculation of the total top quark cross section and the relevant uncertainty using  different  PDF sets for LHC and Tevatron center of mass energies are presented in Table \ref{tab:theory}. 
\begin{table}[!htb]
    \begin{center}
    {\scriptsize 
        \newcolumntype{C}[1]{>{\hsize=#1\centering\arraybackslash}X}
        \begin{adjustbox}{width=1.25\textwidth}
            \begin{tabularx}{.93\linewidth}{lccccc*{2}{C{2.5cm}}C{2.cm}}    
                \hhline{=======}
                PDF Sets &    & \multicolumn{4}{c}{\ \ LHC [TeV] }& Tevatron [TeV] \\
                \cline{3-8}
                &           & ~~~5.02~~~&~~~7~~~&~~~8~~~&~~~13~~~&~~~1.96~~~\\
                \hhline{=======}
                CT14    
                &  &66.16 $^{+6.4}_{-5.1}$ &172.45 $^{+12.68}_{-10.65}$&246.36 $^{+16.14}_{-14.1}$ &806.52 $^{+34.94}_{-35.5}$ & 7.24 $^{+0.41}_{-0.26}$ \\\hhline{-------}
                MMHT2014    
                &  &66.36 $^{+1.8}_{-2.62}$ &172.07 $^{+3.83}_{-5.37}$&245.62 $^{+5.1}_{-7}$ &804.21 $^{+13.33}_{-17.08}$ & 7.33 $^{+0.206}_{-0.191}$ \\\hhline{-------}
                NNPDF3.0
                &  &64.88 $\pm$ 2.14  &170.16 $\pm$ 4.26&243.66 $\pm$ 5.53 &803.26 $\pm$ 14.19 & 7.16 $\pm$ 0.132 \\ 
                \hhline{========}
            \end{tabularx}
        \end{adjustbox}
    }
        
    \caption{\label{tab:theory} The total NNLO top quark total cross section [pb] prediction and total theoretical uncertainties calculated by Hathor at LHC and Tevatron run II energies for CT14 \cite{Dulat:2015mca}, MMHT2014 \cite{Harland-Lang:2014zoa} and NNPDF3.0 \cite{Ball:2014uwa} PDF sets.}
    \end{center}
\end{table}

 To apply the profiling technique for a PDF set, only the new top quark measurements which are not included in that PDF sets are considered. The compatibility of the new measurements with the CT14, MMHT2014 and NNPDF3.0 sets is tested by computing the $\chi^2$ function of Eq.~(1).

\begin{figure}[!htb]
	\begin{center}

		\includegraphics[scale = 0.5]{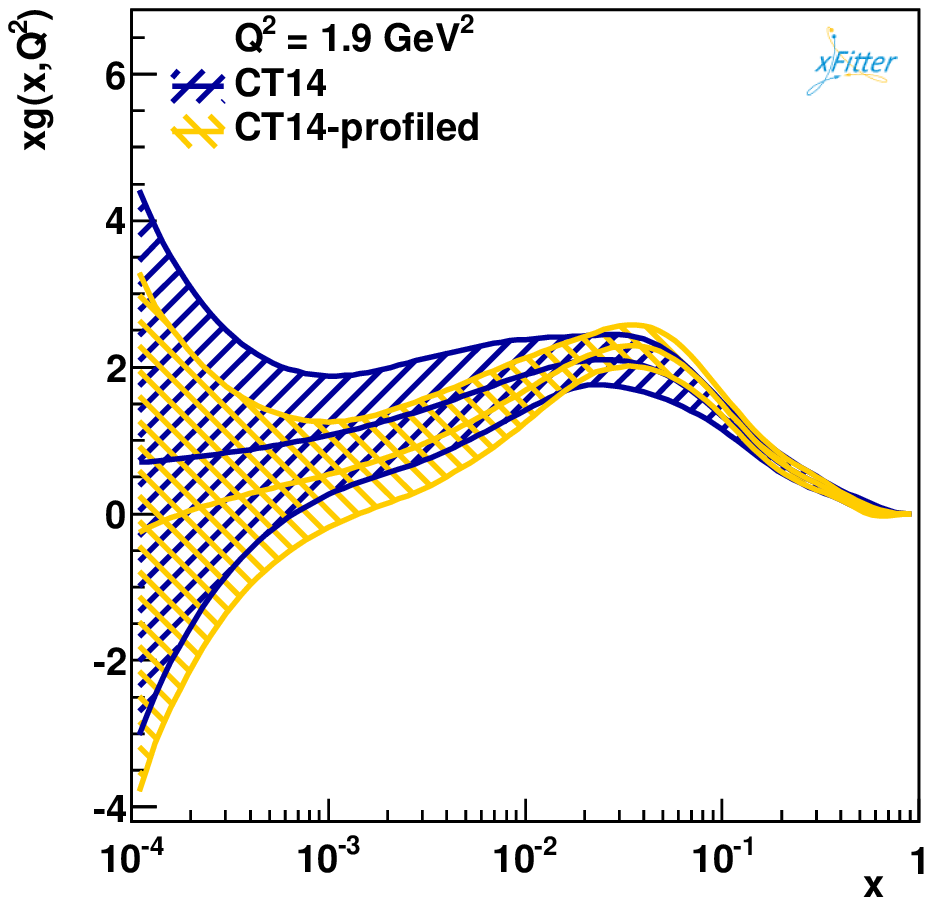}
		\includegraphics[scale = 0.5]{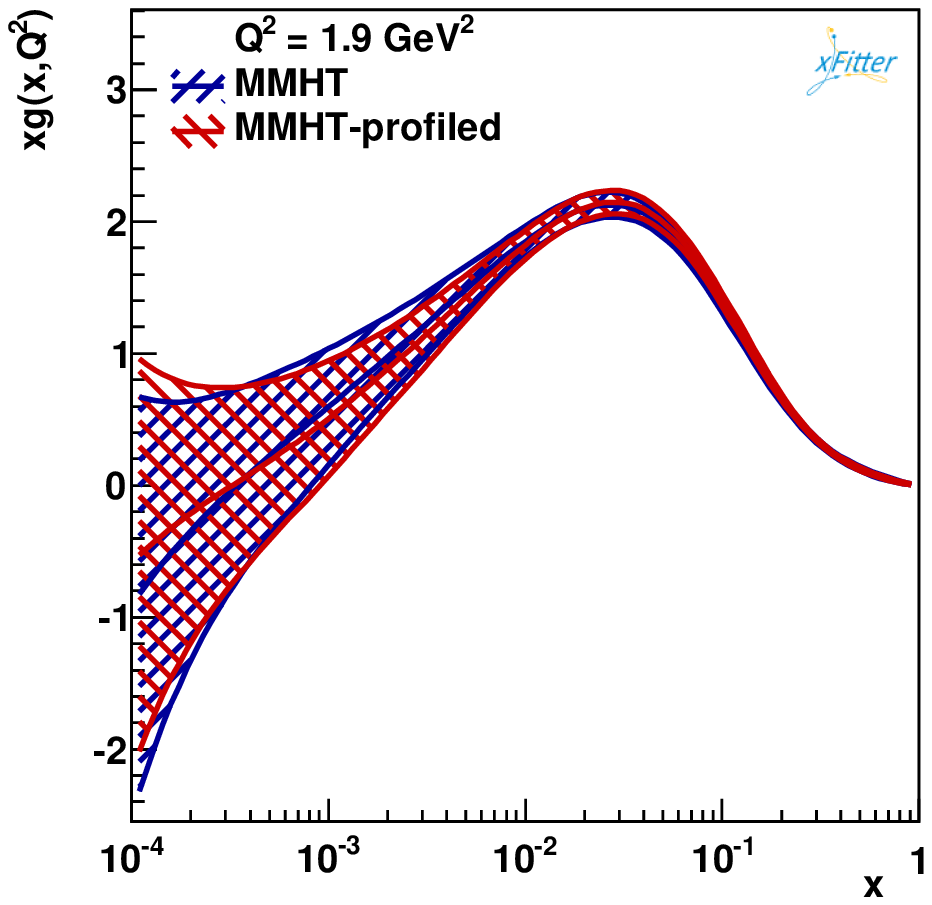}
		\includegraphics[scale = 0.5]{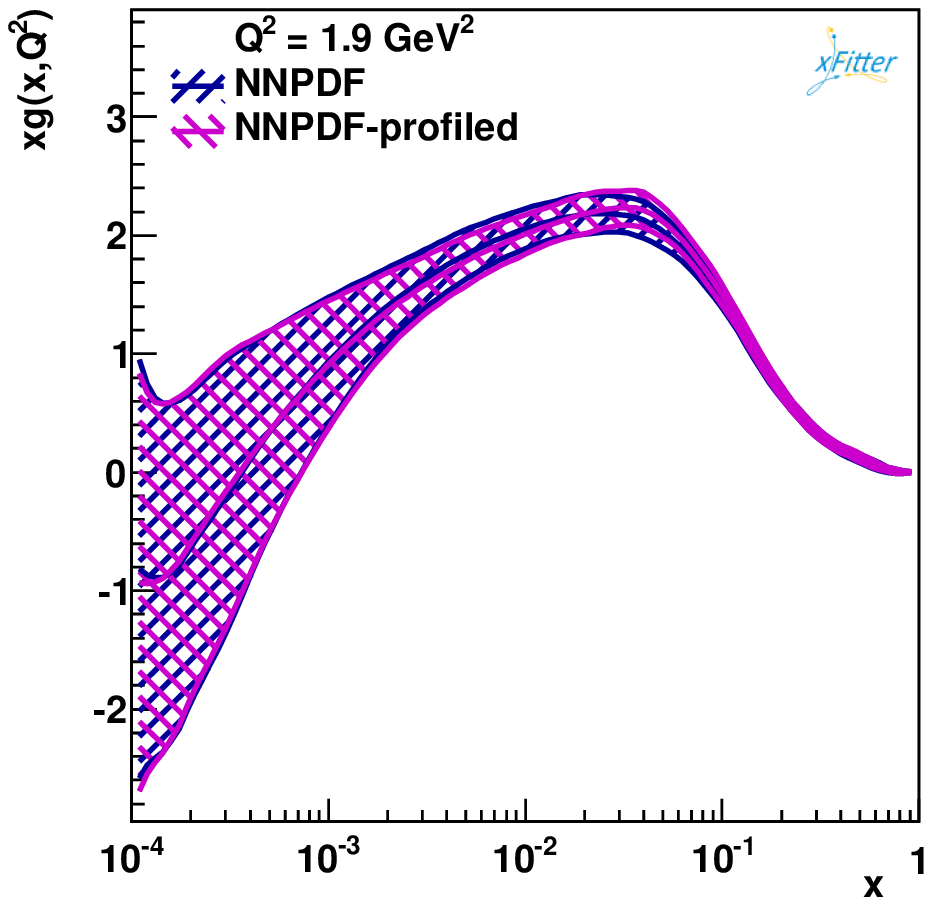}
		\includegraphics[scale = 0.5]{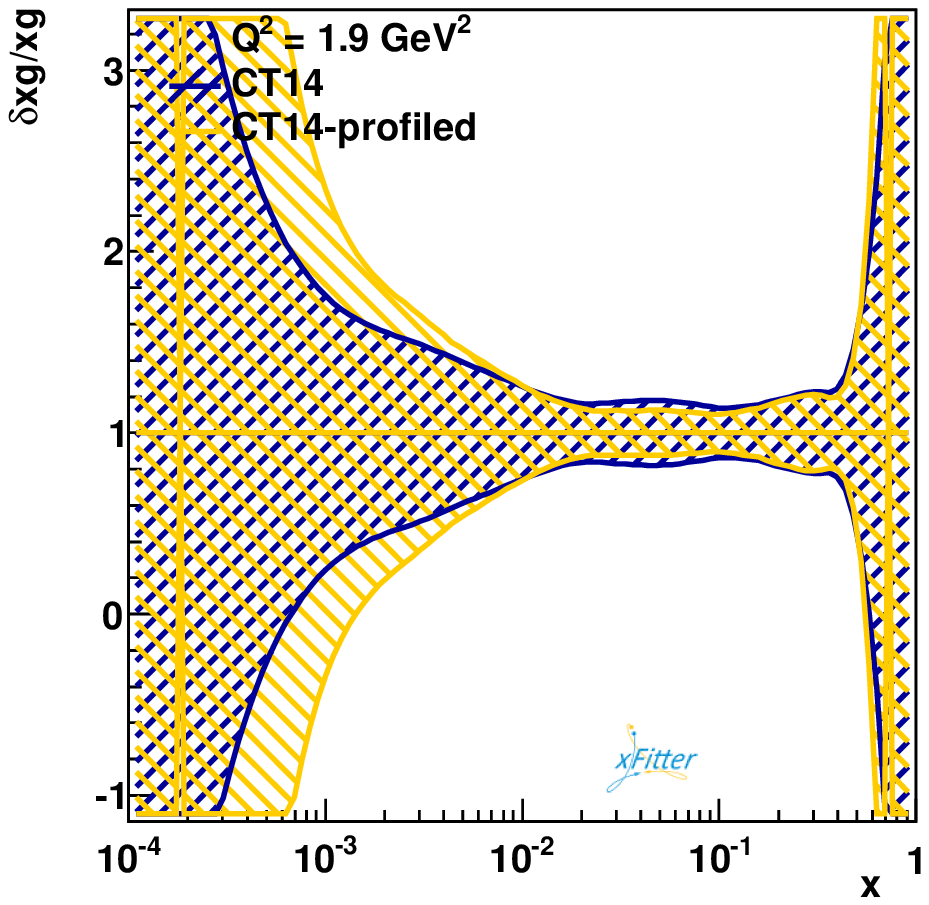}
		\includegraphics[scale = 0.5]{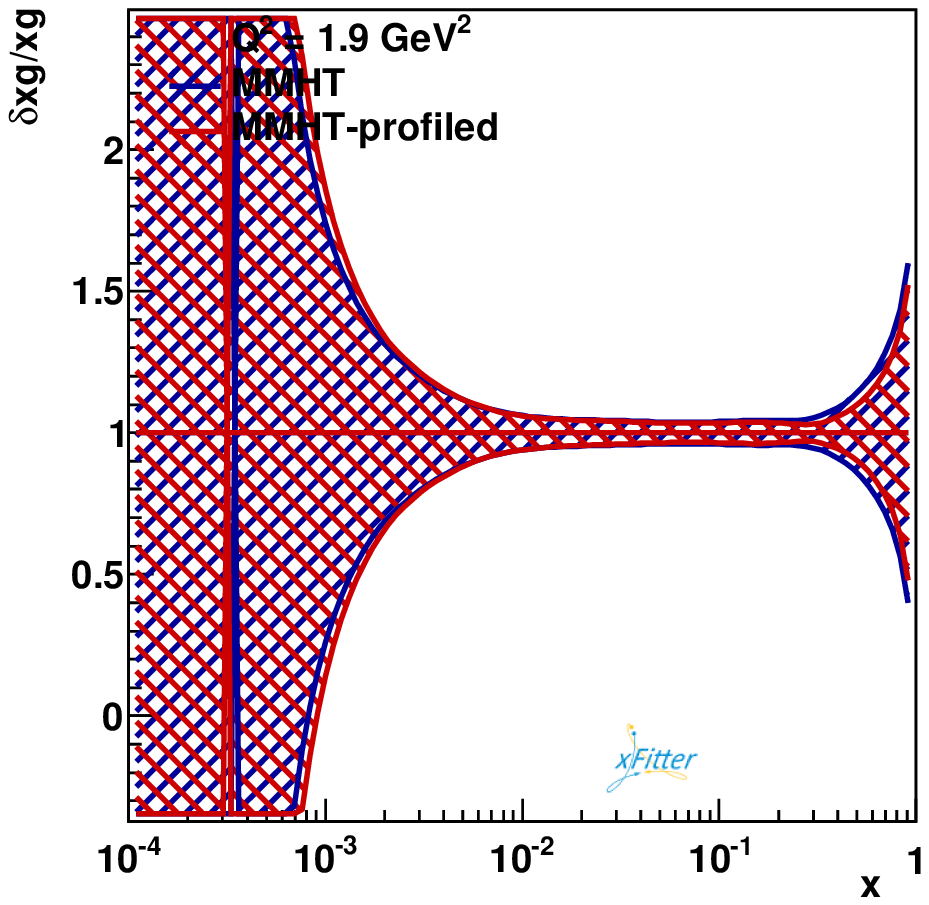}
		\includegraphics[scale = 0.5]{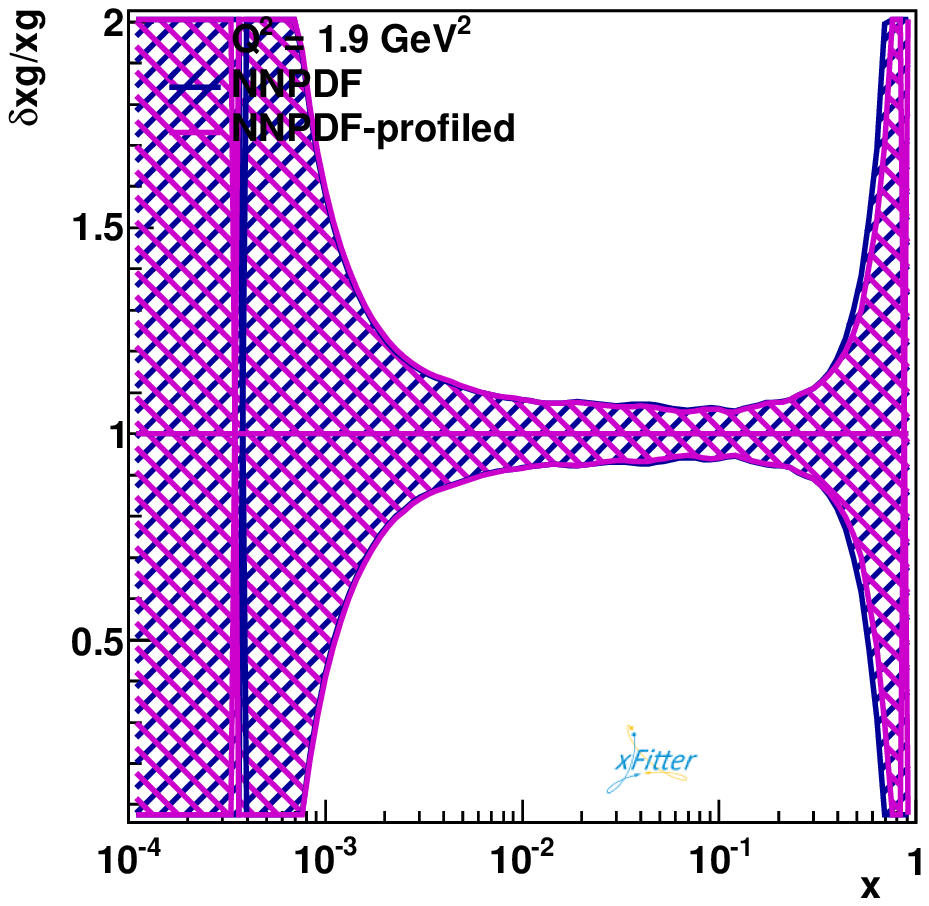}

		\caption{The gluon PDFs and relative uncertainties extracted from profiled CT14 \cite{Dulat:2015mca}, MMHT2014 \cite{Harland-Lang:2014zoa}, and NNPDF3.0\cite{Ball:2014uwa}  PDF sets at 1.9 GeV$^2$ as a function of $x$. The results obtained after the profiling procedure compared with corresponding same features before profiling. }
		\label{fig:partonDis}
	\end{center}
\end{figure}

In Fig.~\ref{fig:partonDis},  the original and profiled CT14, MMHT2014 and NNPDF3.0 parton distribution functions for $xg$ gluon PDF at the NNLO  are presented.  It can be seen that  the recent top quark  measurements at the LHC and Tevatron provide significant constraints in particular on the central value and the uncertainties of $xg$ for CT14, MMHT2014.  There is no impact on the NNPDF3.0 PDF set. So, we study the impact of  these data on the CT14, MMHT2014 PDF sets only.

\begin{figure}[!htb]
	\begin{center}	

	    \includegraphics[scale = 0.35]{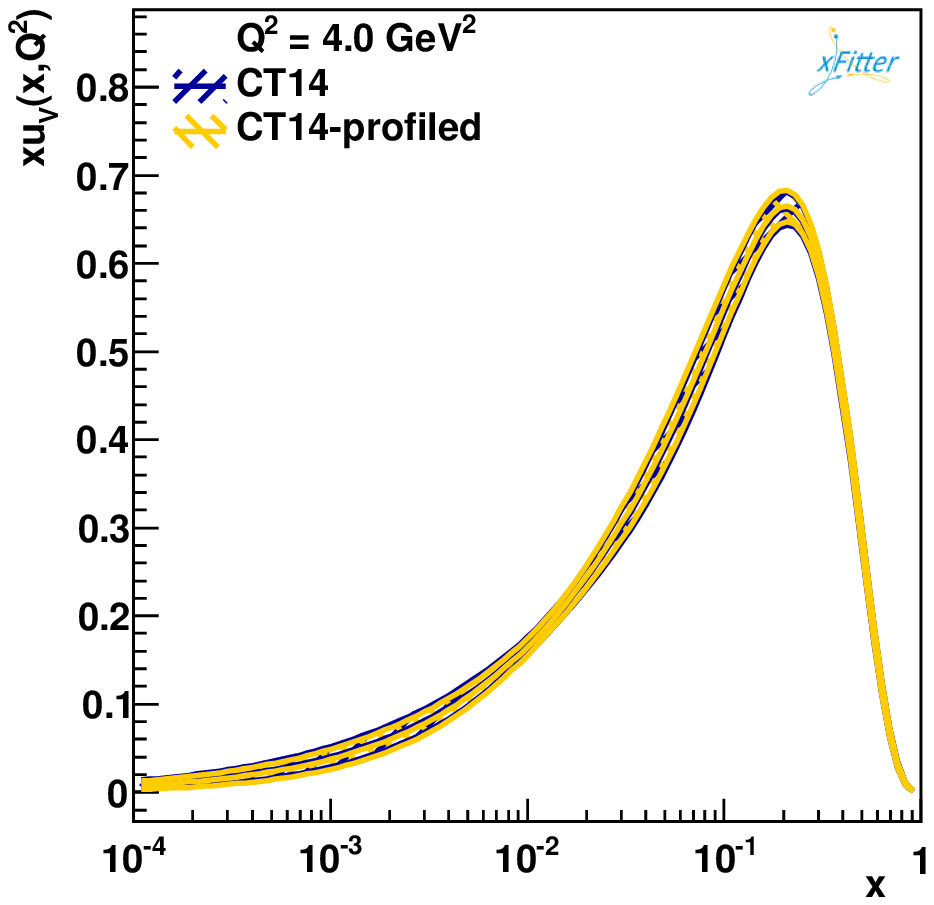}	    	    
	    \includegraphics[scale = 0.35]{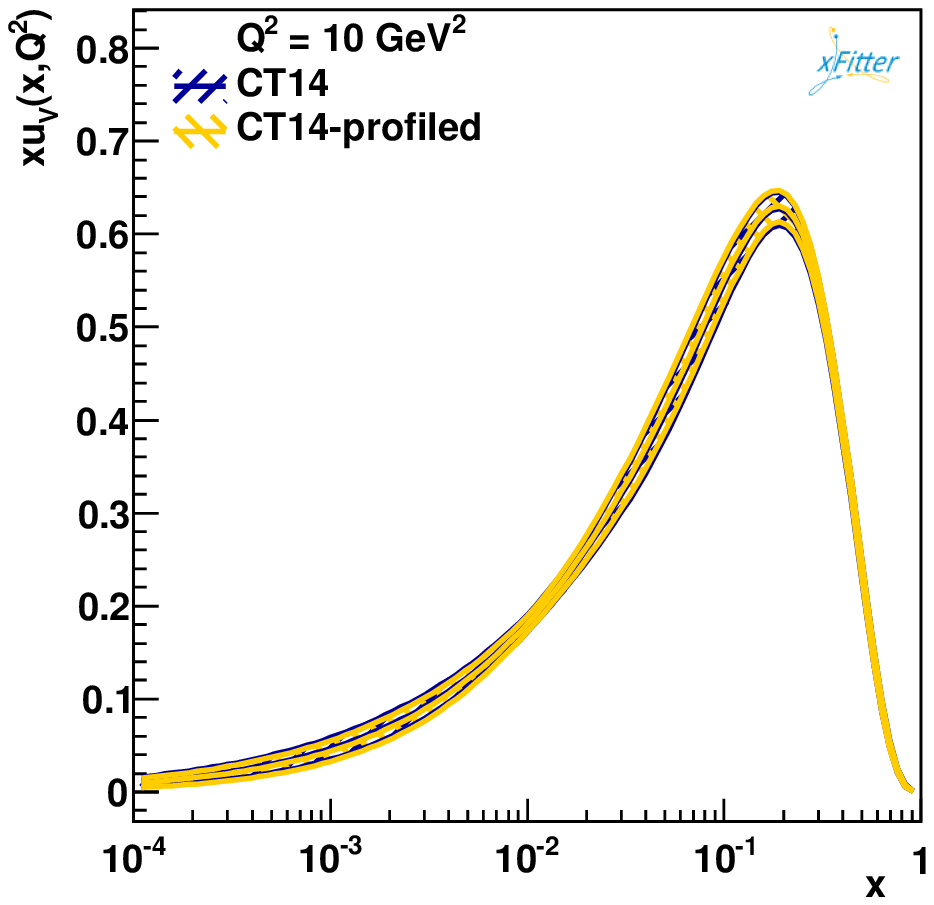}
	    \includegraphics[scale = 0.35]{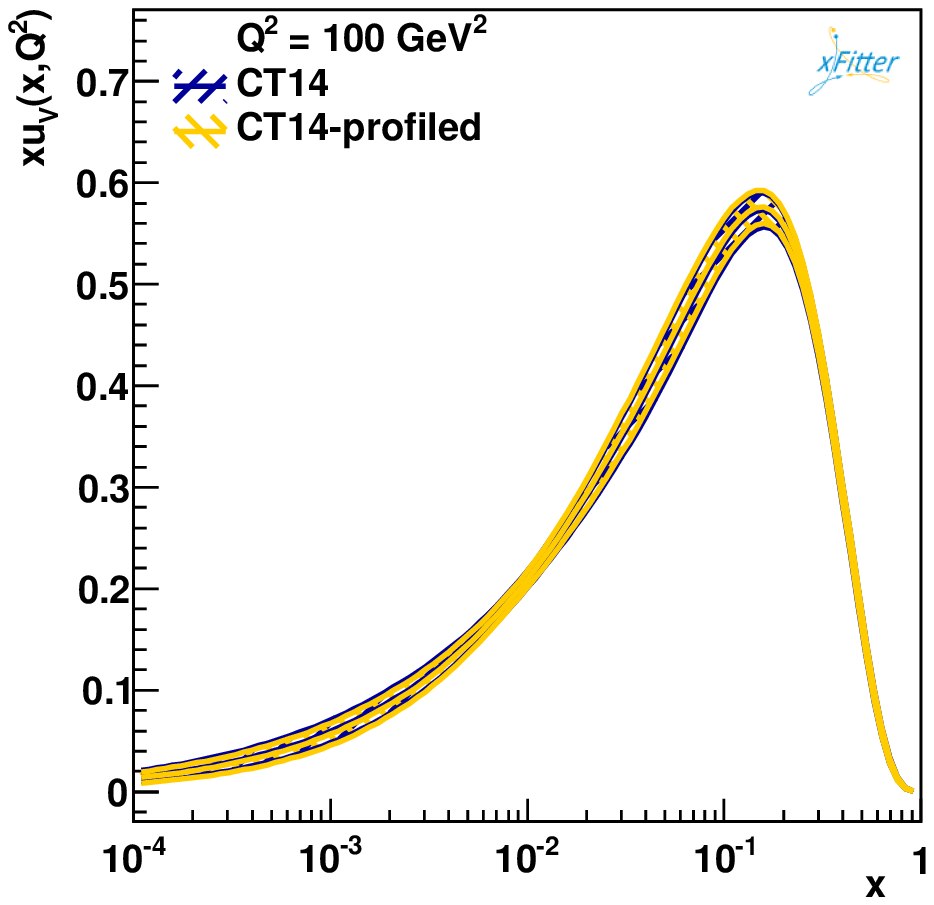}
	    \includegraphics[scale = 0.35]{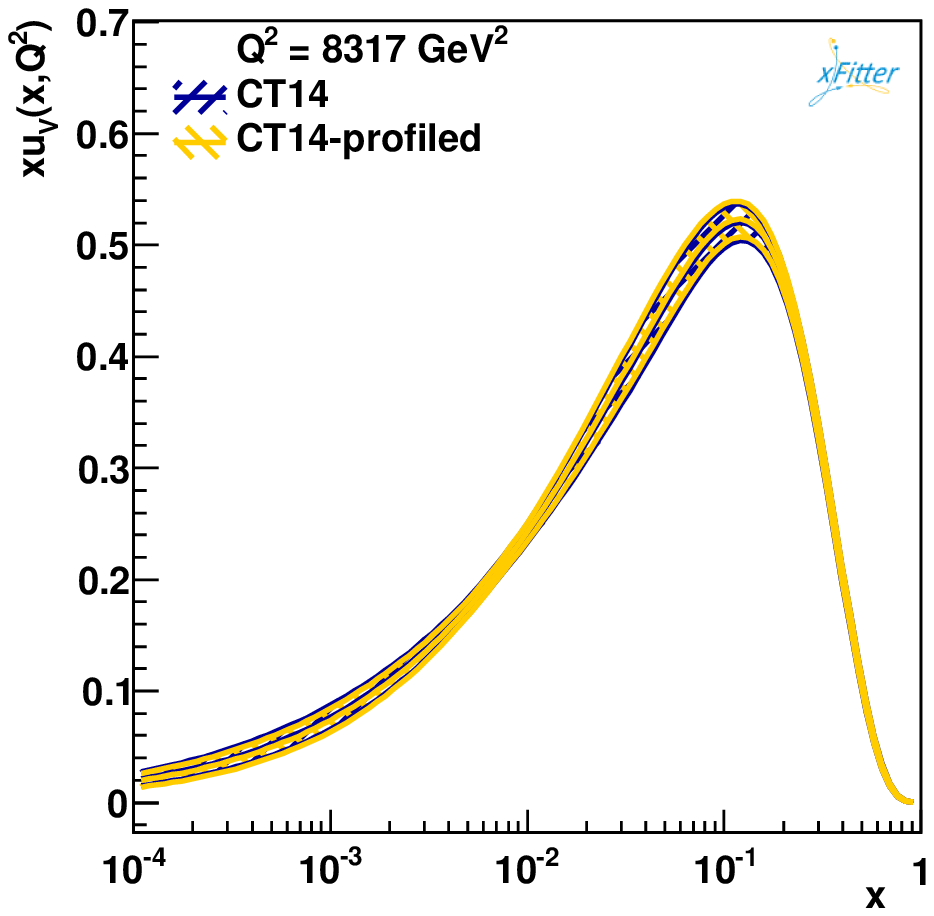}

	    \includegraphics[scale = 0.35]{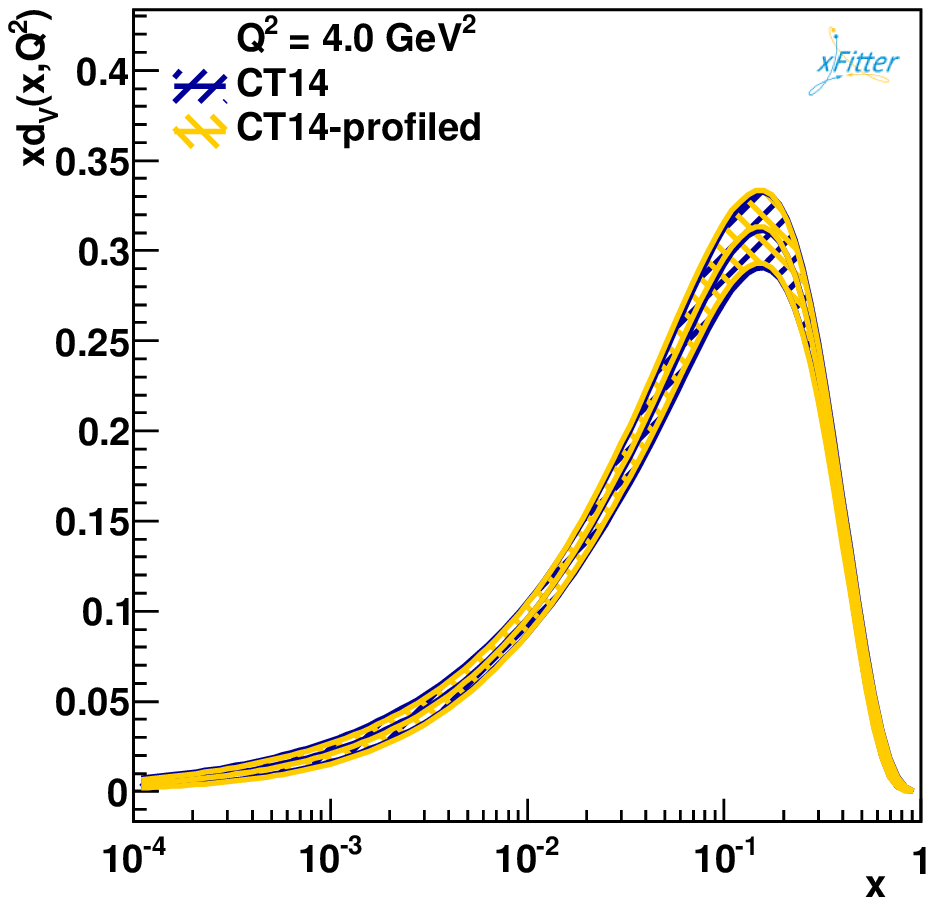}
	    \includegraphics[scale = 0.35]{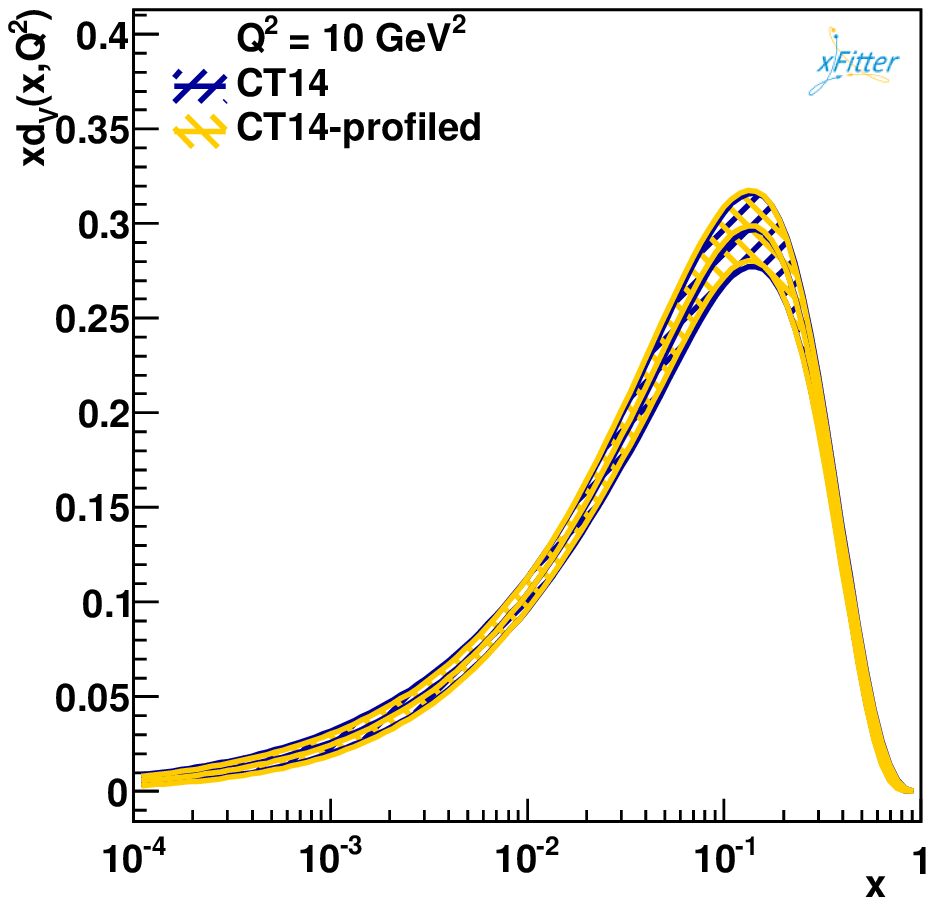}
	    \includegraphics[scale = 0.35]{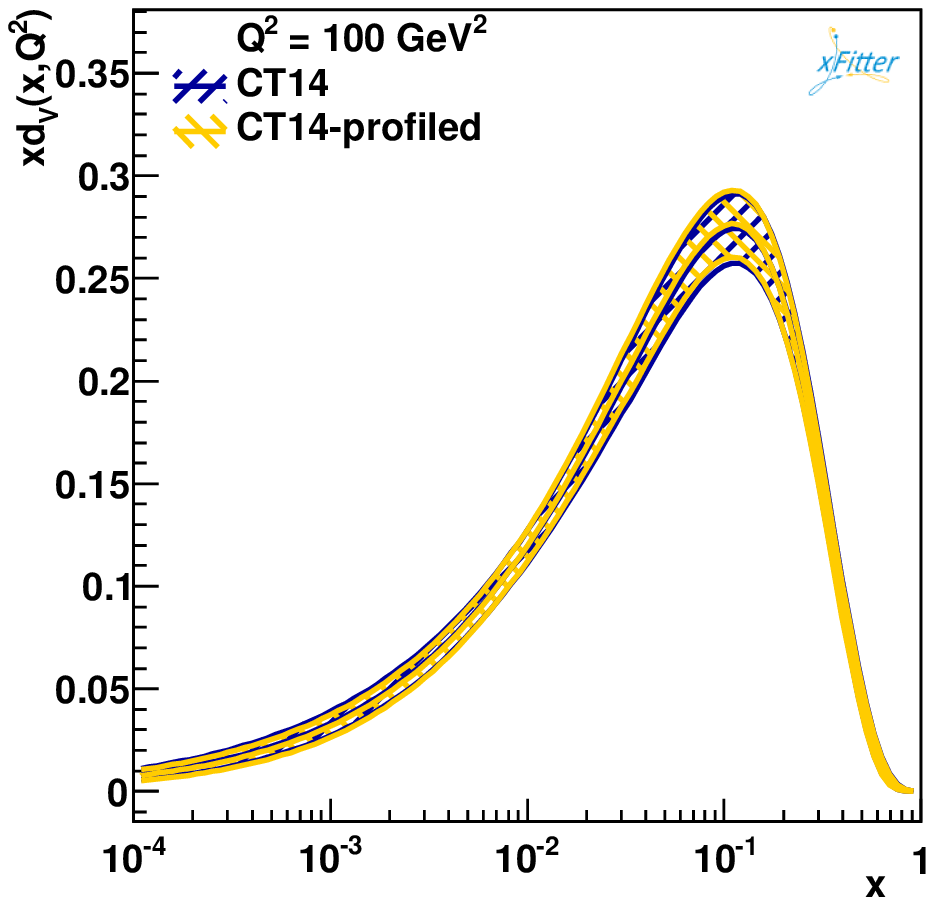}
	    \includegraphics[scale = 0.35]{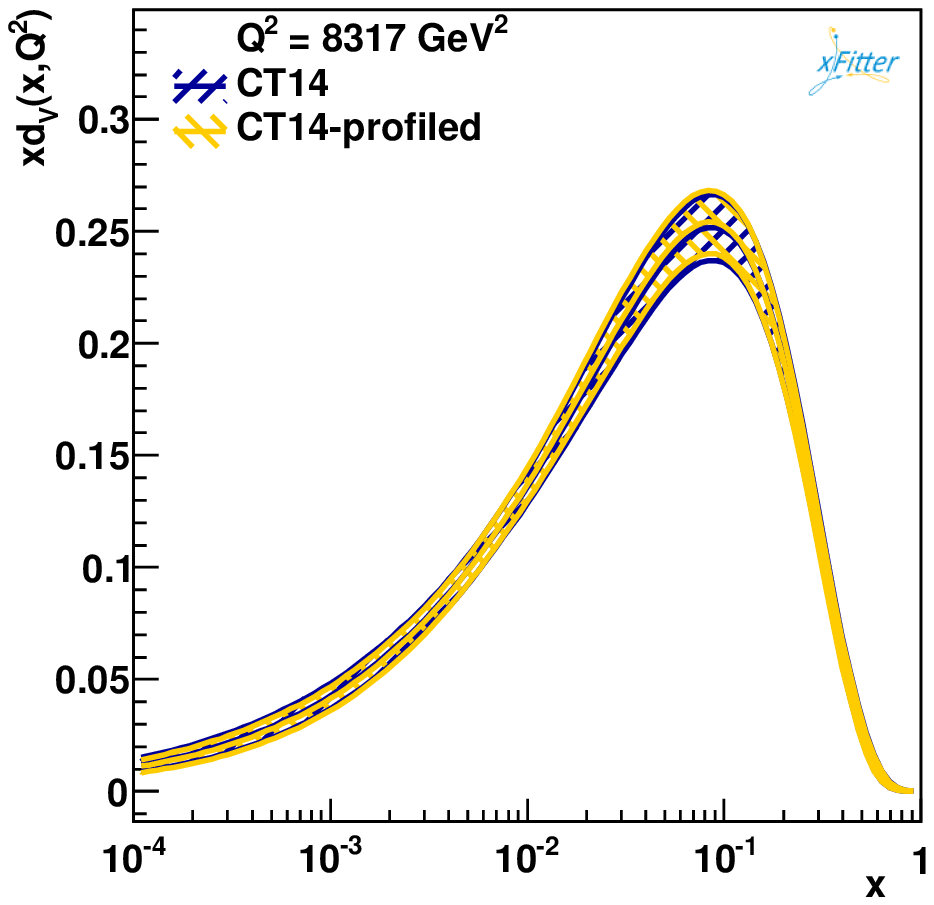}

	    \includegraphics[scale = 0.35]{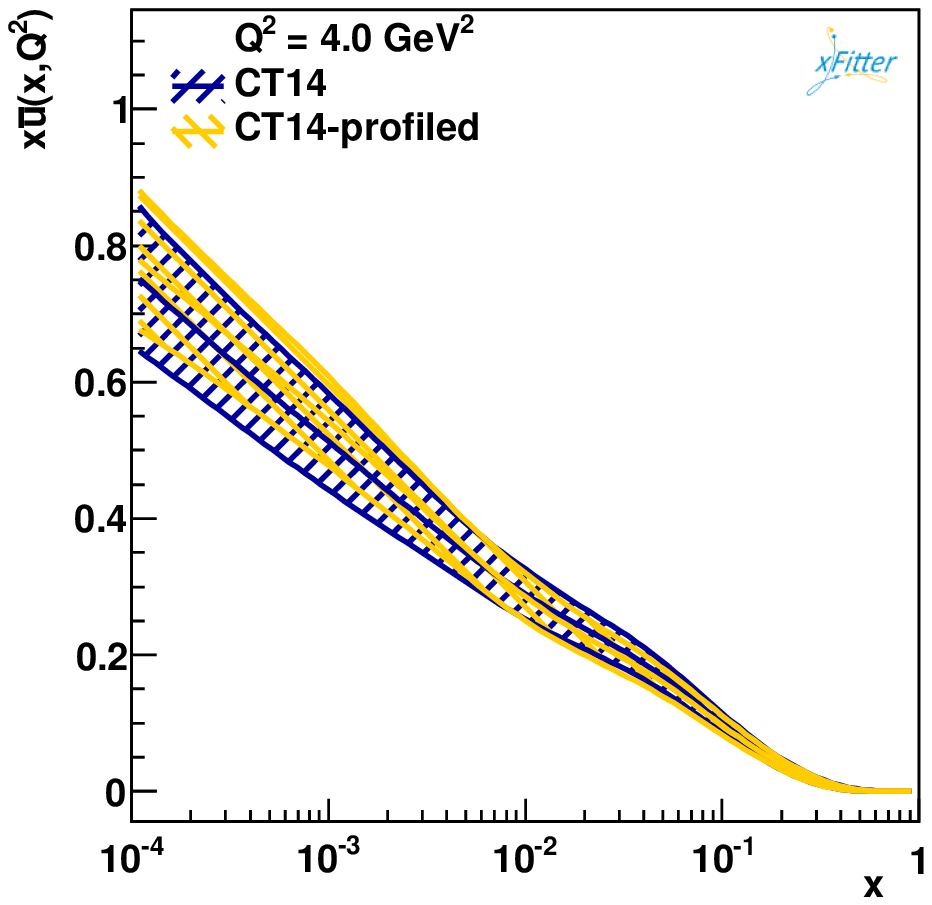}	    	    
	    \includegraphics[scale = 0.35]{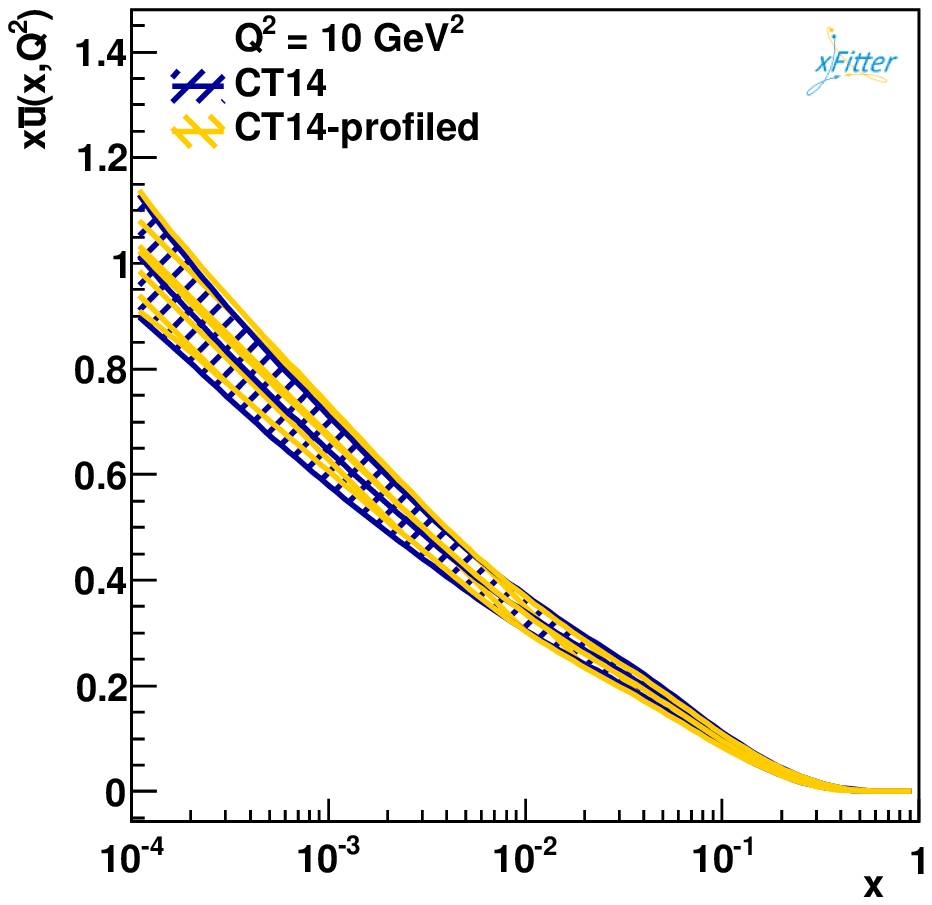}
	    \includegraphics[scale = 0.35]{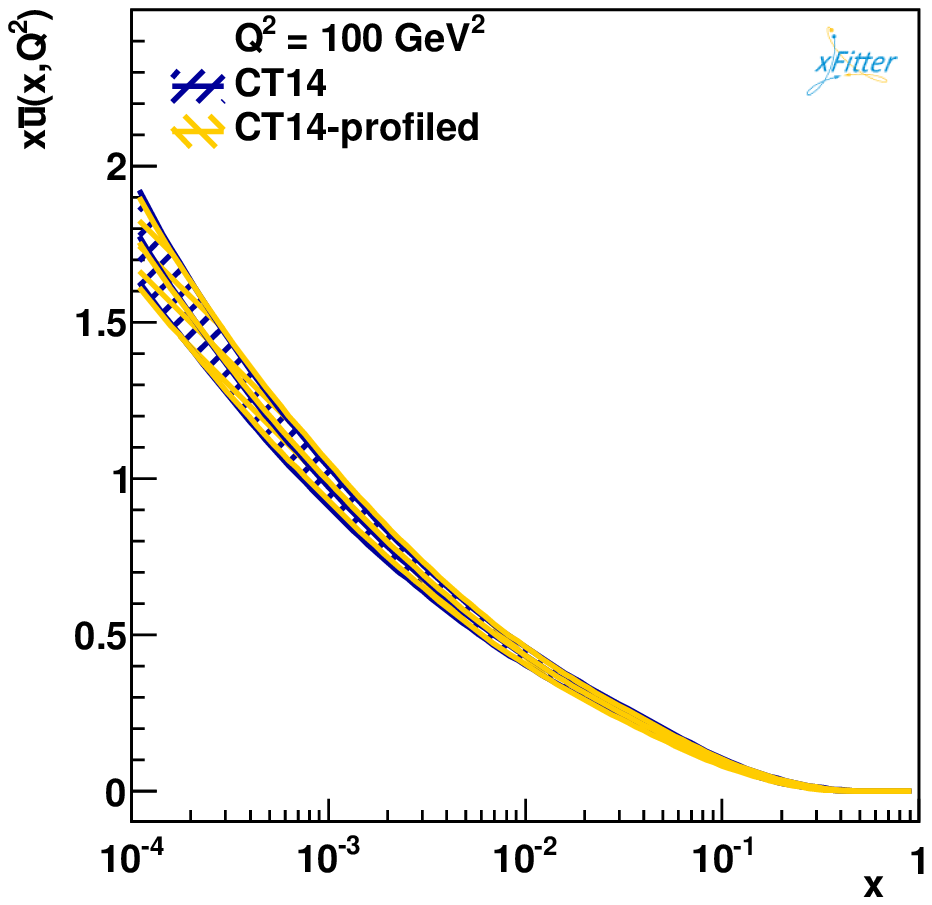}
	    \includegraphics[scale = 0.35]{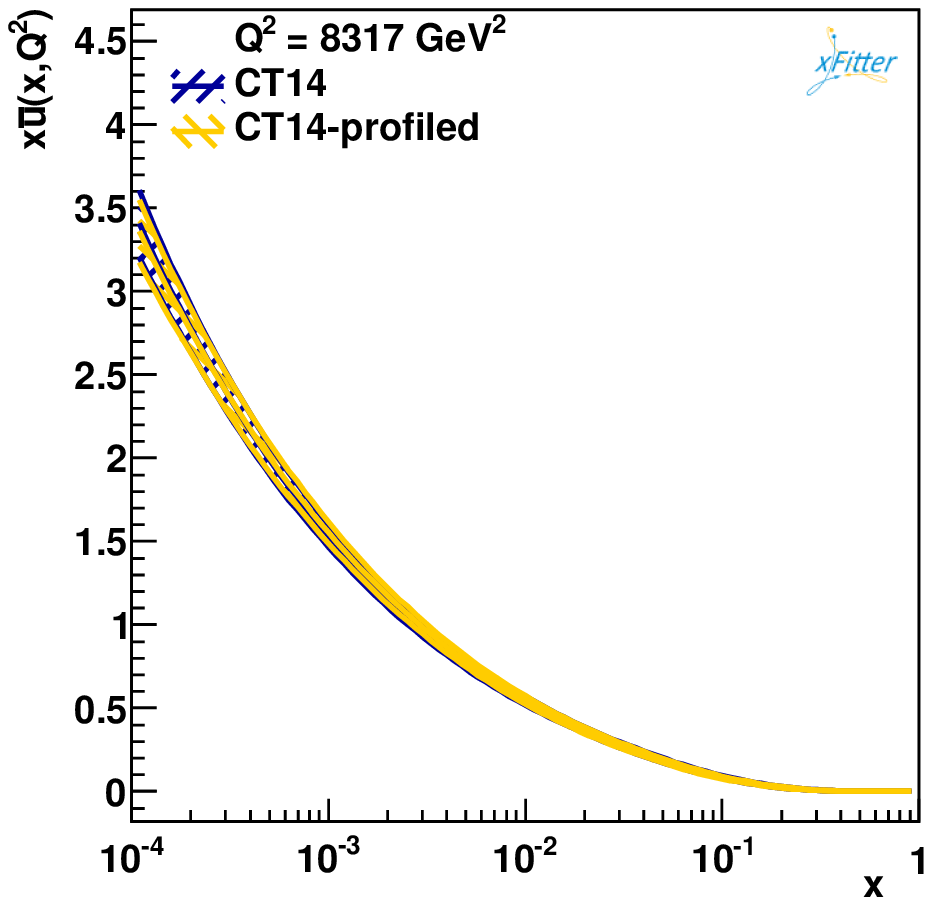}

	    \includegraphics[scale = 0.35]{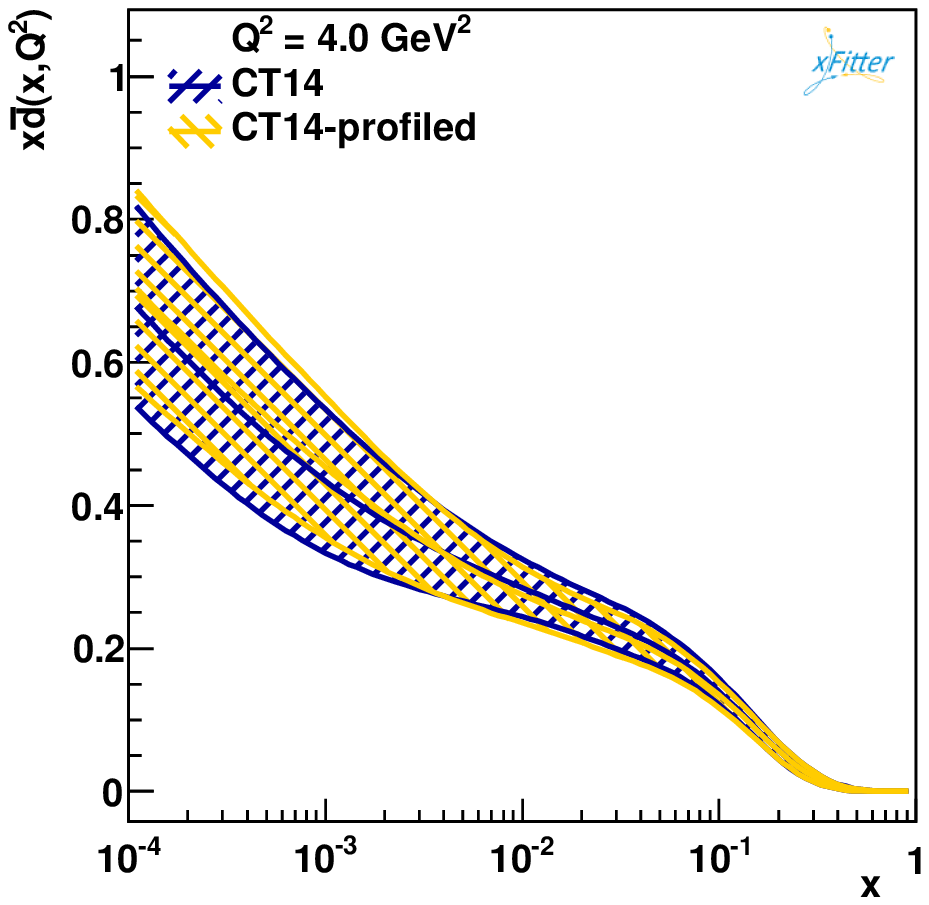}	    	    
	    \includegraphics[scale = 0.35]{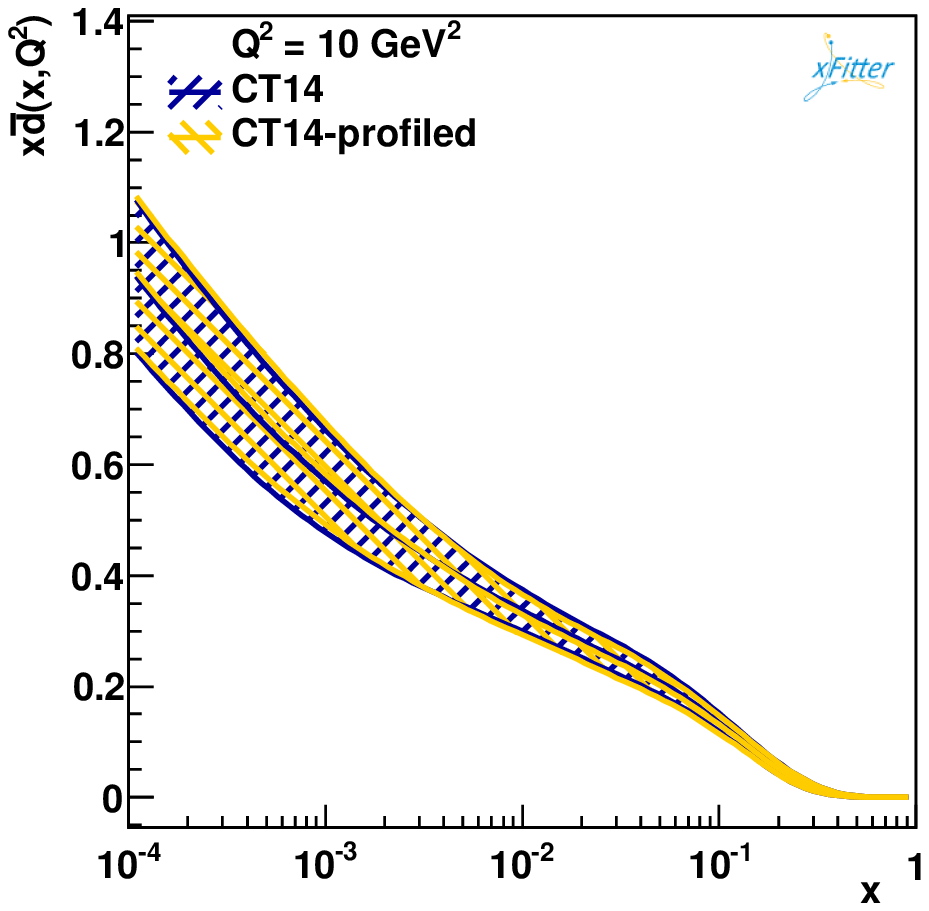}
	    \includegraphics[scale = 0.35]{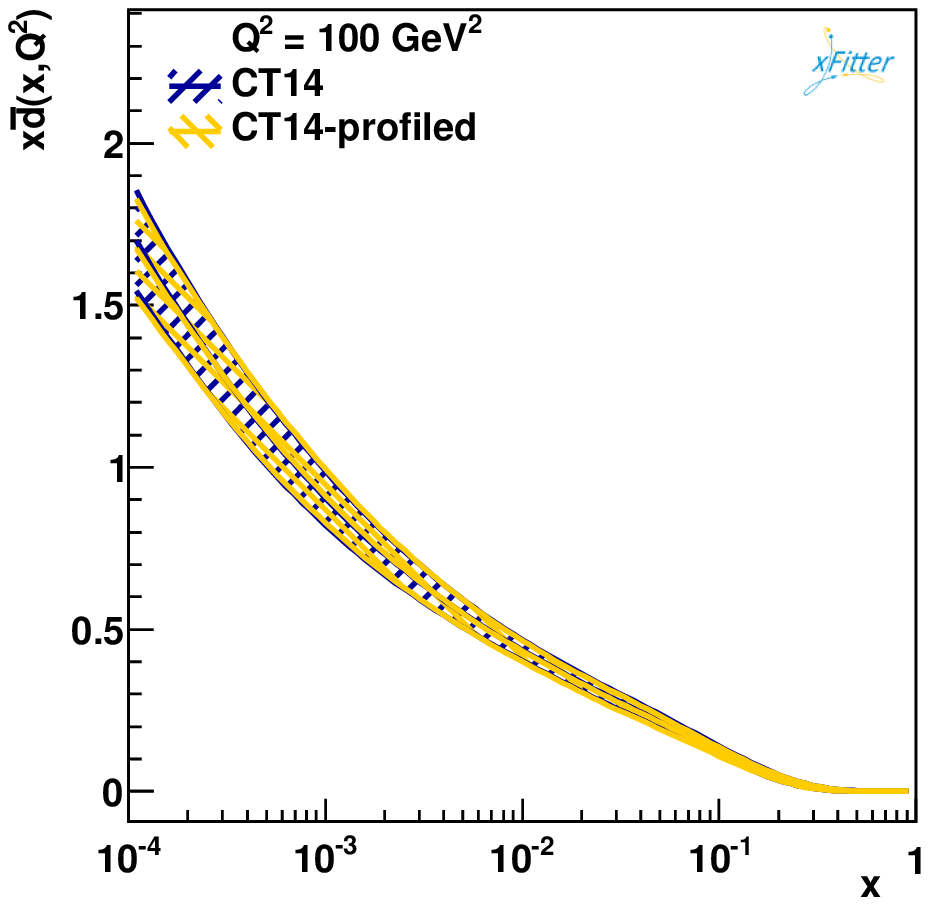}
	    \includegraphics[scale = 0.35]{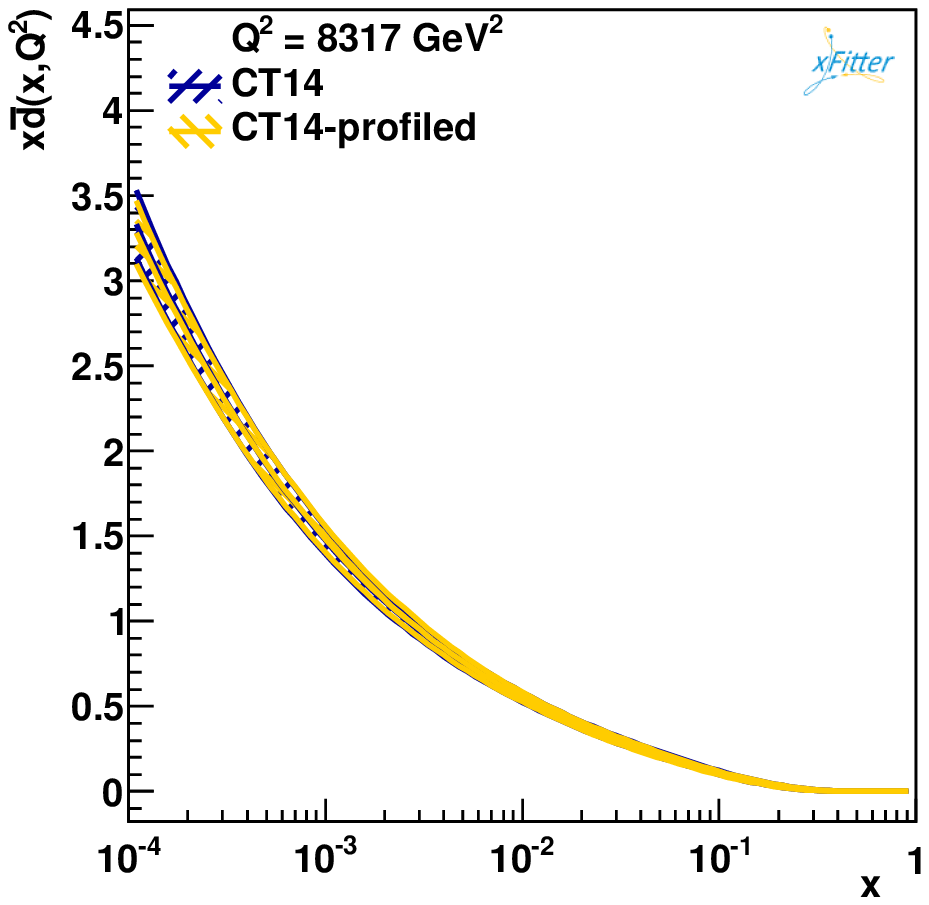}

	    \includegraphics[scale = 0.35]{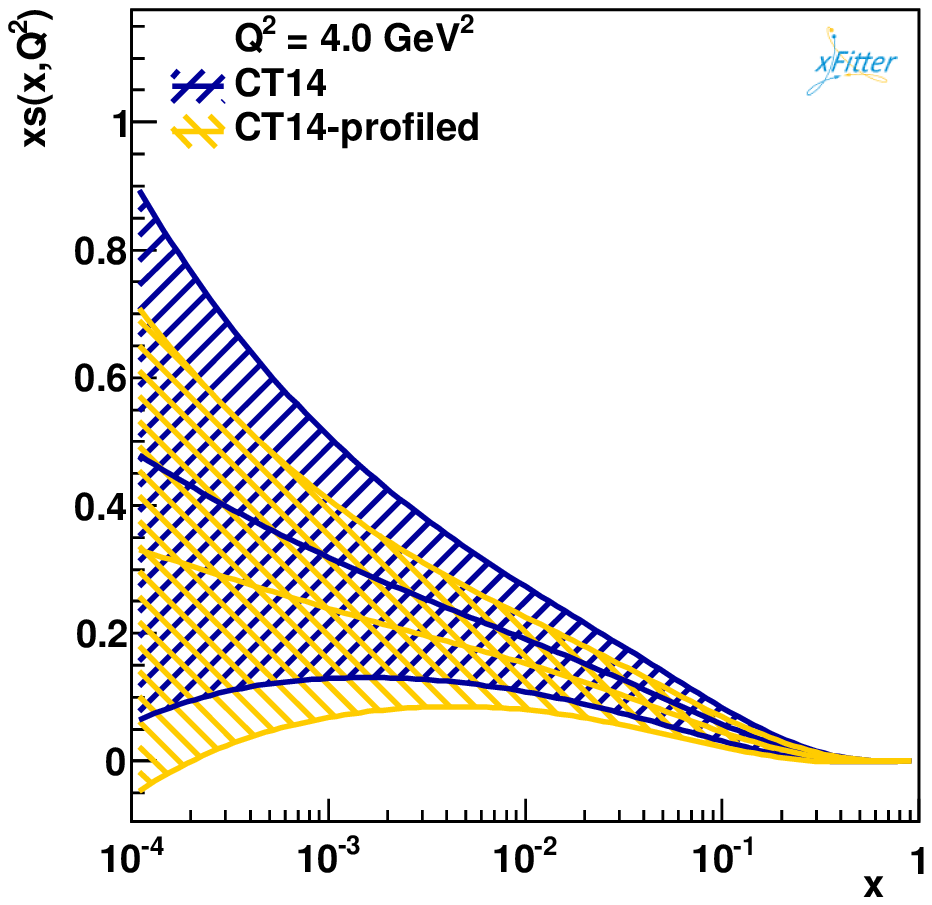}        
        	\includegraphics[scale = 0.35]{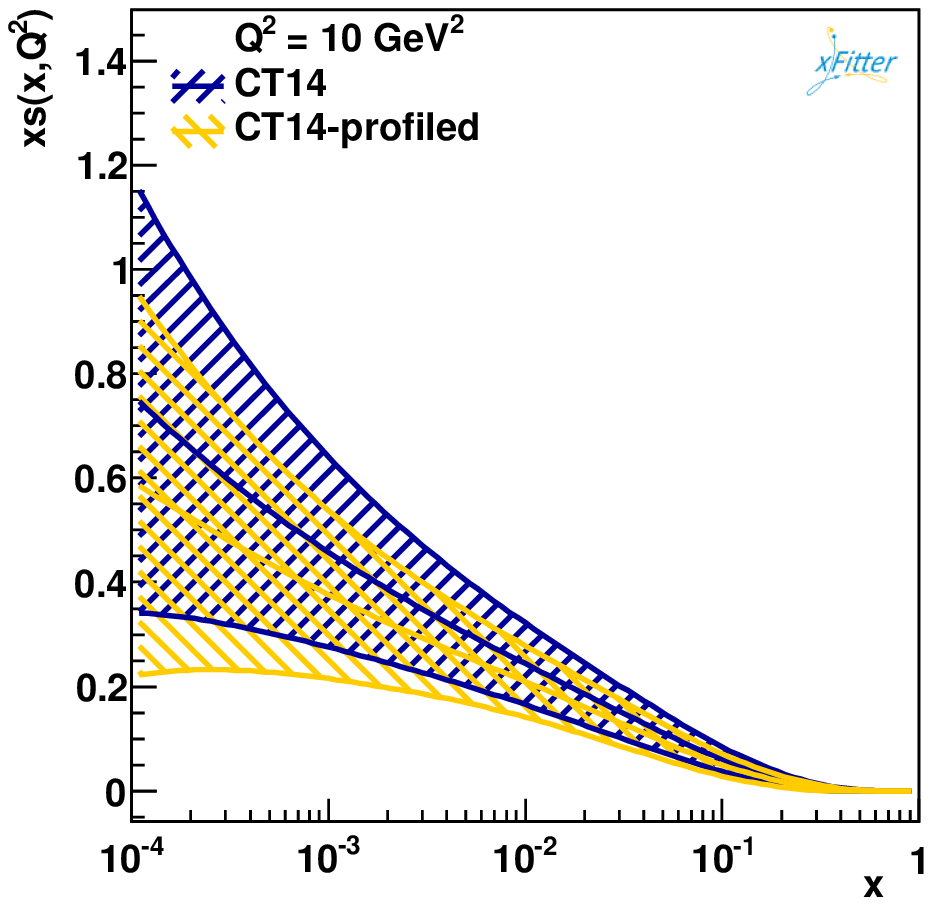}
	    \includegraphics[scale = 0.35]{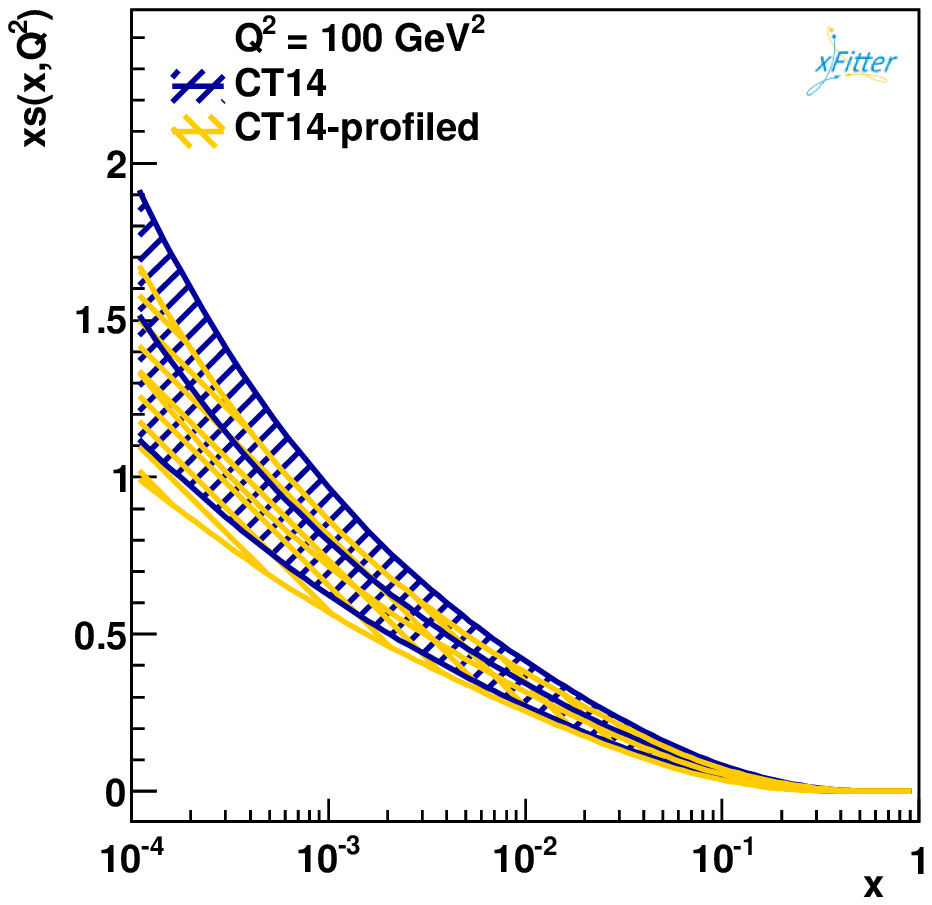}
	    \includegraphics[scale = 0.35]{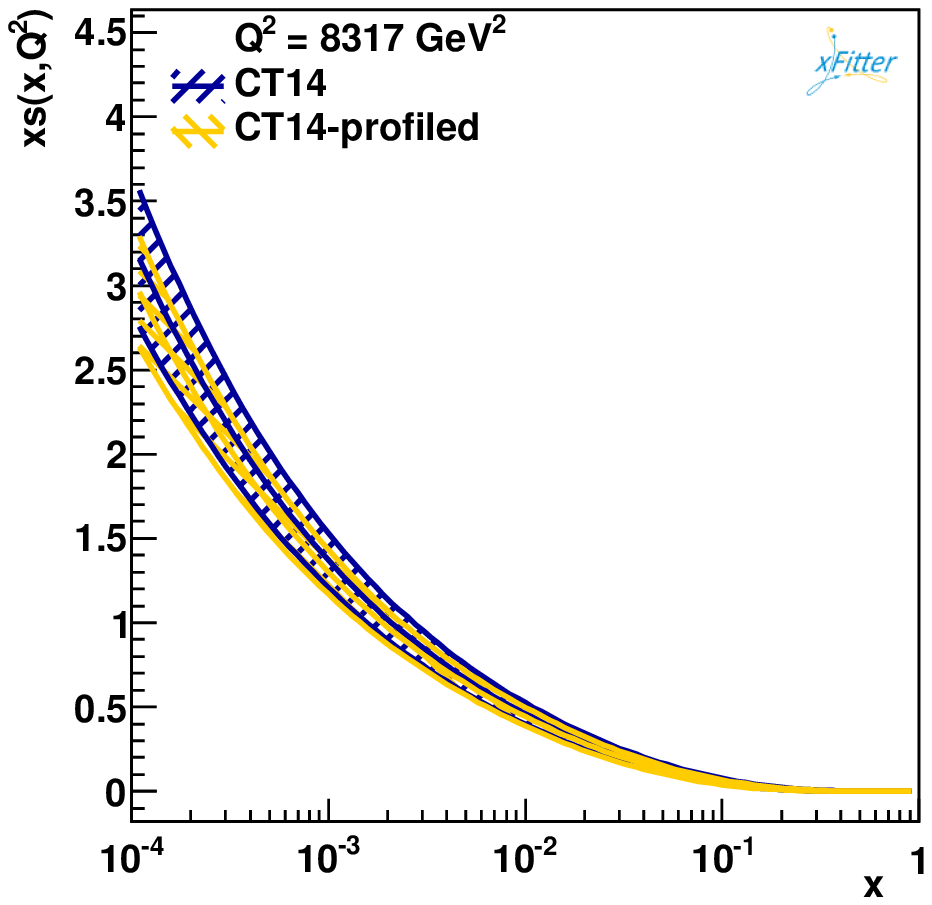}

	    \includegraphics[scale = 0.35]{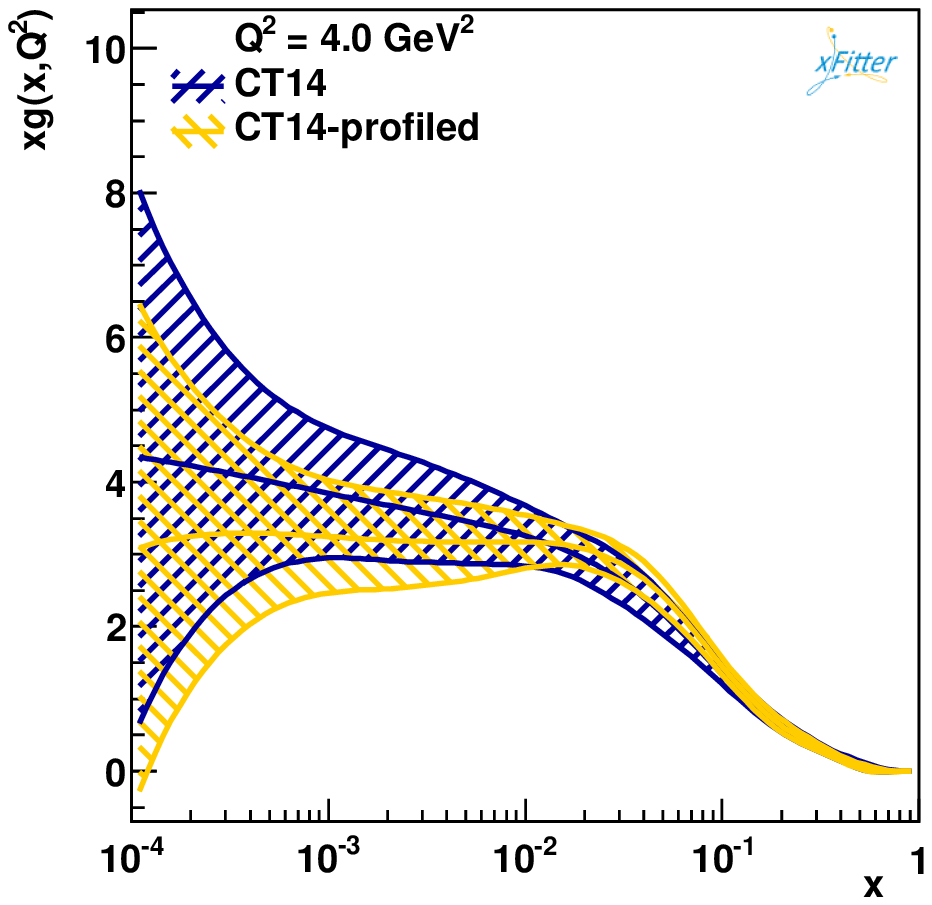}        
        	\includegraphics[scale = 0.35]{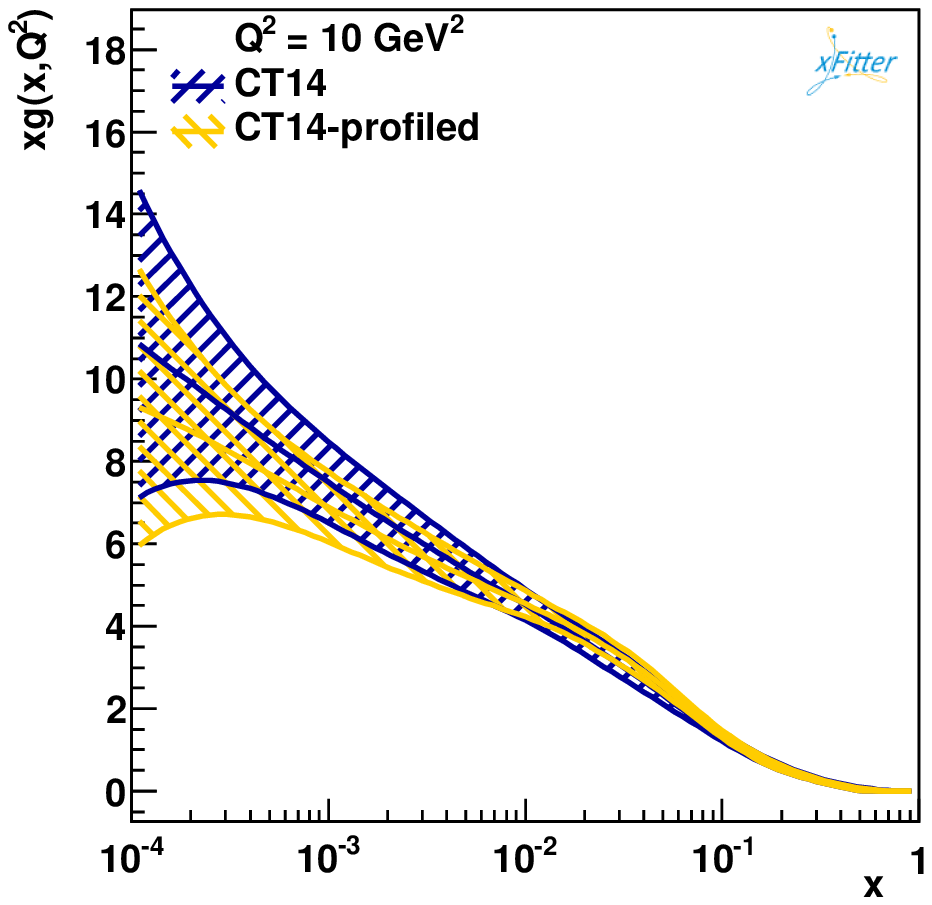}
	    \includegraphics[scale = 0.35]{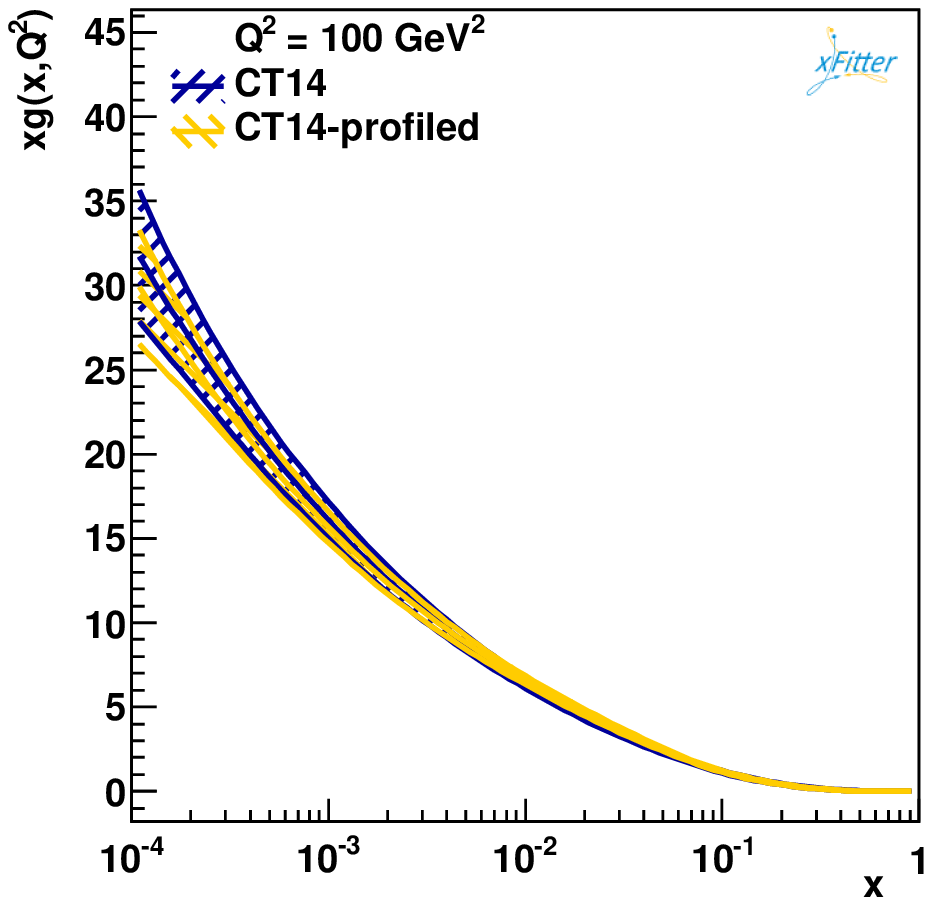}
	    \includegraphics[scale = 0.35]{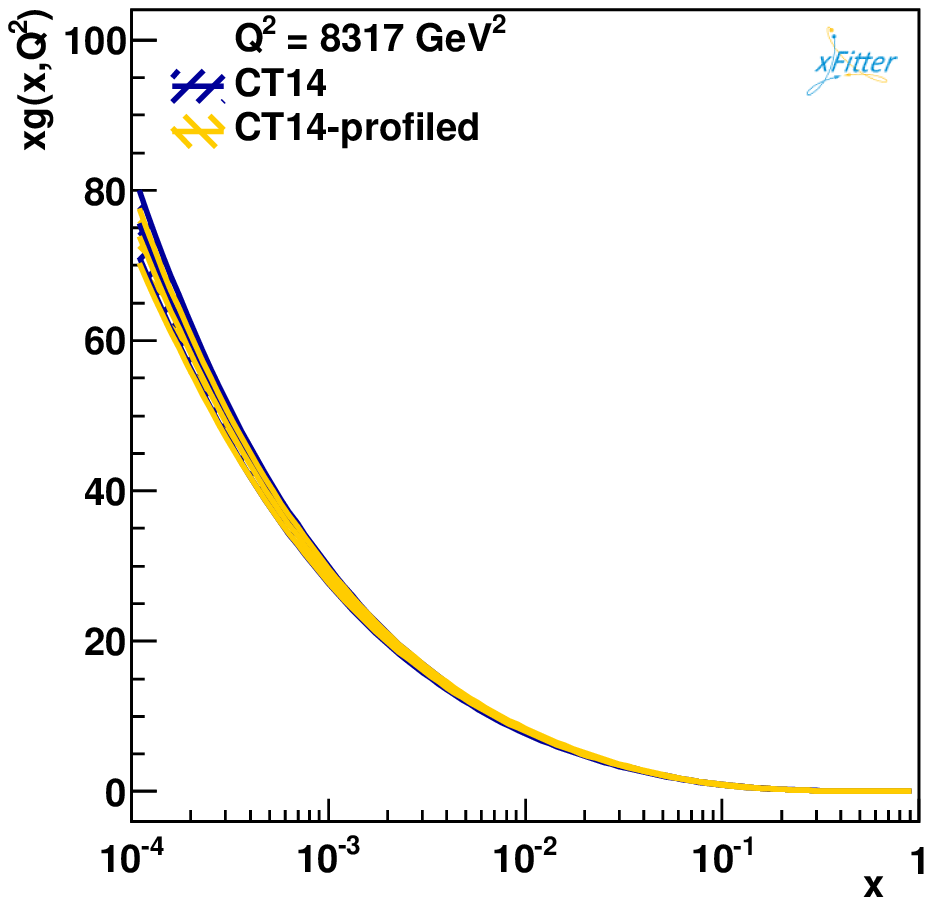}

		\caption{The parton distribution of $xu_v, xd_v, x\bar{u}, x\bar{d}, xs$, and $xg$ extracted from CT14 \cite{Dulat:2015mca} PDFs as a function of $x$ at 4, 10, 100, and 8317 GeV$^2$. The results obtained after the profiling procedure compared with corresponding same features before profiling. Newly added top quark data obviously affected distributions of $xs$, and $xg$.}
		\label{fig:partonDisCT14}
	\end{center}
\end{figure}

\begin{figure}[!htb]
	\begin{center}

 	    \includegraphics[scale = 0.35]{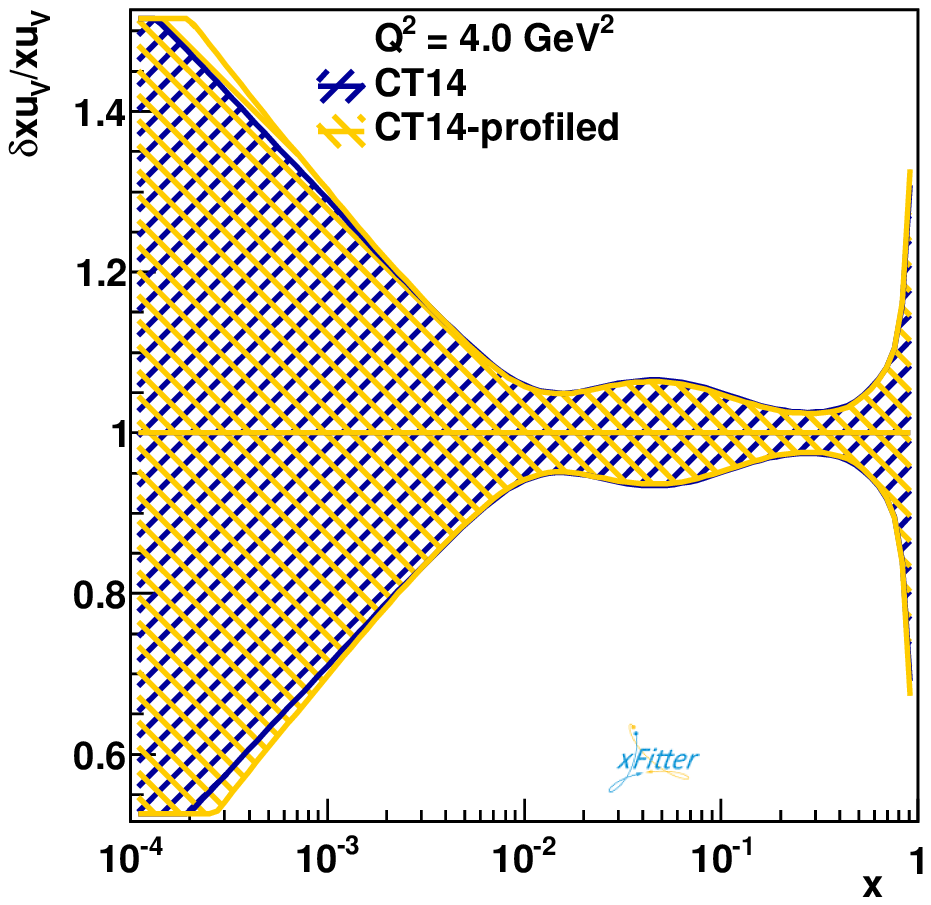}	    	    
	    \includegraphics[scale = 0.35]{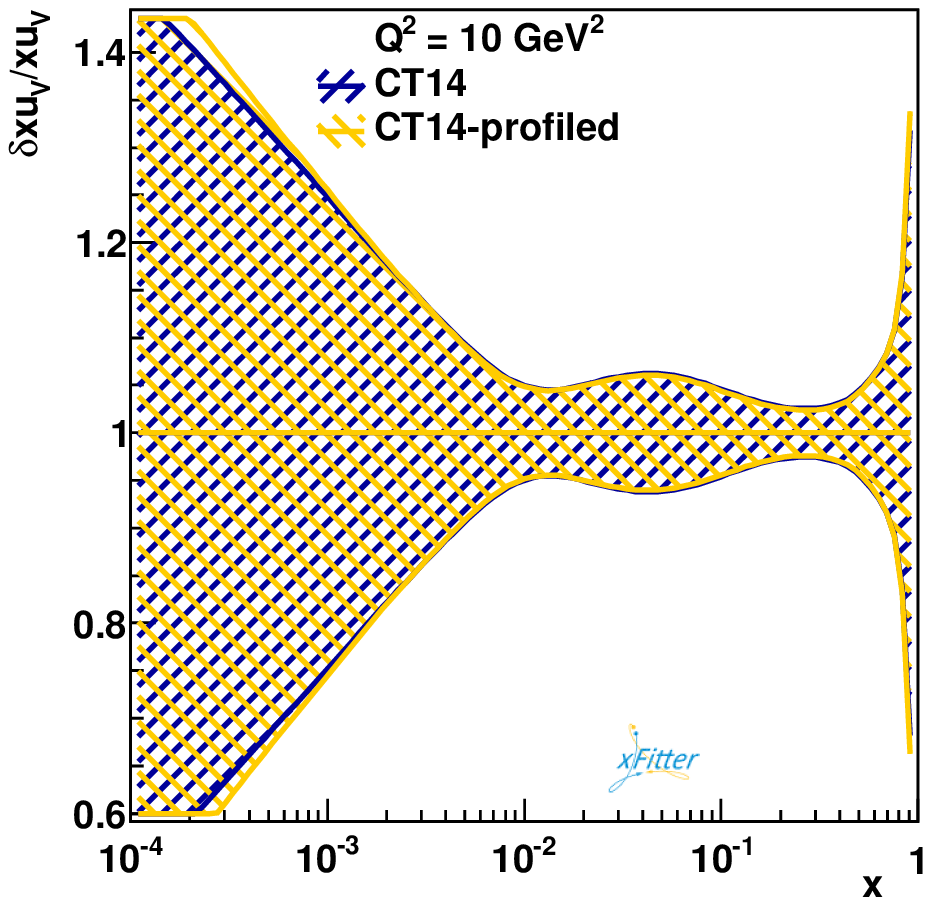}
	    \includegraphics[scale = 0.35]{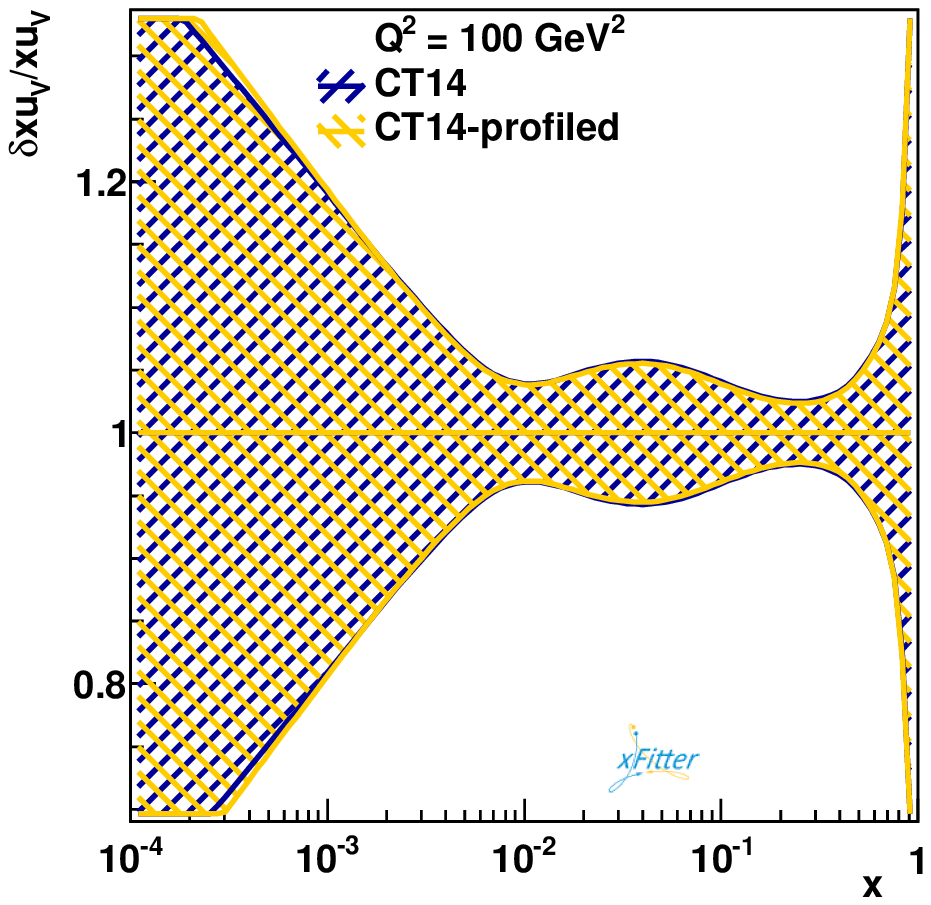}
	    \includegraphics[scale = 0.35]{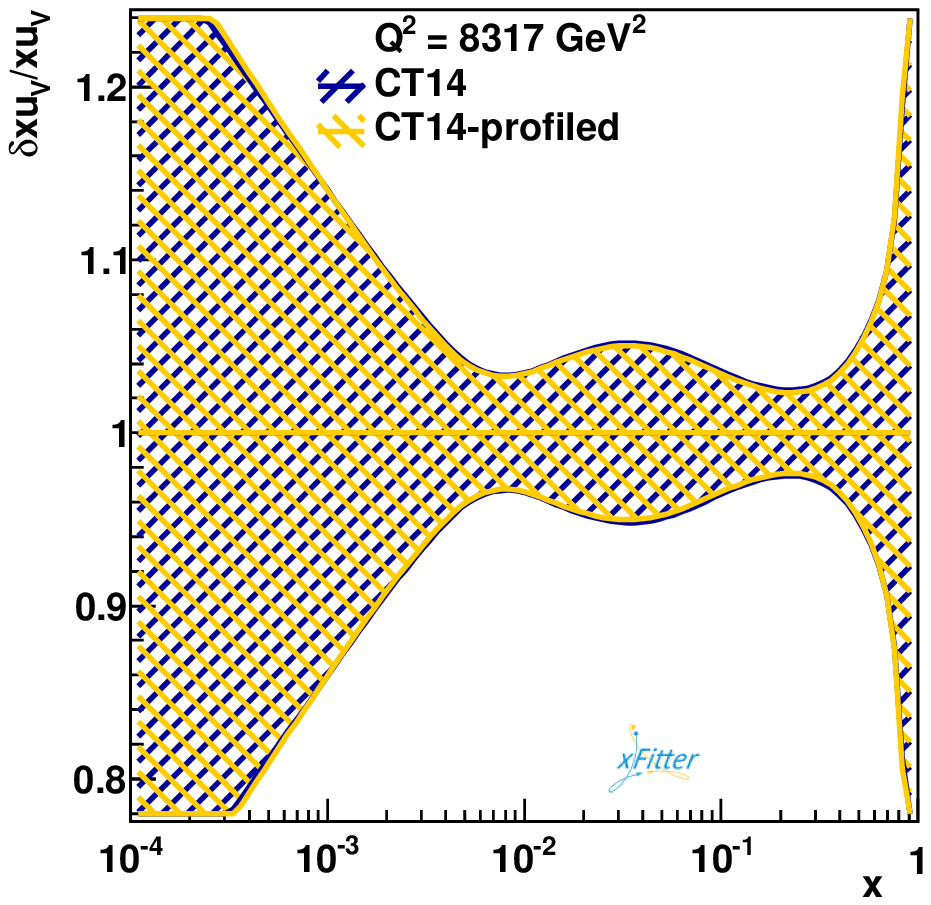}

	    \includegraphics[scale = 0.35]{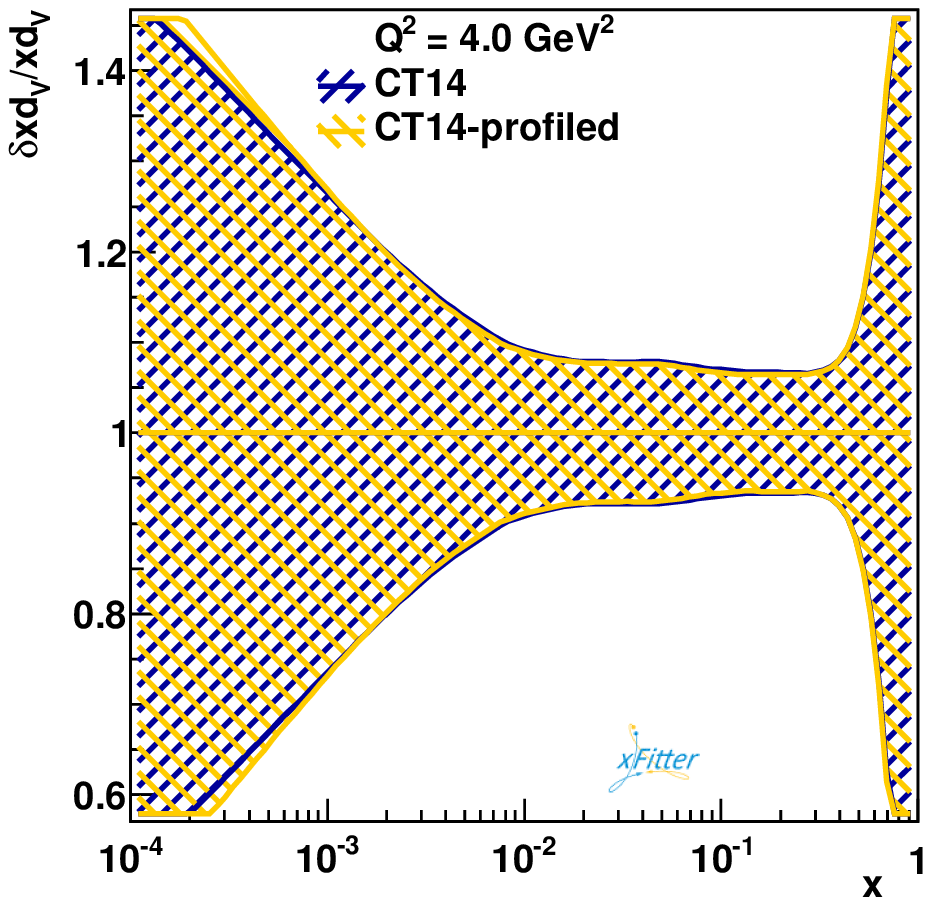}
	    \includegraphics[scale = 0.35]{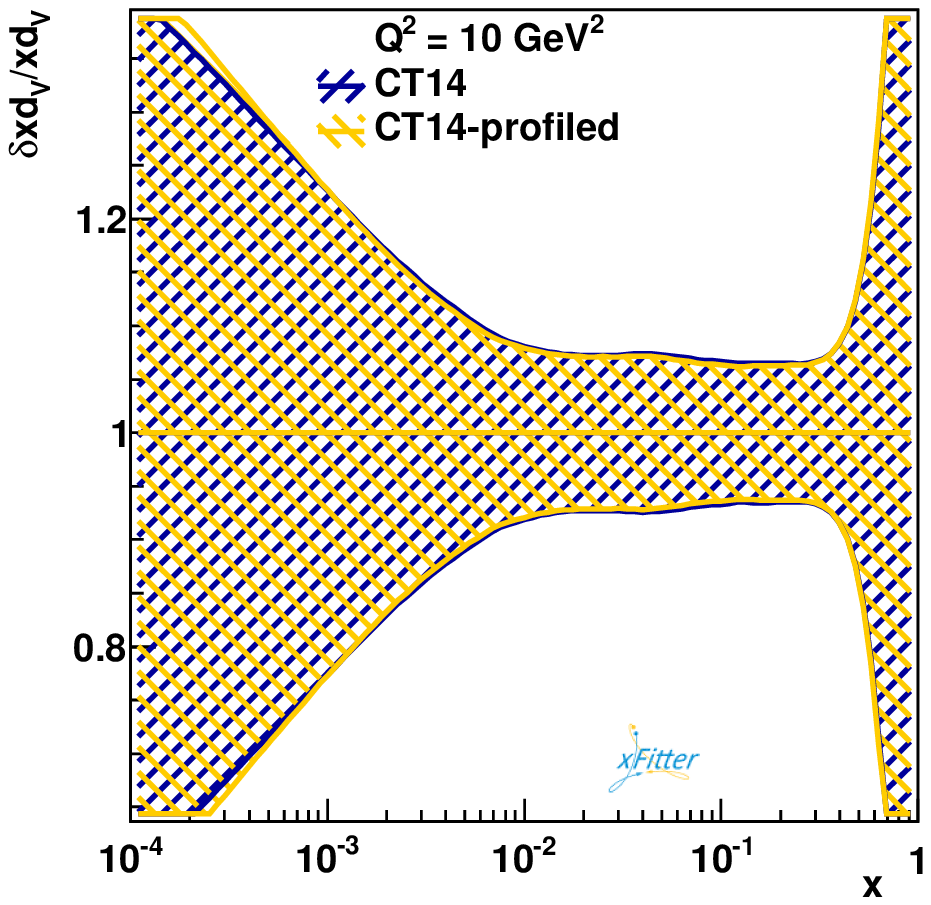}
	    \includegraphics[scale = 0.35]{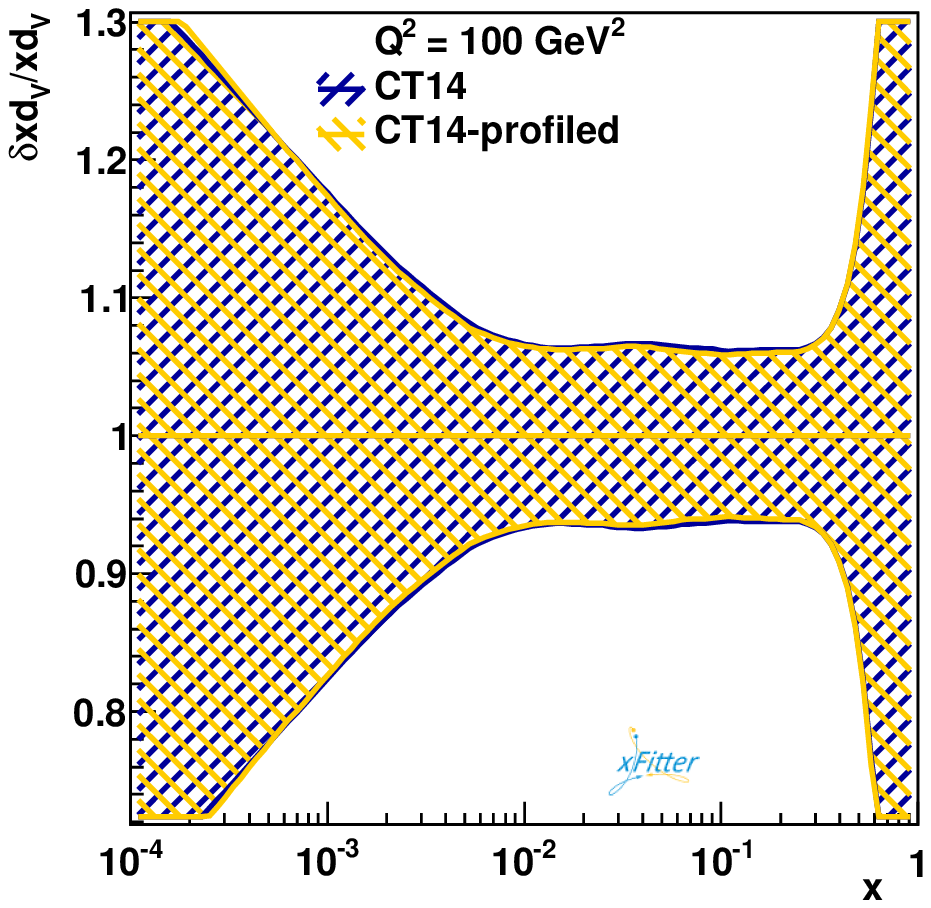}
	    \includegraphics[scale = 0.35]{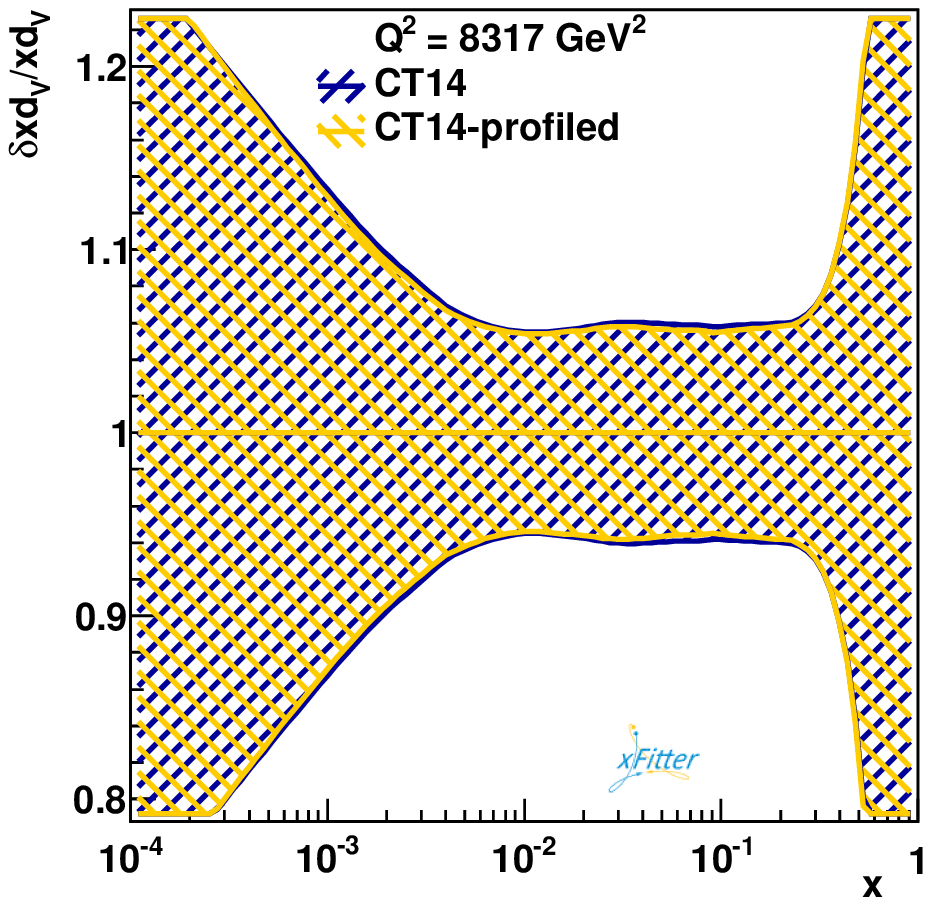}

	    \includegraphics[scale = 0.35]{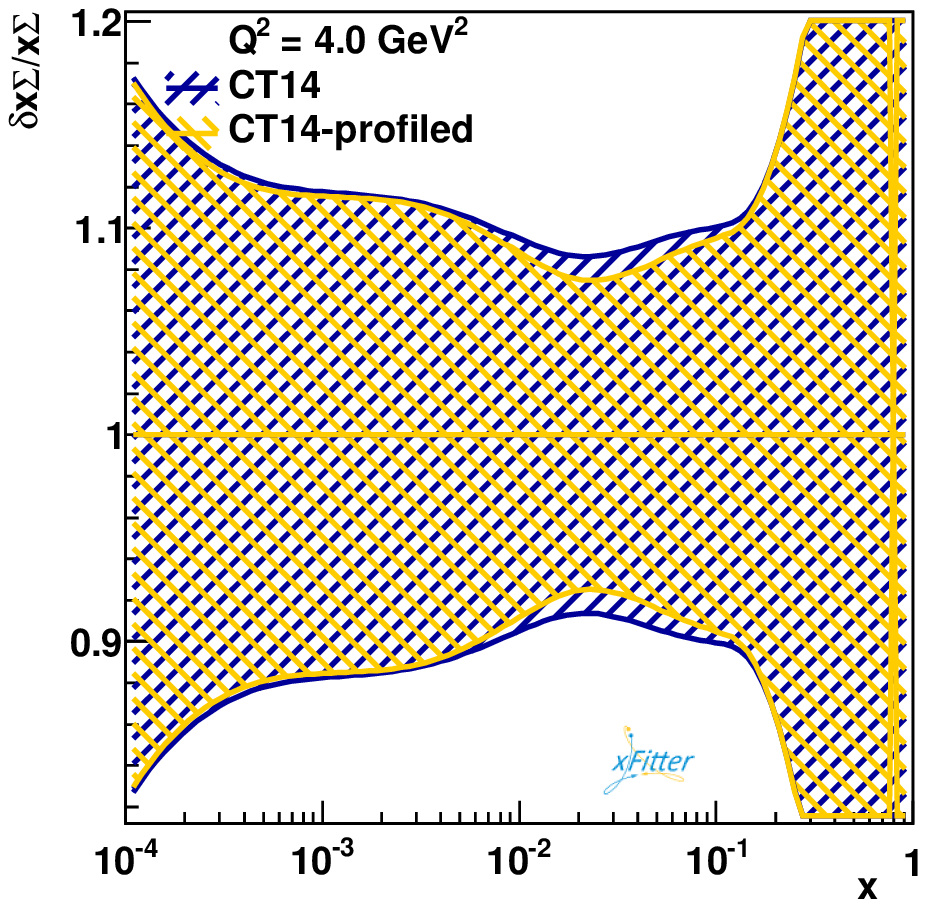}        
      	\includegraphics[scale = 0.35]{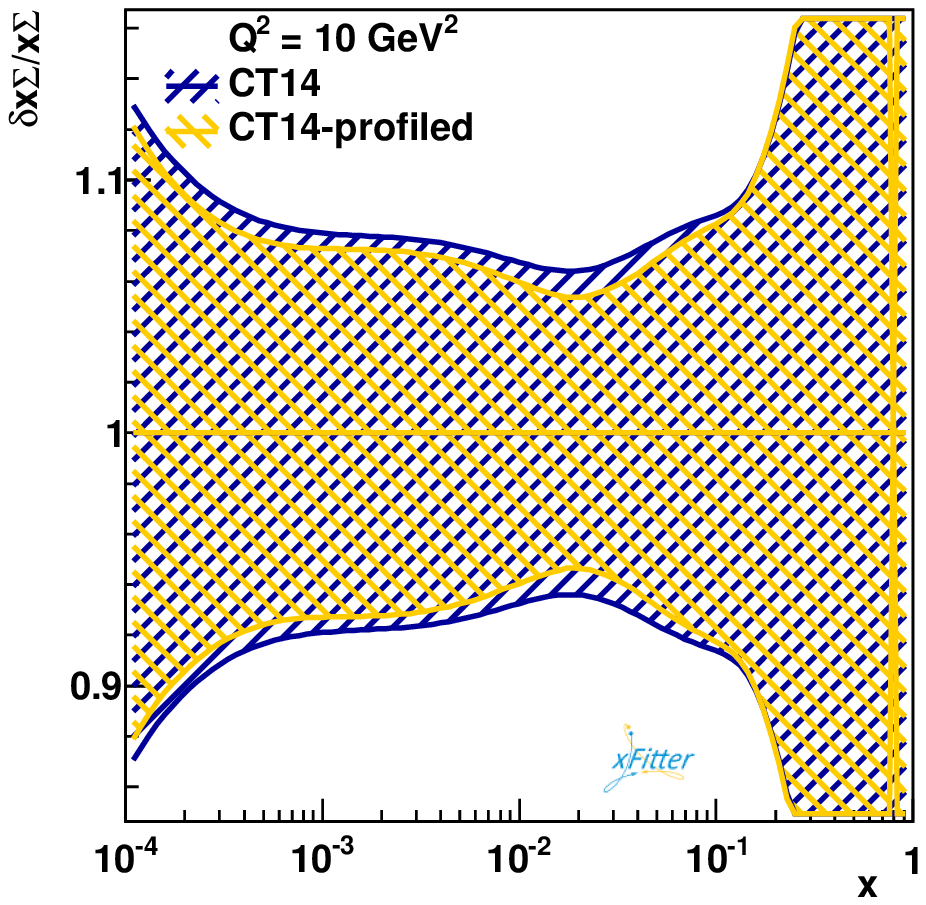}
	    \includegraphics[scale = 0.35]{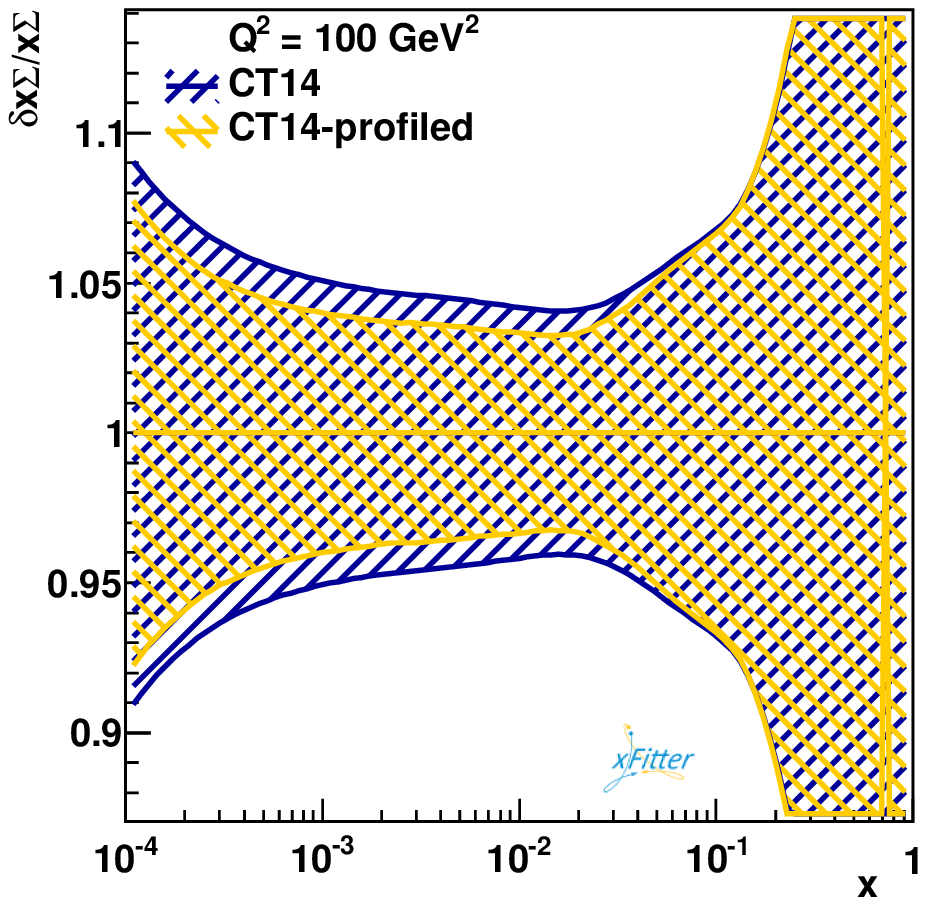}
	    \includegraphics[scale = 0.35]{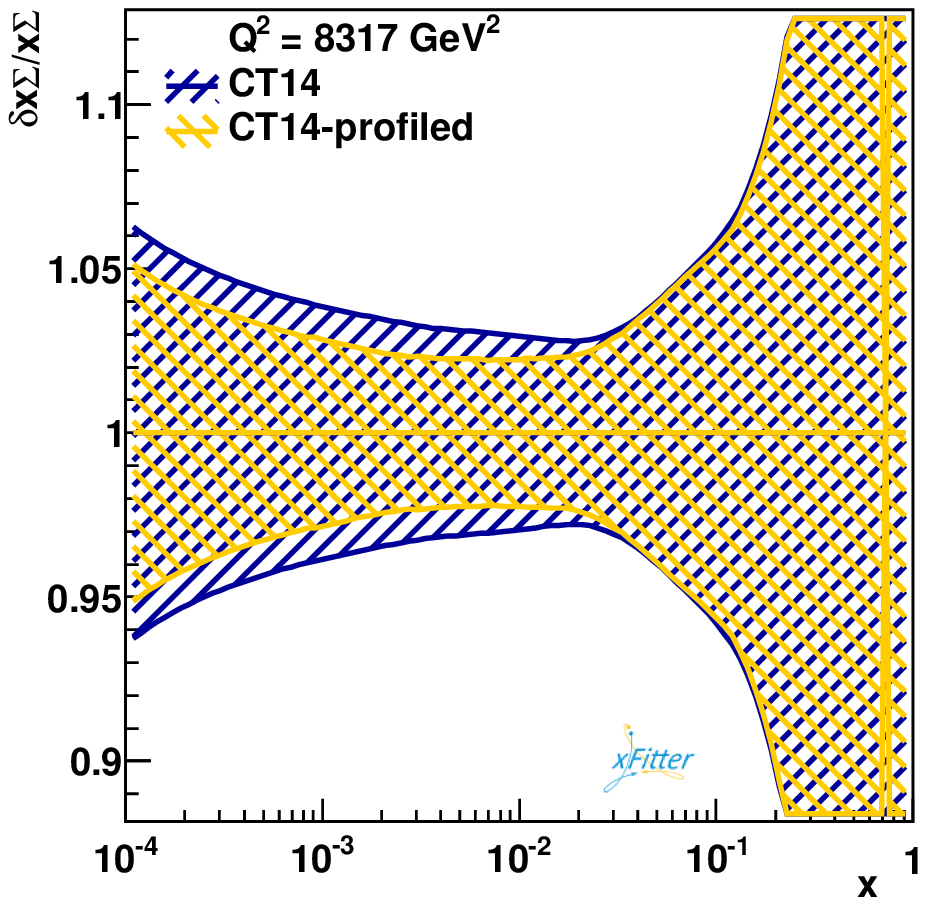}

	    \includegraphics[scale = 0.35]{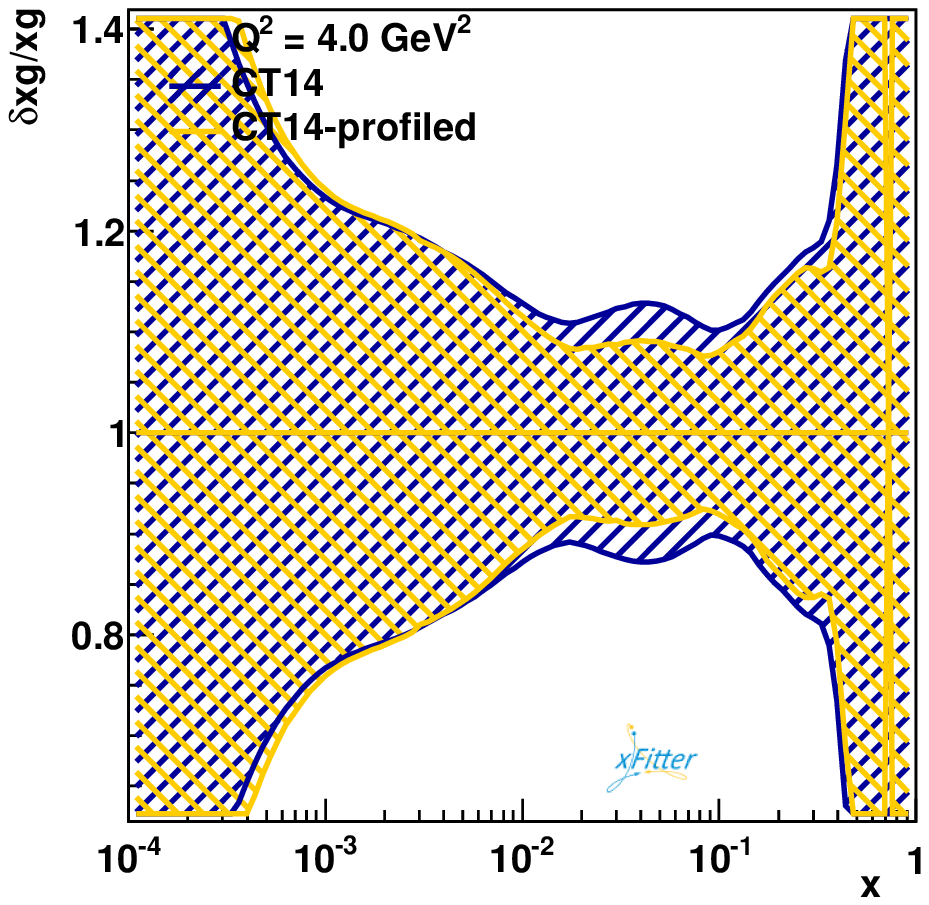}        
        	\includegraphics[scale = 0.35]{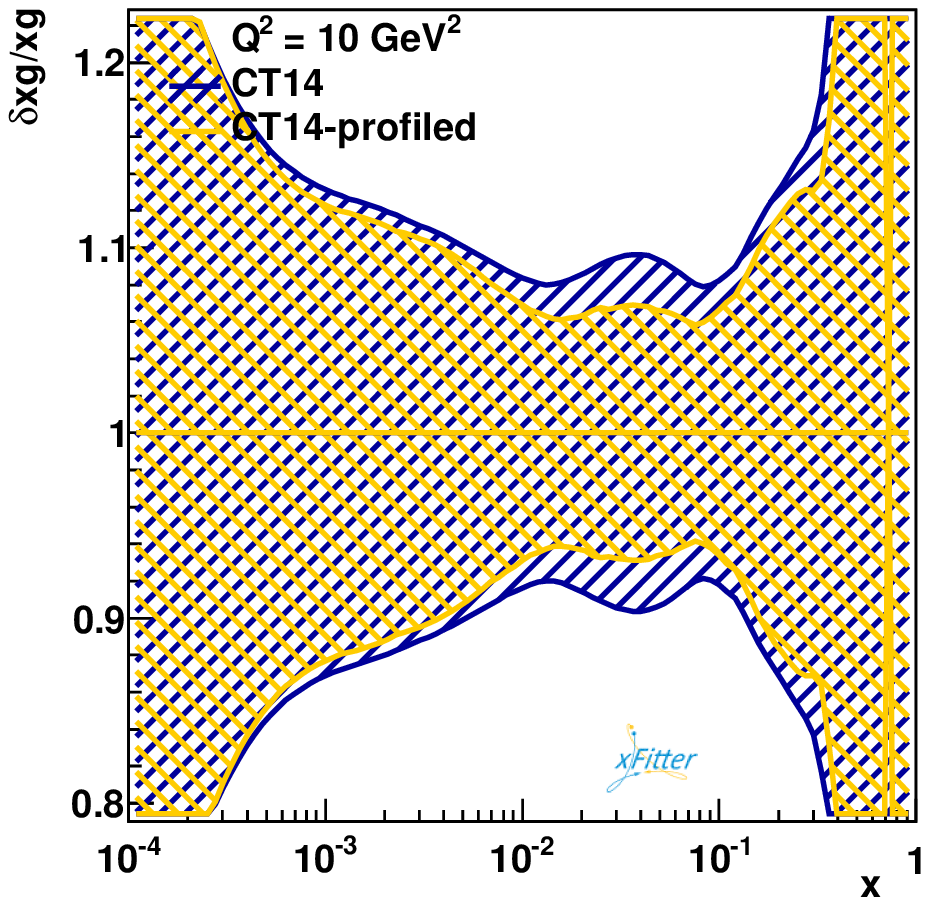}
	    \includegraphics[scale = 0.35]{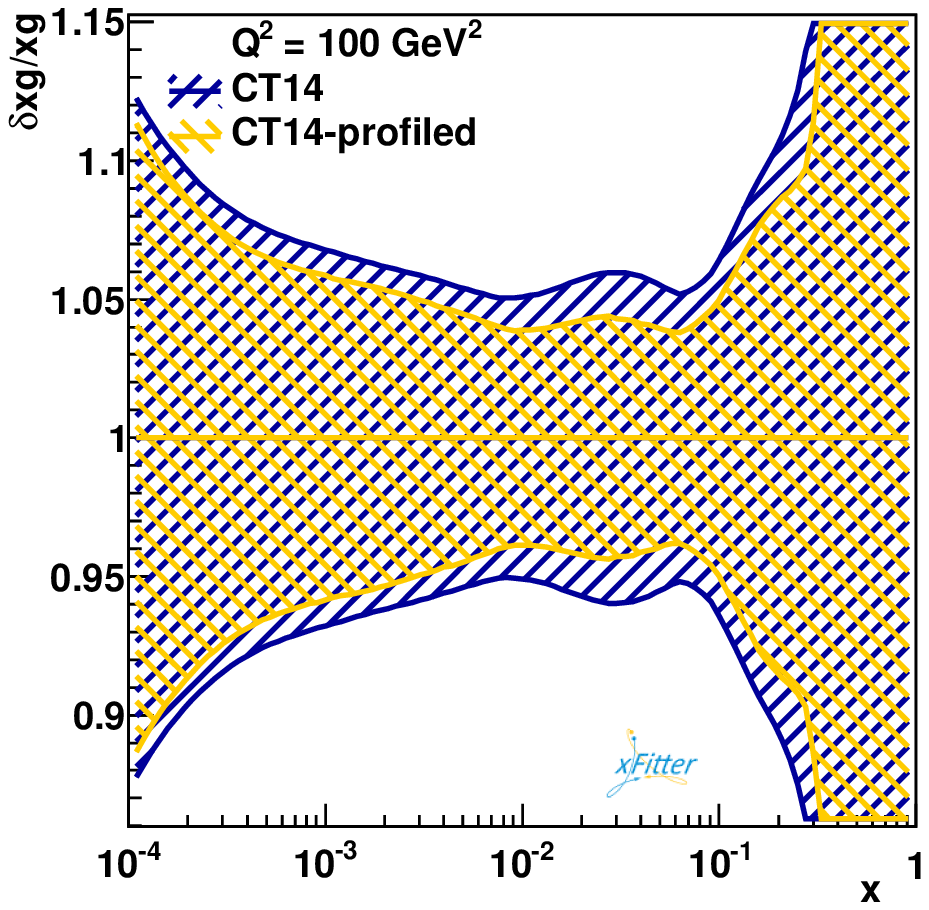}
	    \includegraphics[scale = 0.35]{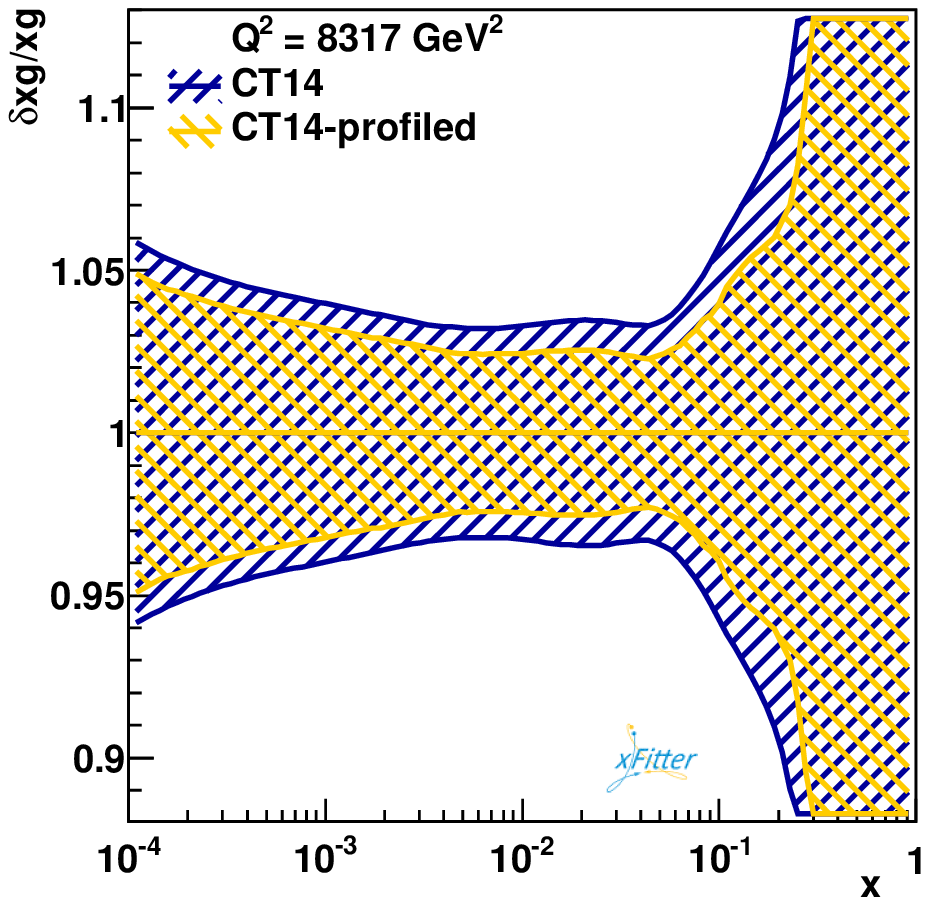}
		
\caption{The relative PDF uncertainties $\delta x u_v/xu_v$, $\delta x d_v/xd_v$, $\delta x \Sigma/x\Sigma$, and $\delta x g/xg$ extracted from CT14 \cite{Dulat:2015mca} PDFs as a function of $x$ at 4, 10, 100, and 8317 GeV$^2$. The results obtained after the profiling procedure compared with corresponding same features  before profiling. Newly added top quark data obviously constrained distributions of $\delta x \Sigma/x\Sigma$, and $\delta x g/xg$.}
		\label{fig:partonRatioCT14}
	\end{center}
\end{figure}

The comparison between original and profiled parton distribution of $xu_v, xd_v, x\bar{u}, x\bar{d}, xs$, and $xg$ extracted from CT14 \cite{Dulat:2015mca} PDFs  are presented in Fig.~\ref{fig:partonDisCT14}. According to this figure, the new top quark cross section data provide significant constraints on the central values and their uncertainties  of $xs$, and $xg$ PDFs. 
In Fig.~\ref{fig:partonRatioCT14}, the most significant impact of new measurements  are observed only on the gluon PDF ratio. 

\begin{figure}[!htb]
	\begin{center}     	
        \includegraphics[scale = 0.35]{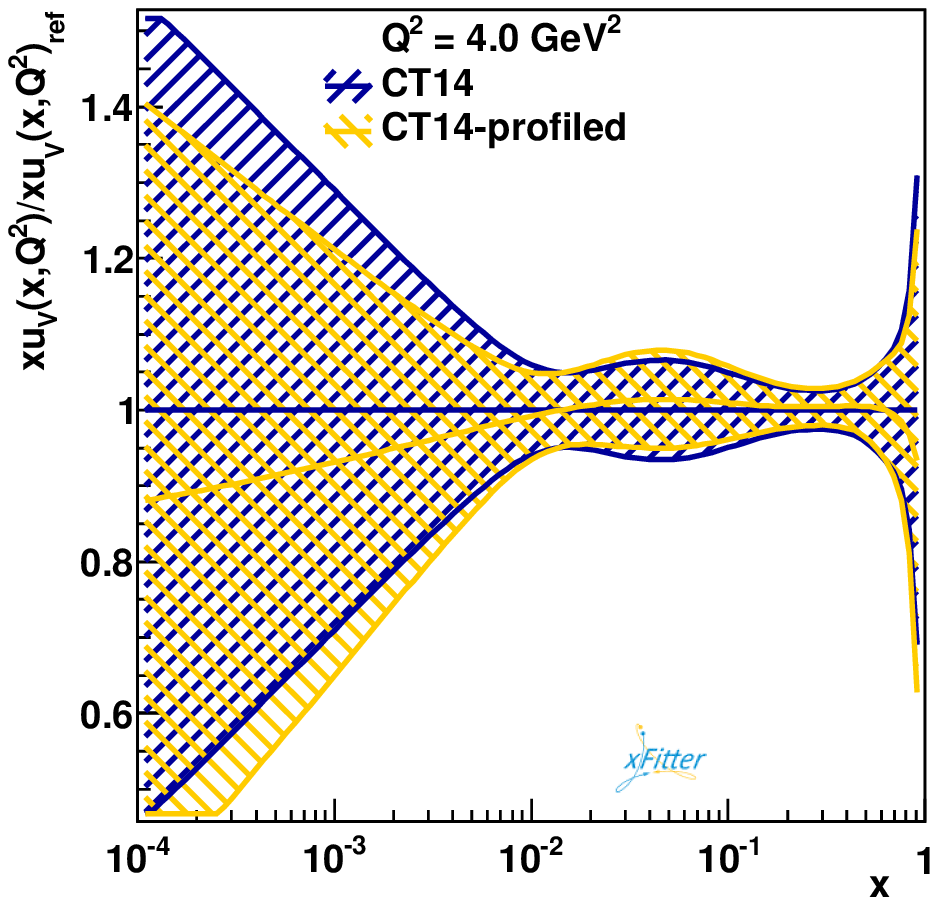}                
        \includegraphics[scale = 0.35]{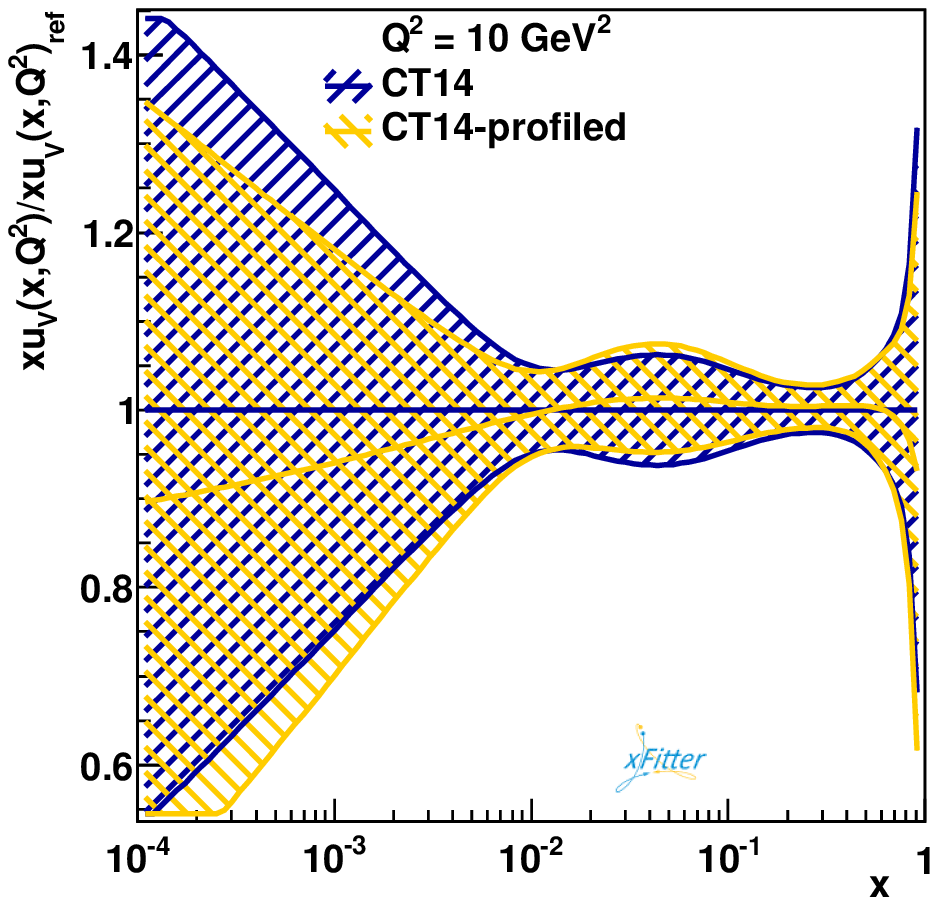}
        \includegraphics[scale = 0.35]{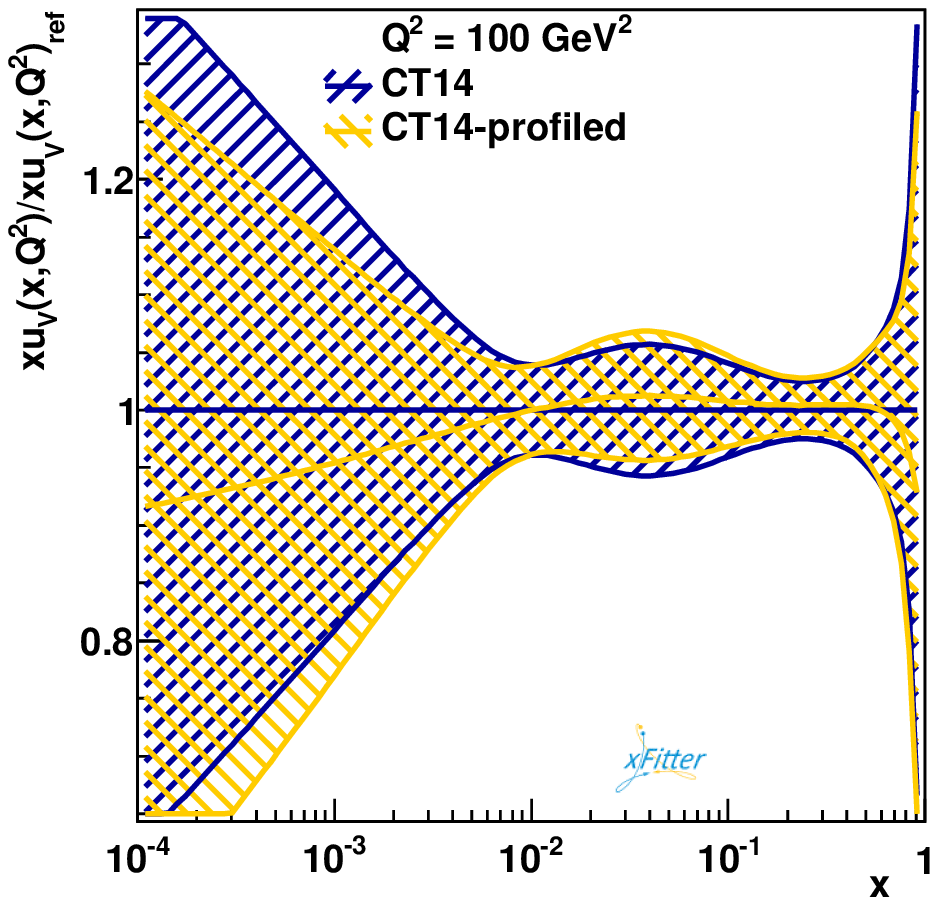}
        \includegraphics[scale = 0.35]{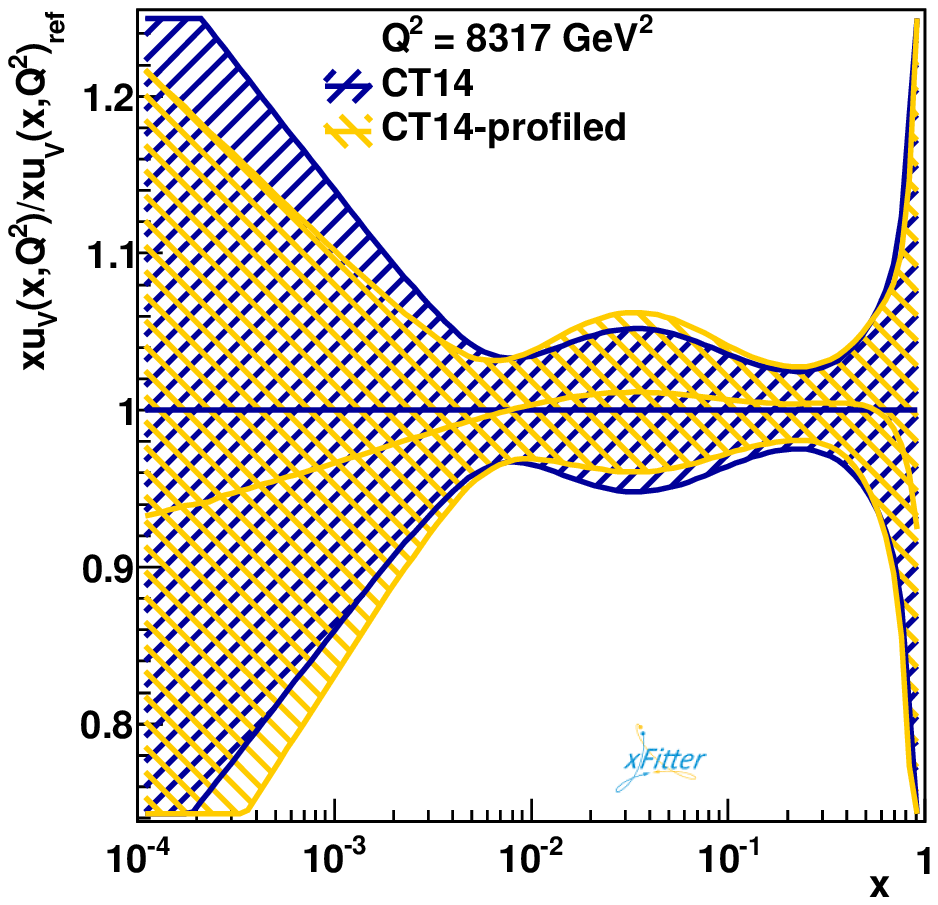}

        \includegraphics[scale = 0.35]{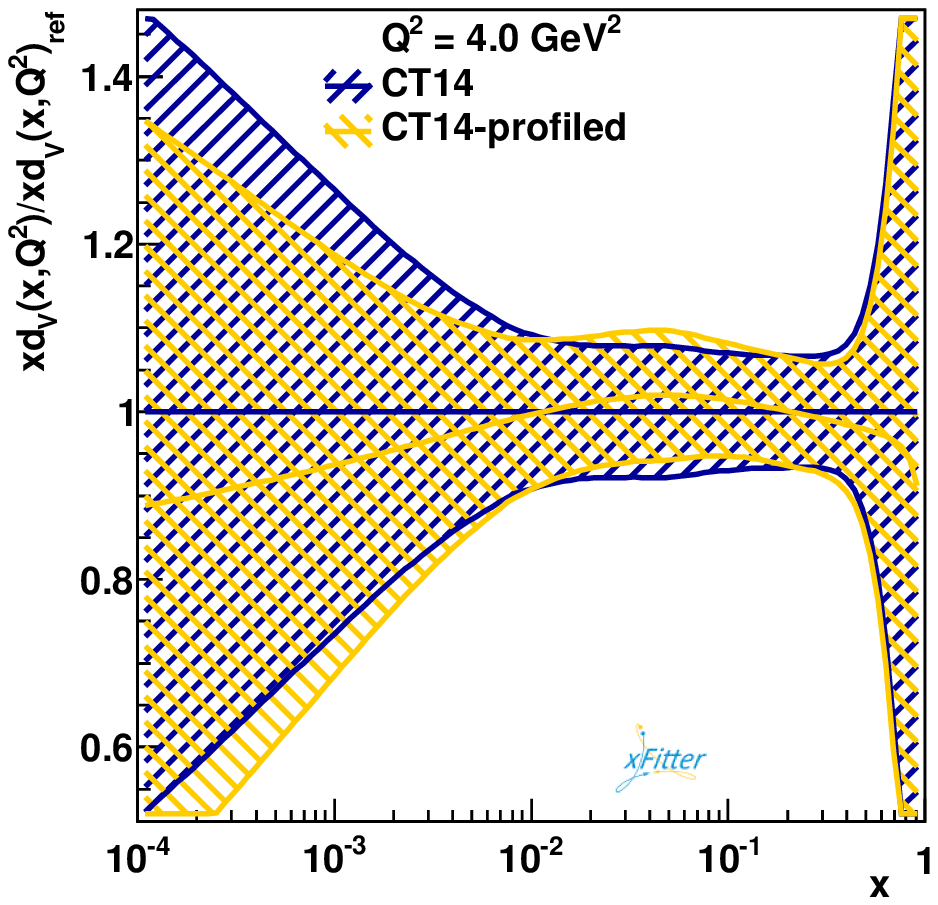}
        \includegraphics[scale = 0.35]{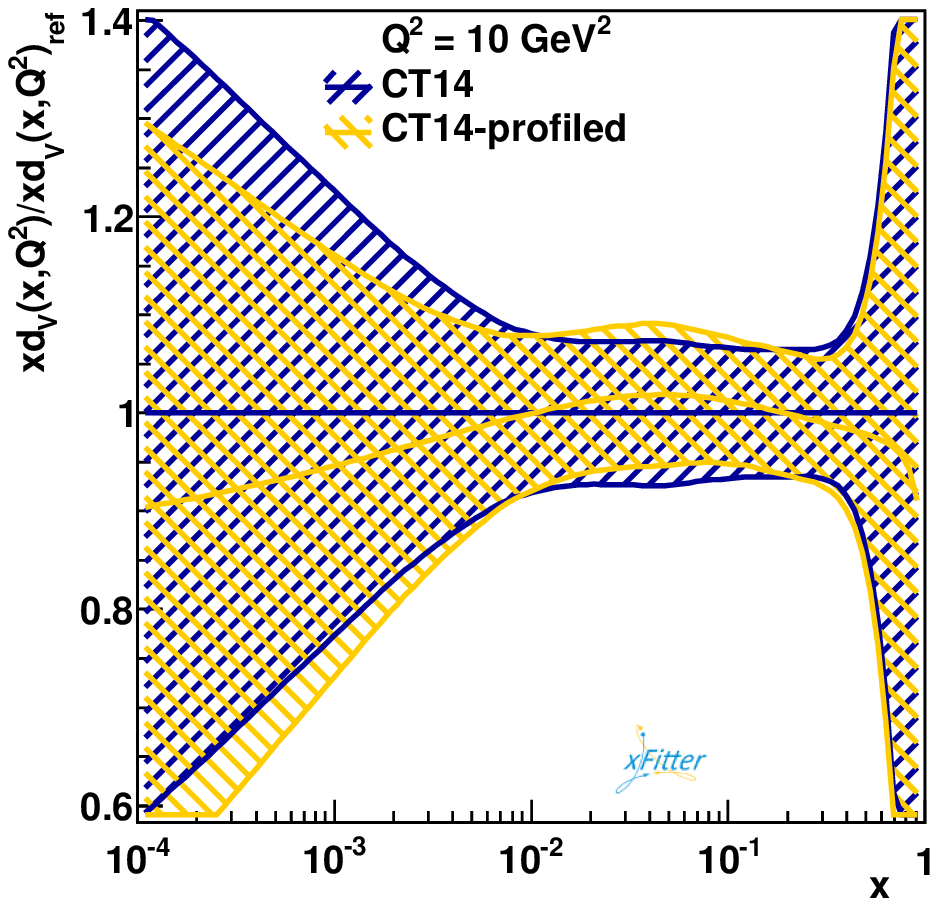}
        \includegraphics[scale = 0.35]{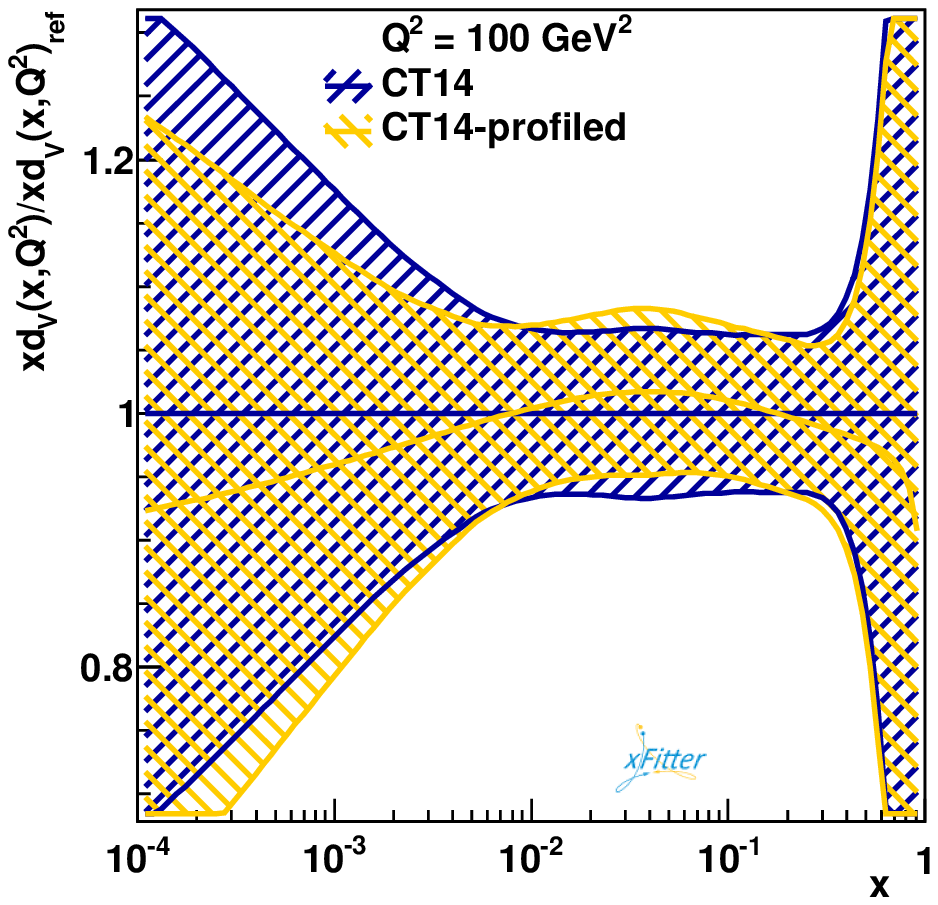}
        \includegraphics[scale = 0.35]{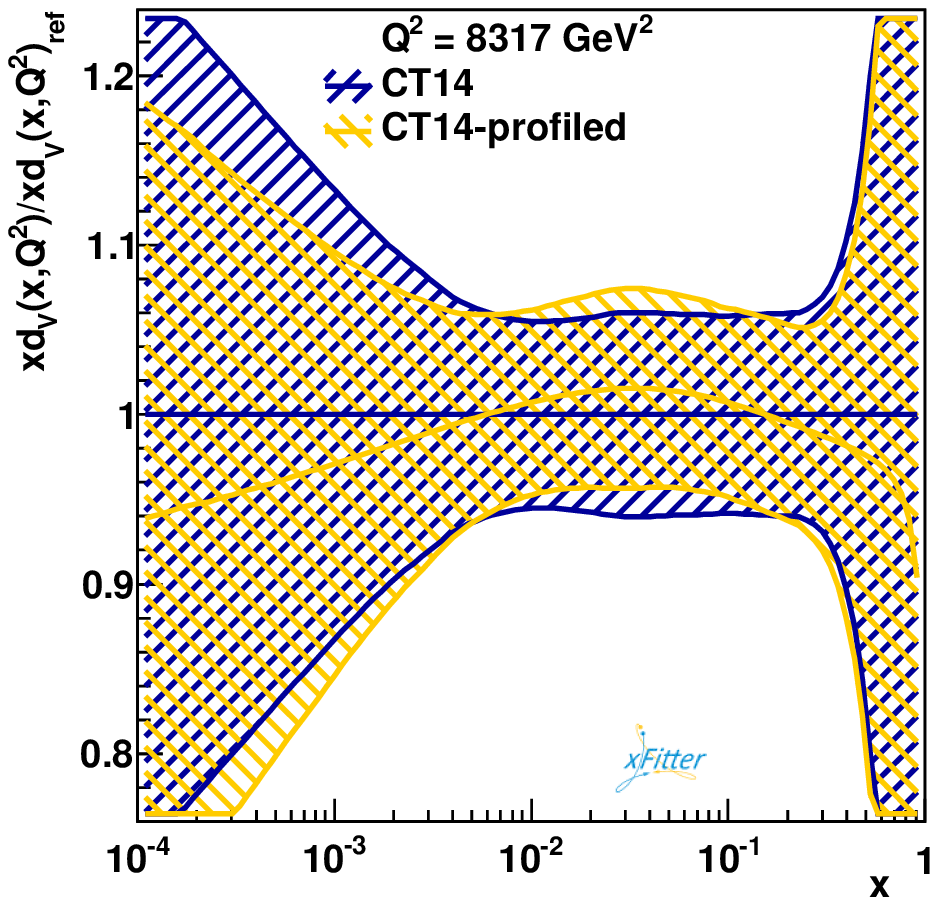}

        \includegraphics[scale = 0.35]{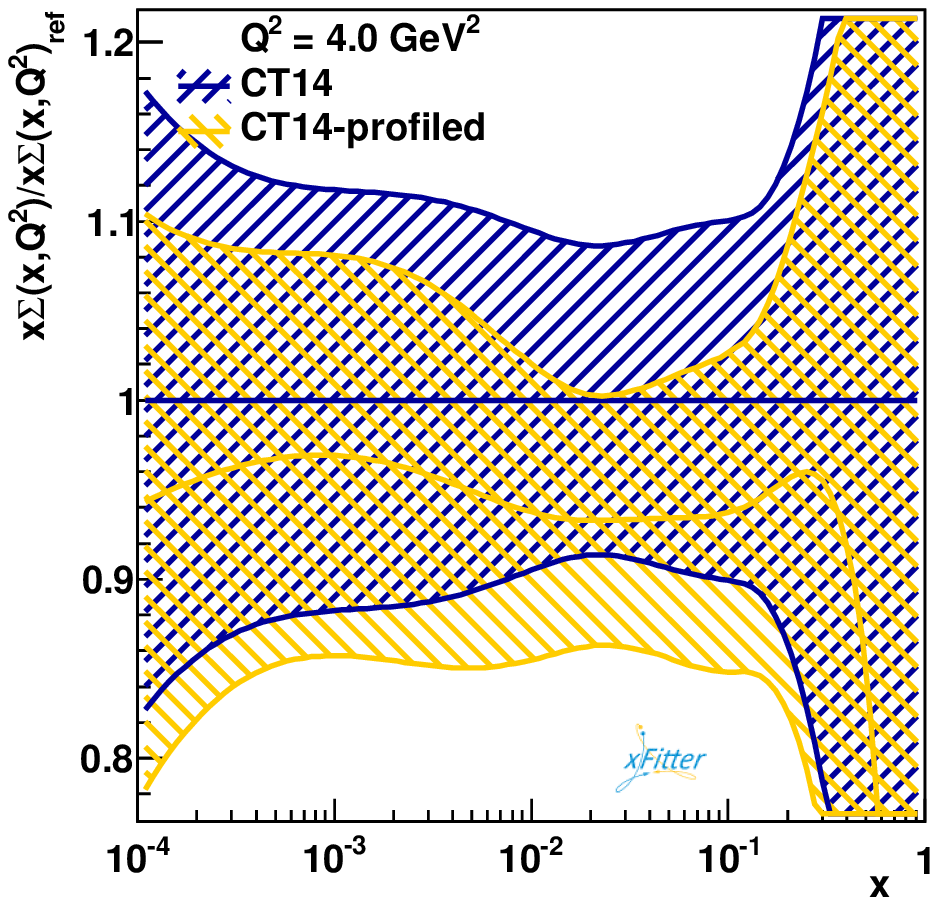}
        \includegraphics[scale = 0.35]{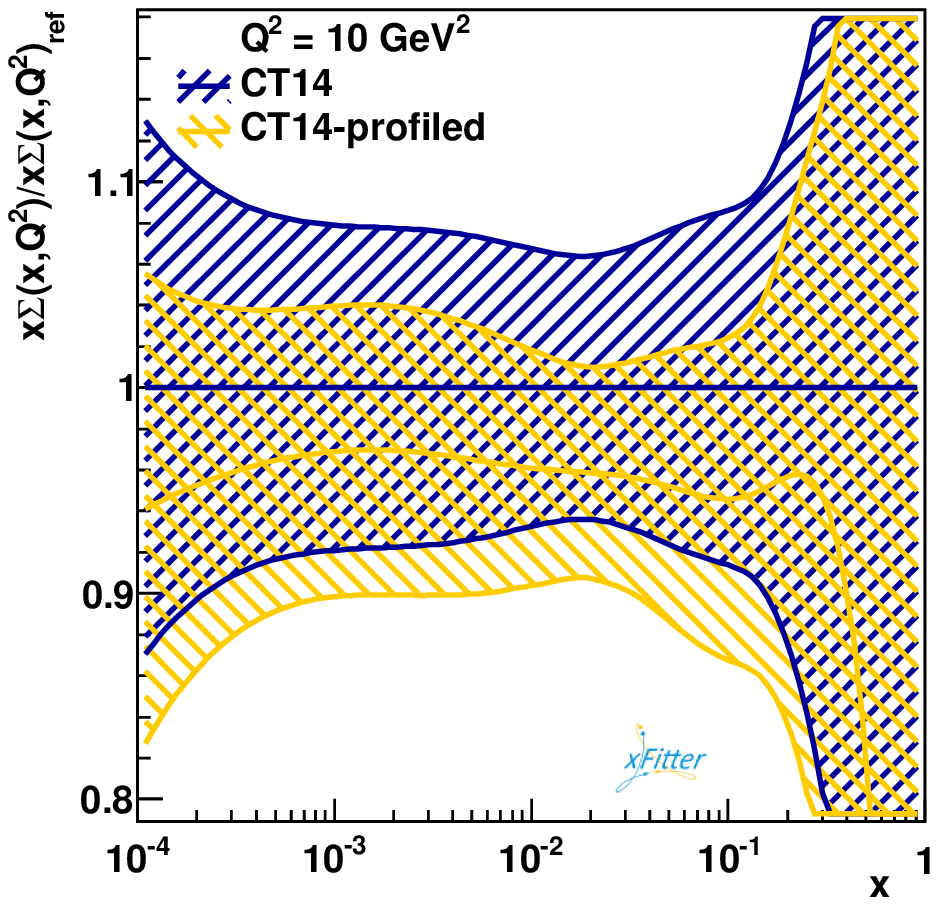}
        \includegraphics[scale = 0.35]{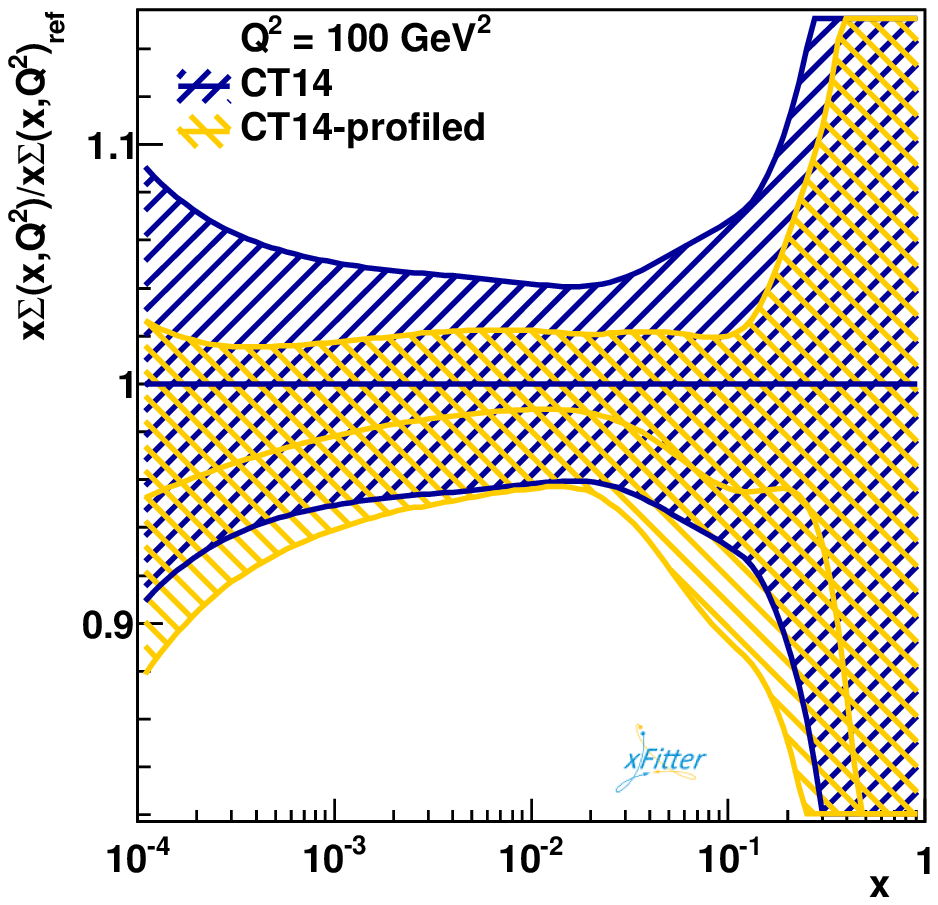}
        \includegraphics[scale = 0.35]{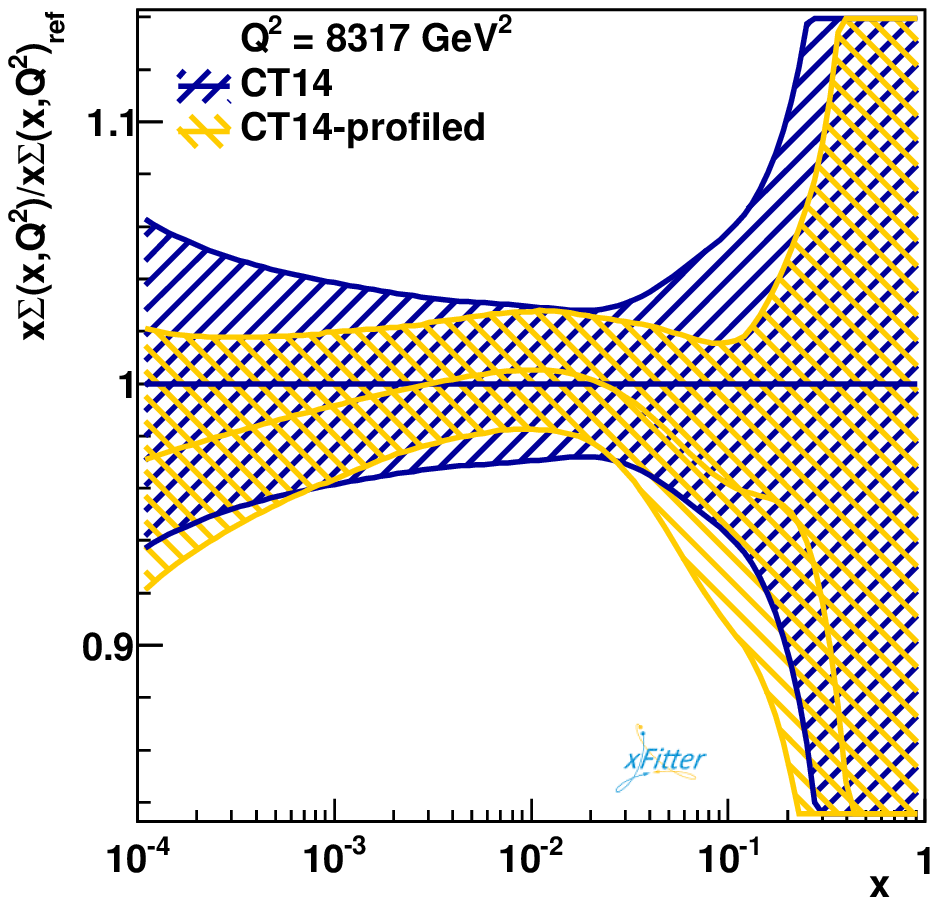}

        \includegraphics[scale = 0.35]{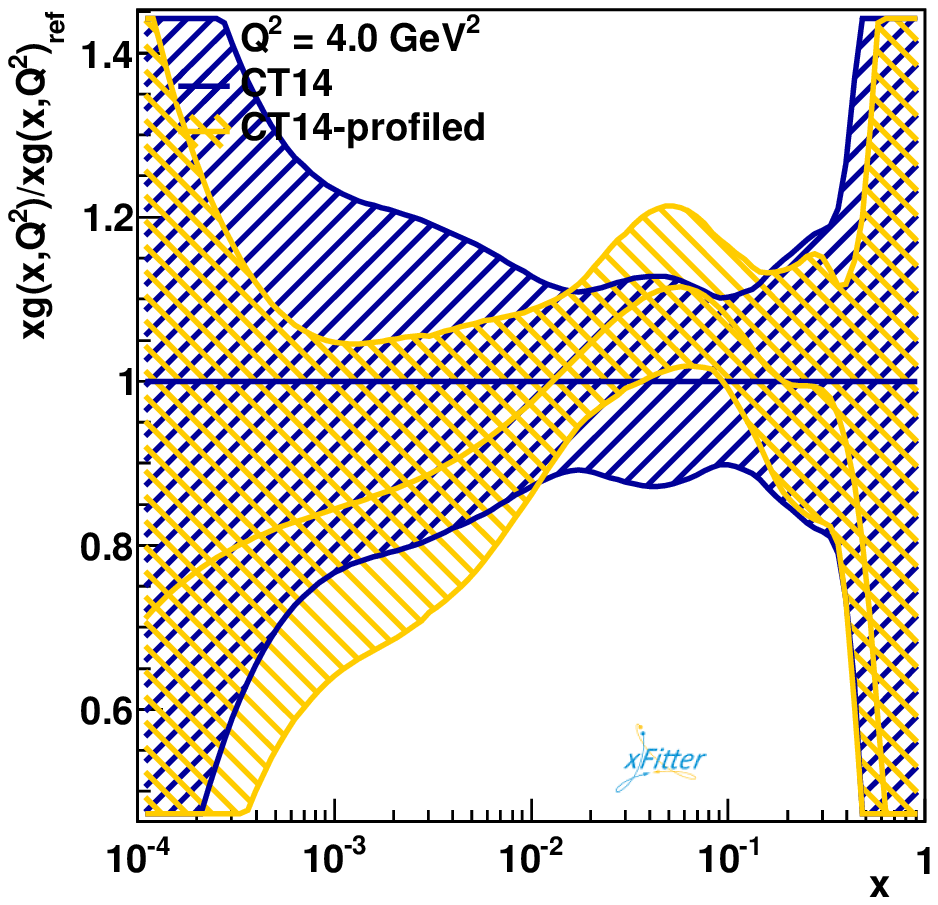}
        \includegraphics[scale = 0.35]{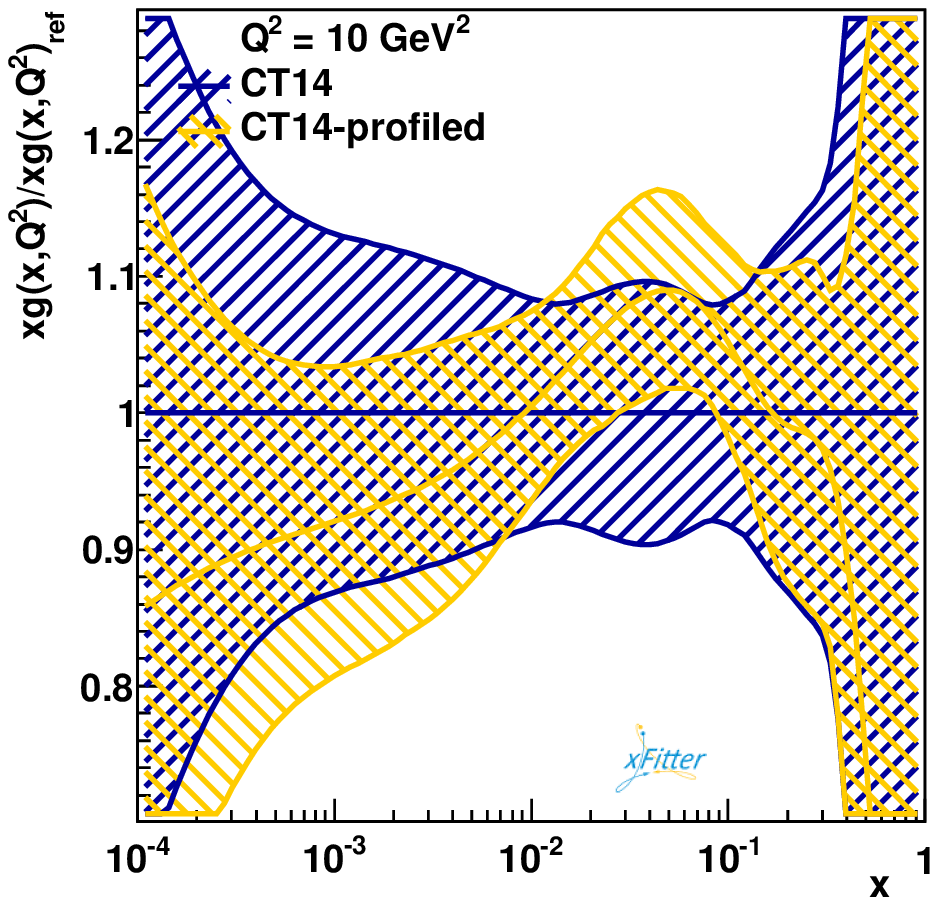}
        \includegraphics[scale = 0.35]{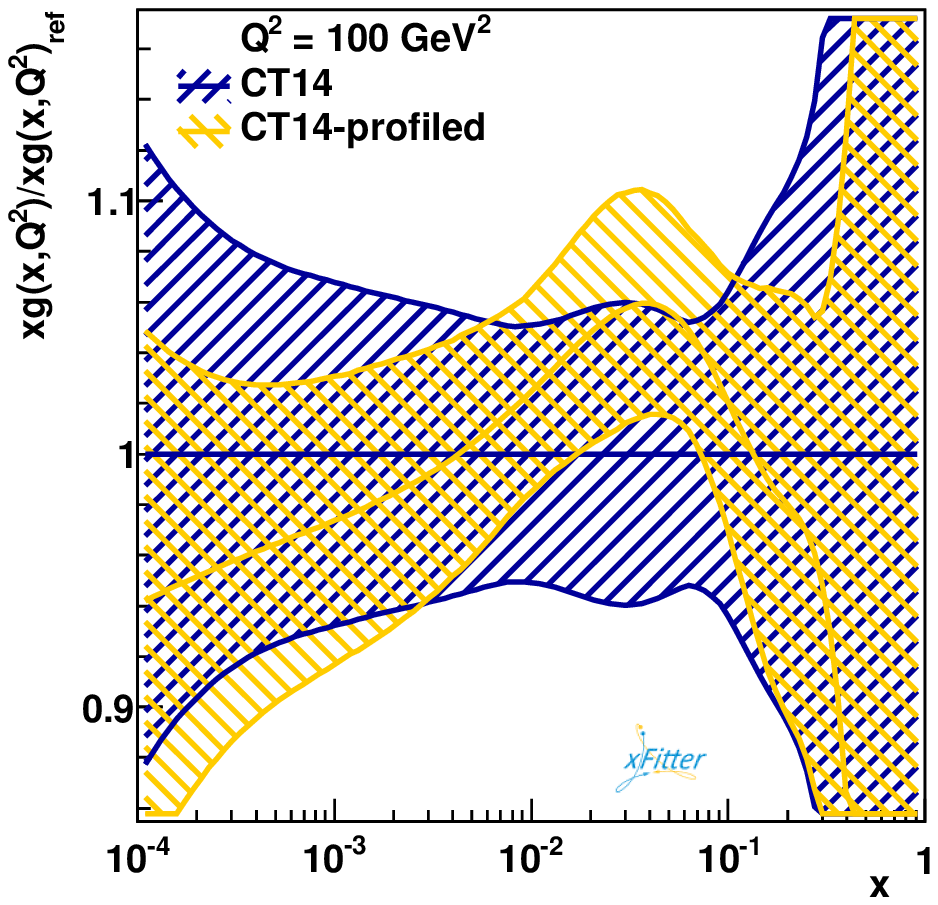}
        \includegraphics[scale = 0.35]{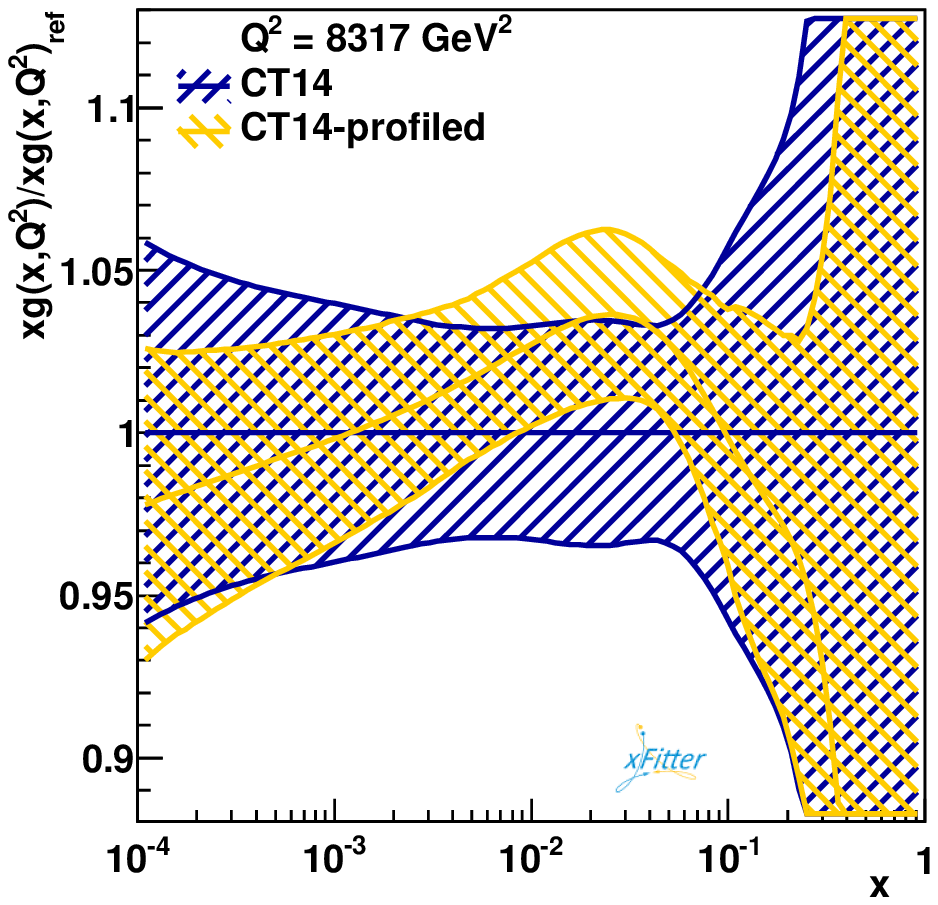}
		
\caption{The parton distribution ratio  $xu_v/xu_{v_{ref}}$, $xd_v/xd_{v_{ref}}$, $x\Sigma/x\Sigma_{ref}$, and $xg/xg_{ref}$ with respect to without profiling procedure, extracted from CT14 \cite{Dulat:2015mca} PDFs as a function of $x$ at  4, 10, 100, and 8317 GeV$^2$. The results obtained after the profiling procedure compared with corresponding same features 
	 before profiling.}
		\label{fig:partonRefCT14}
	\end{center}
\end{figure}

The impact of the recent measurements of top quark cross section on the parton distribution ratio  $xu_v/xu_{v_{ref}}$, $xd_v/xd_{v_{ref}}$, $x\Sigma/x\Sigma_{ref}$, and $xg/xg_{ref}$  for CT14 PDFs are represented in Fig.~\ref{fig:partonRefCT14}.

\begin{figure}[!htb]
	\begin{center}

	    \includegraphics[scale = 0.35]{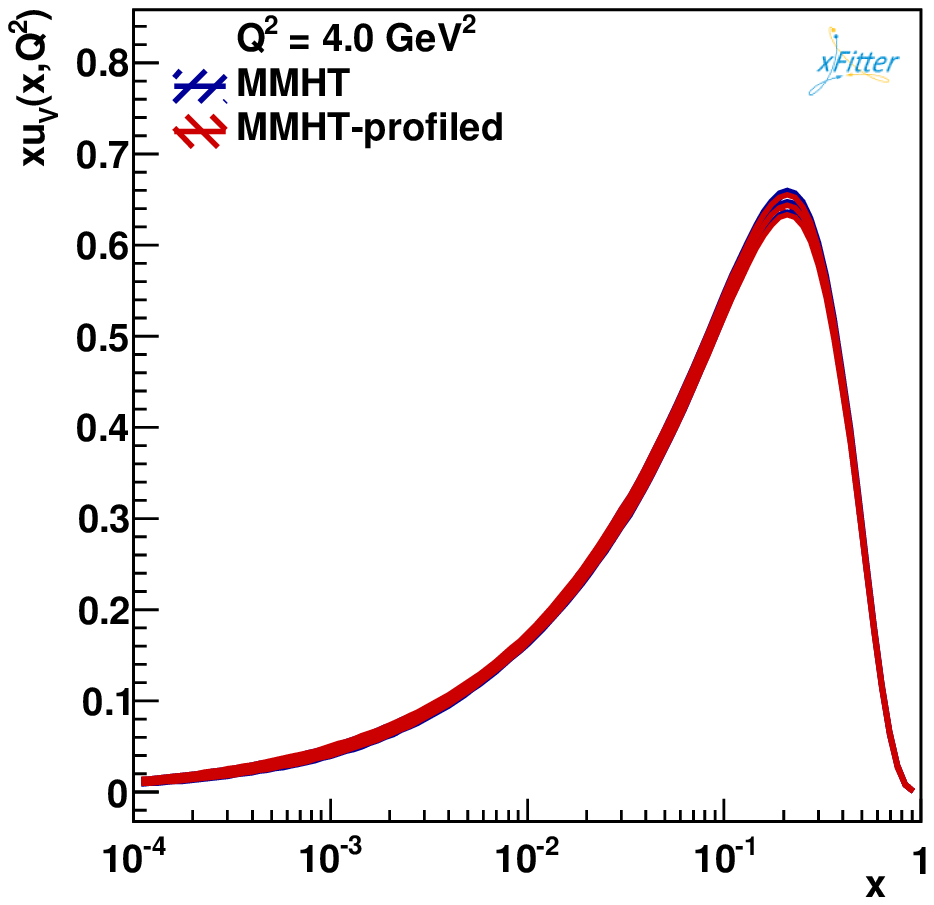}	    	    
	    \includegraphics[scale = 0.35]{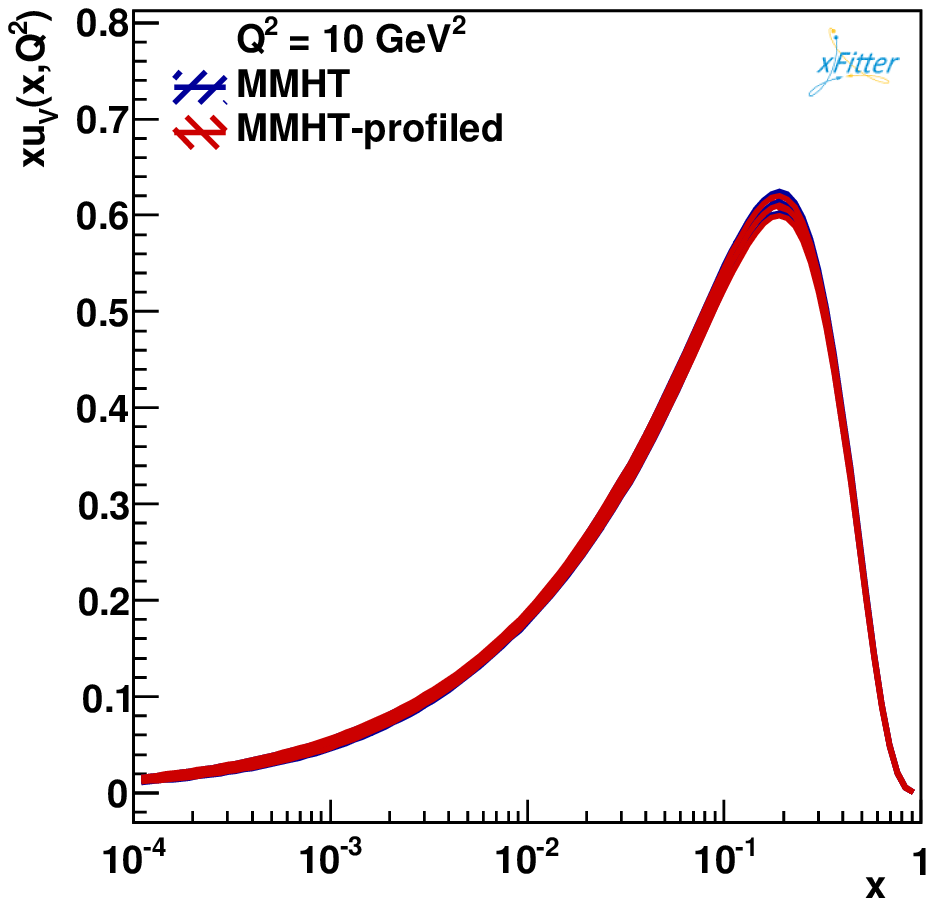}
	    \includegraphics[scale = 0.35]{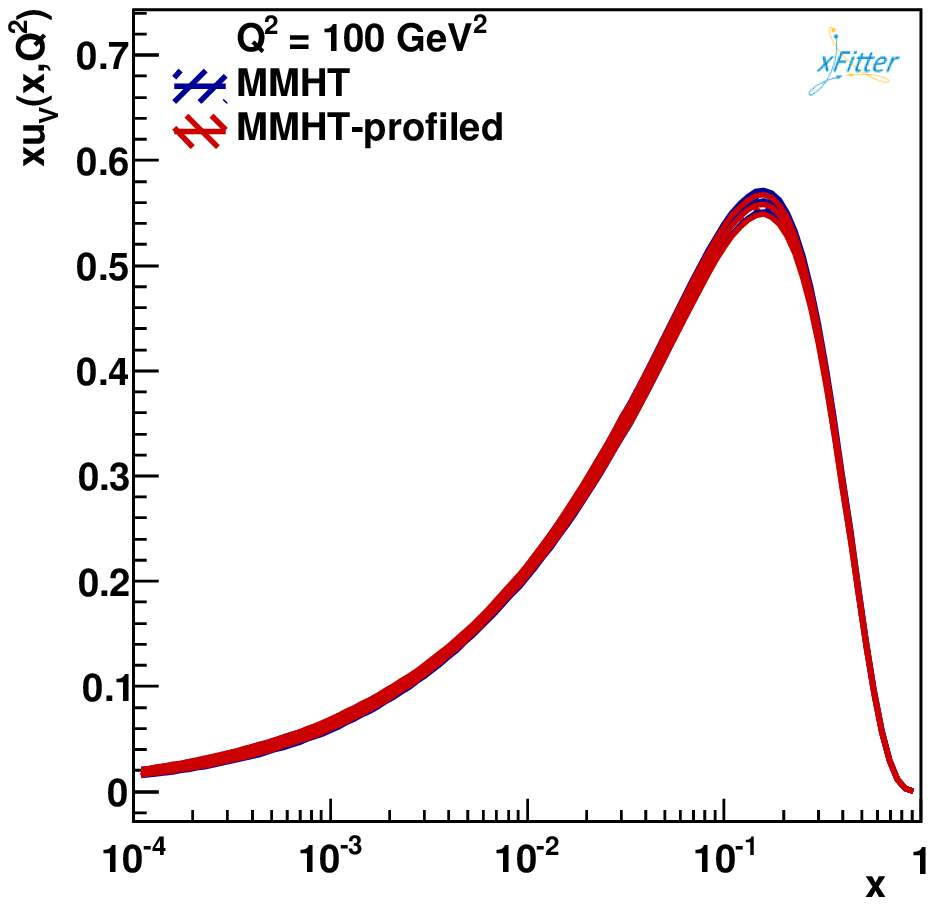}
	    \includegraphics[scale = 0.35]{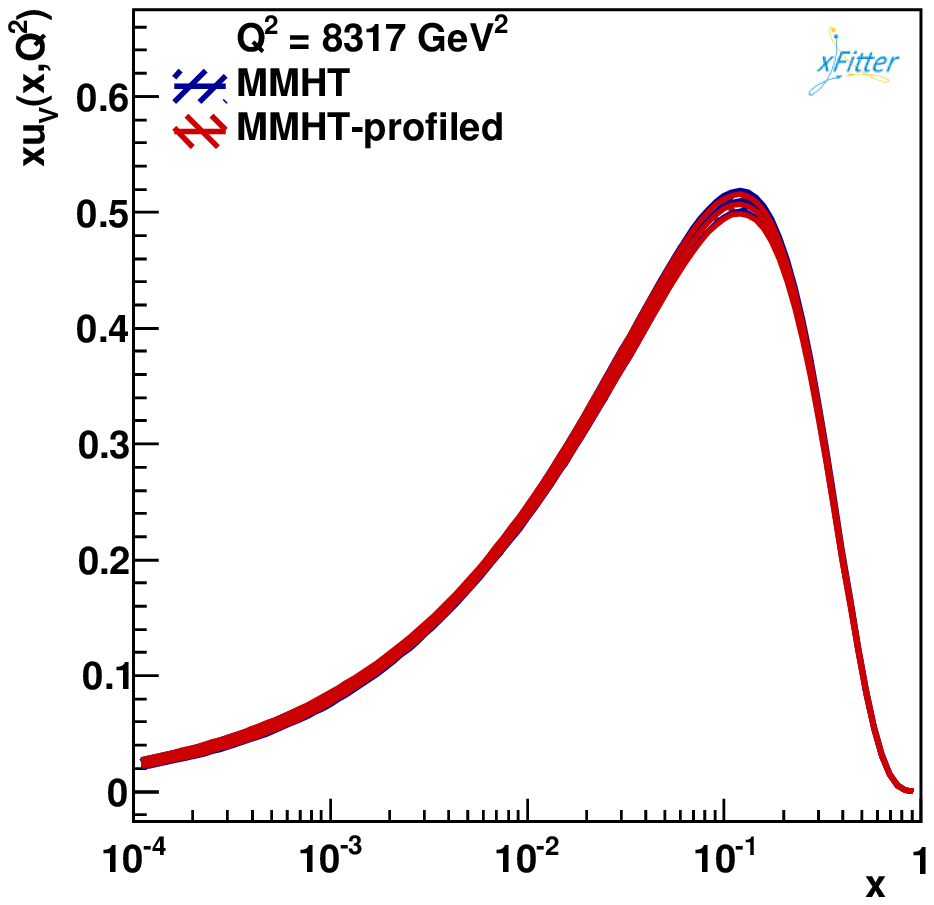}

	    \includegraphics[scale = 0.35]{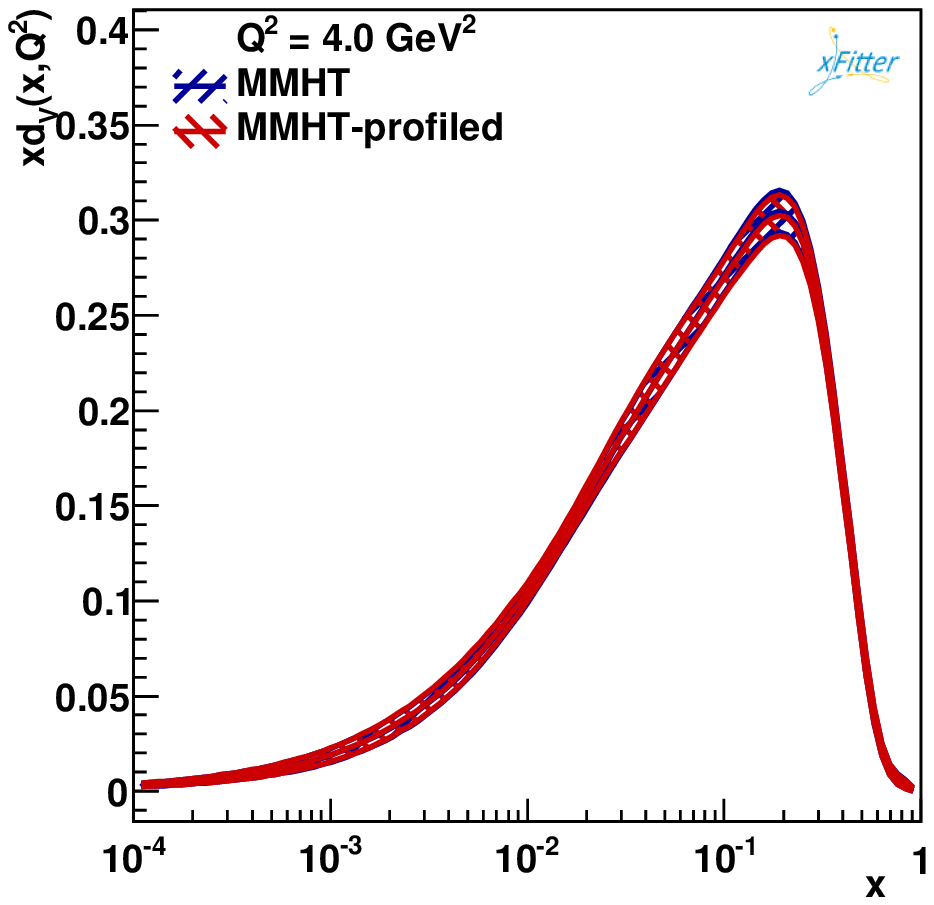}
	    \includegraphics[scale = 0.35]{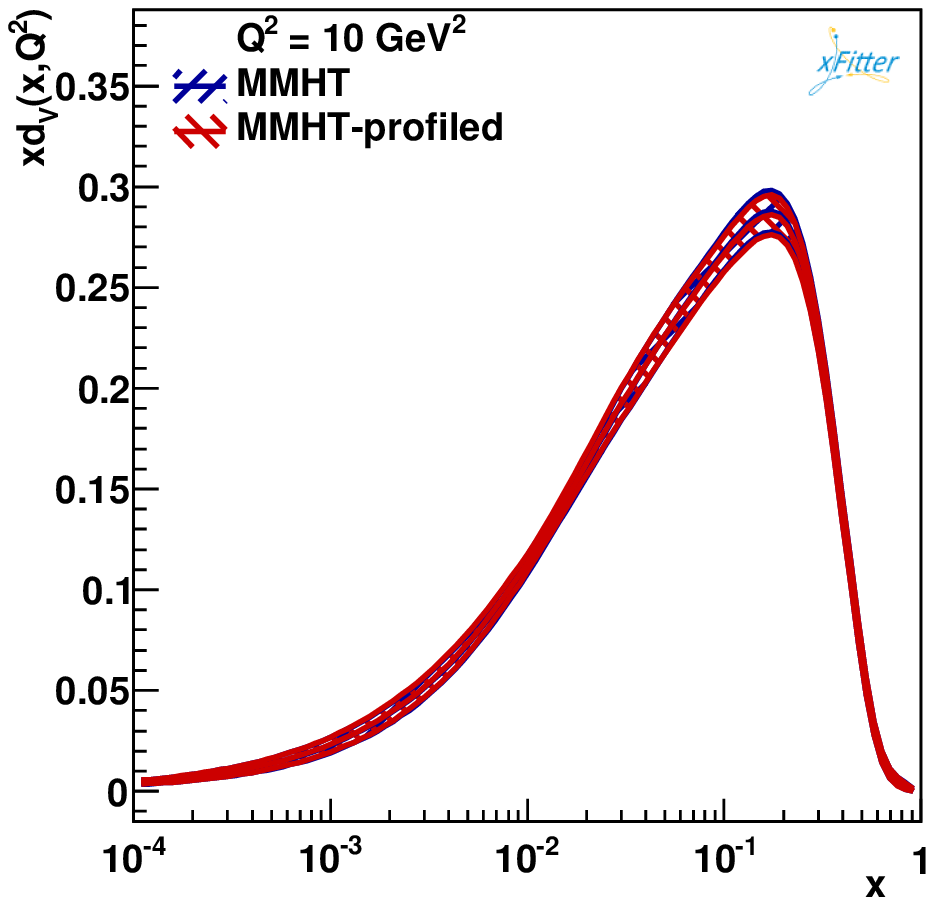}
	    \includegraphics[scale = 0.35]{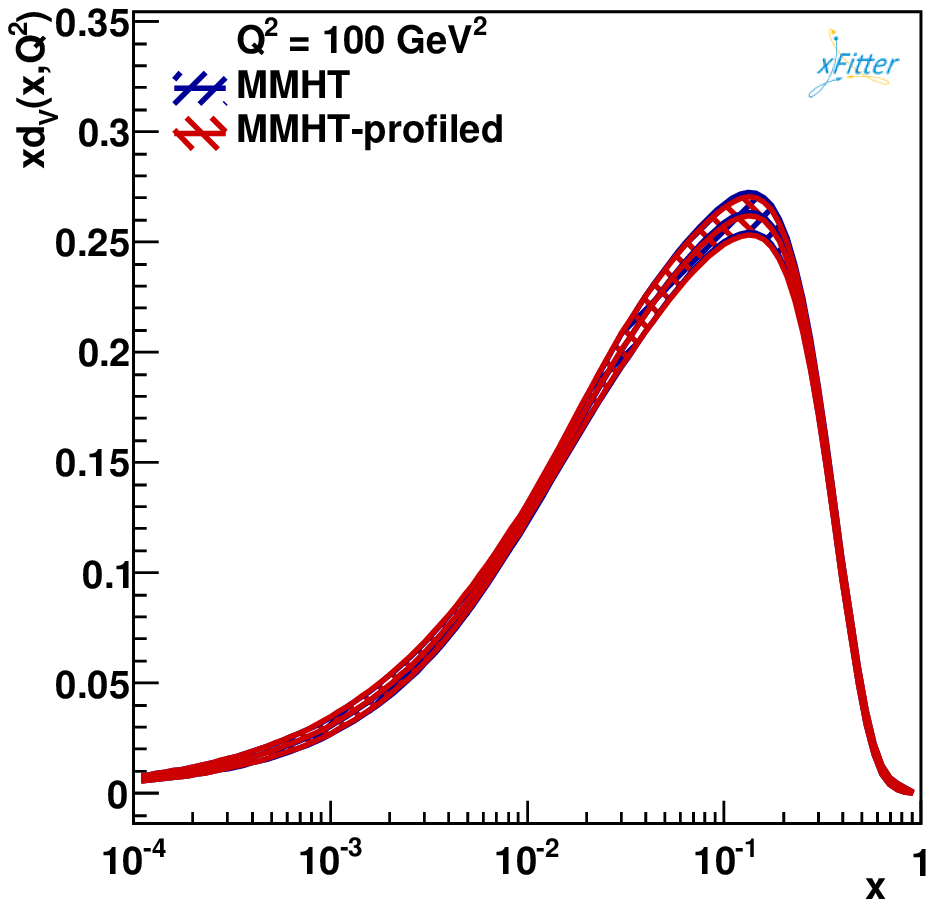}
	    \includegraphics[scale = 0.35]{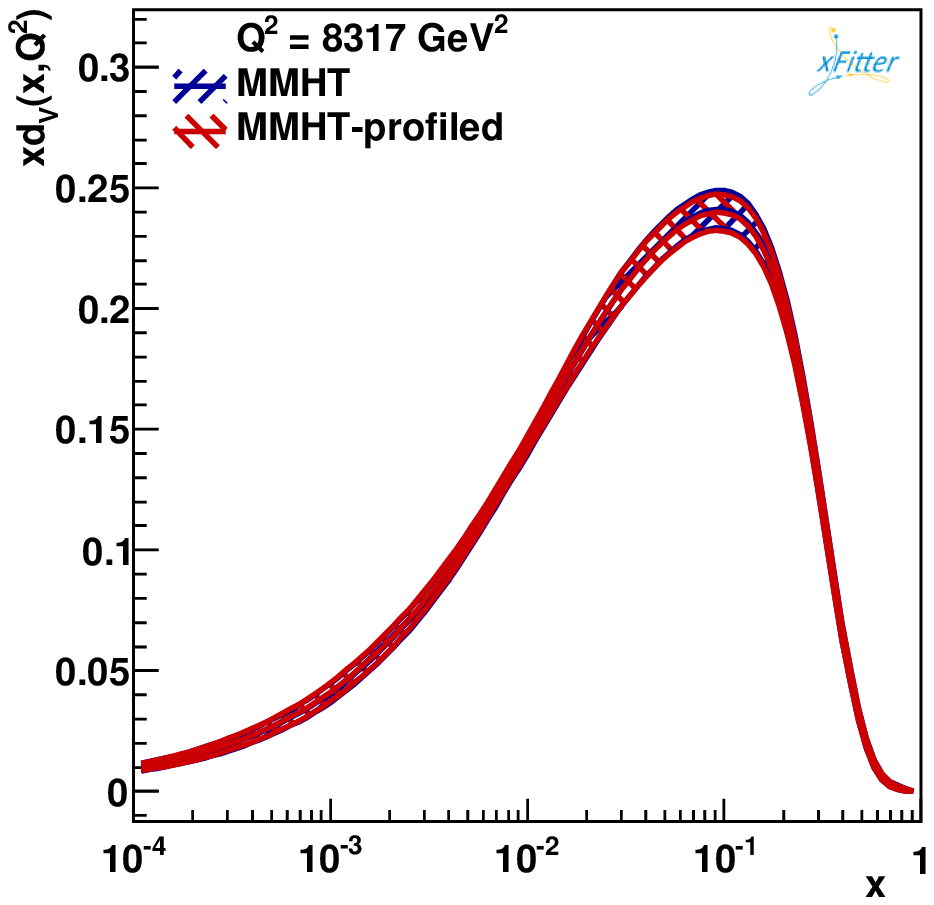}

	    \includegraphics[scale = 0.35]{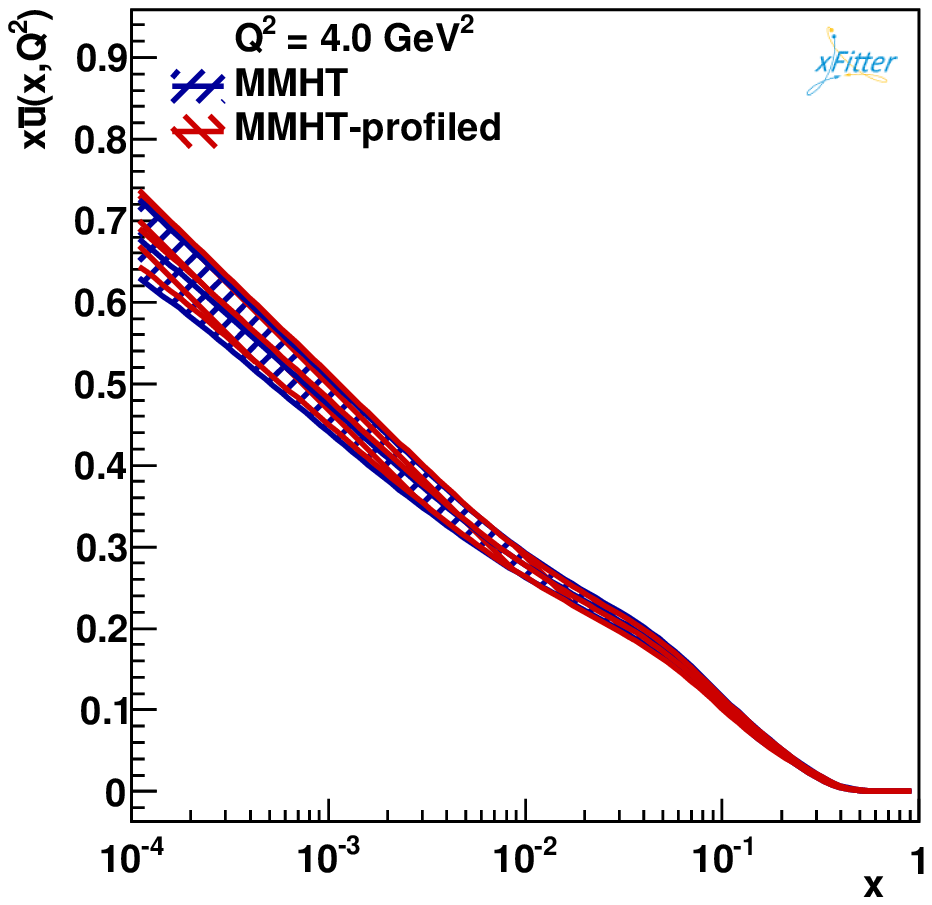}	    	    
	    \includegraphics[scale = 0.35]{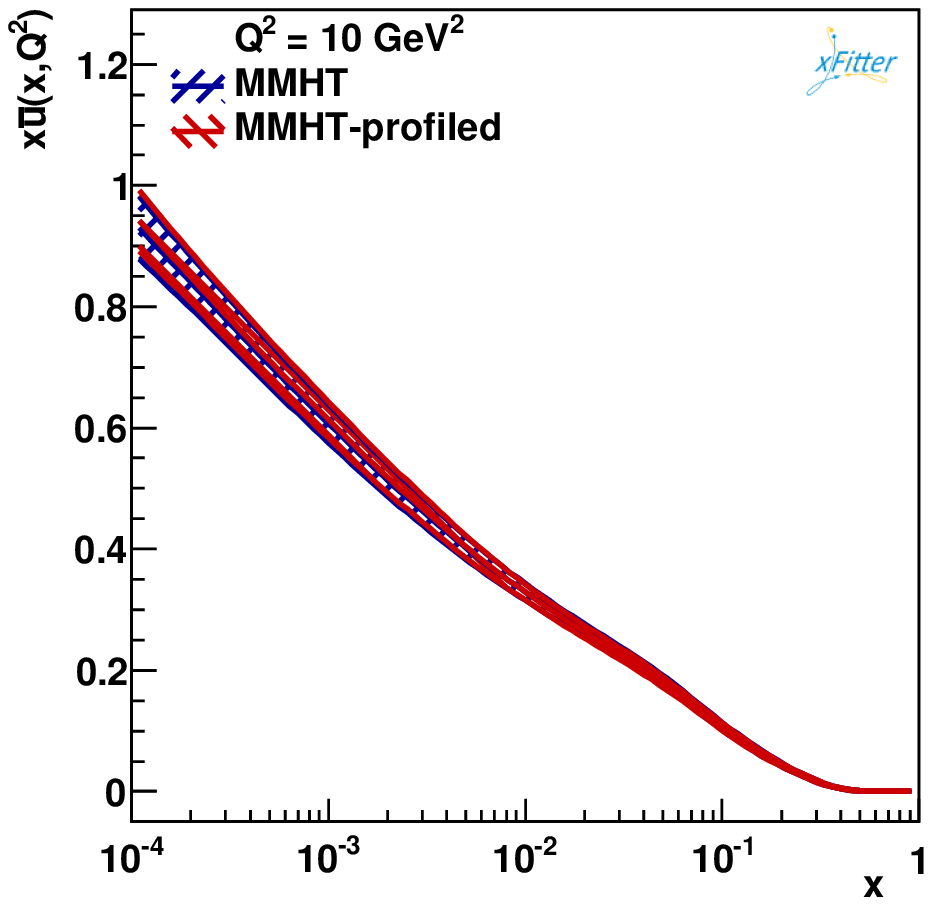}
	    \includegraphics[scale = 0.35]{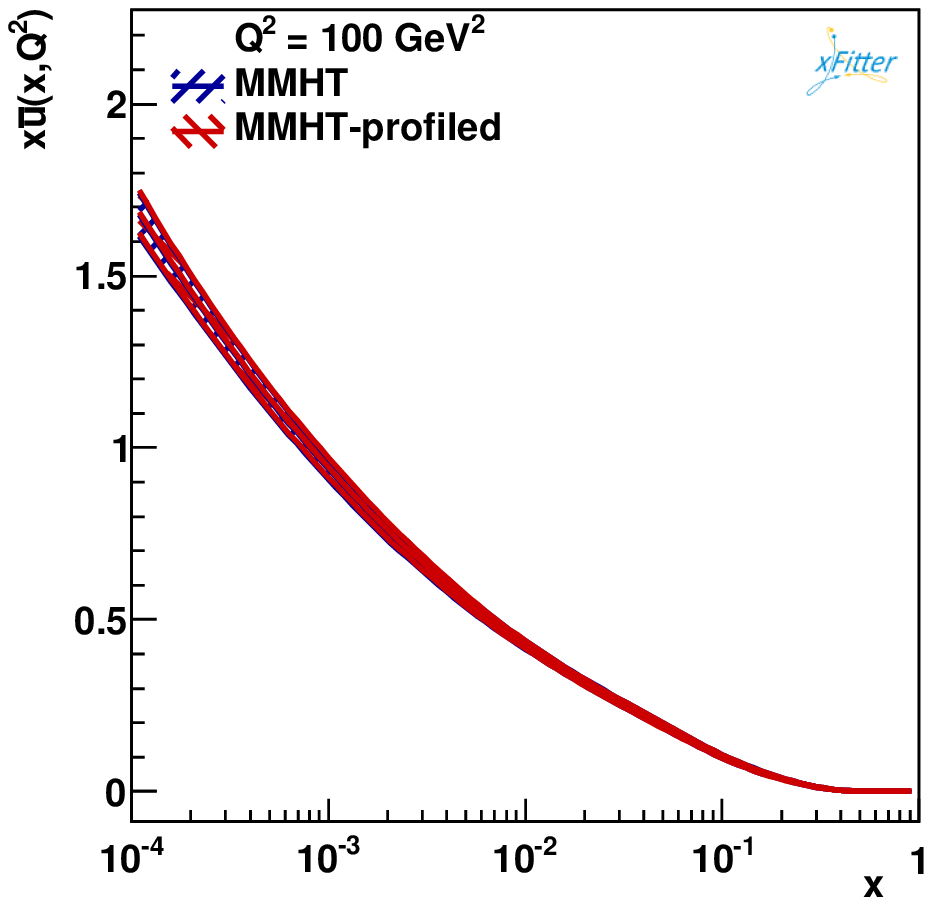}
	    \includegraphics[scale = 0.35]{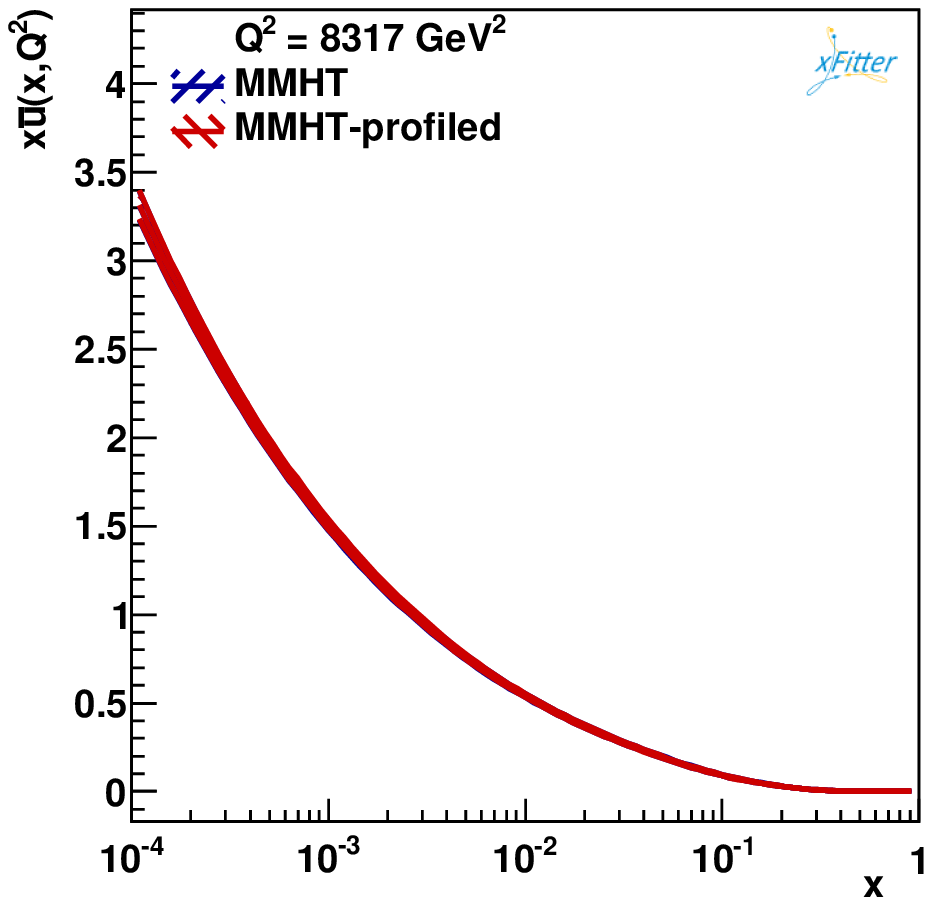}

	    \includegraphics[scale = 0.35]{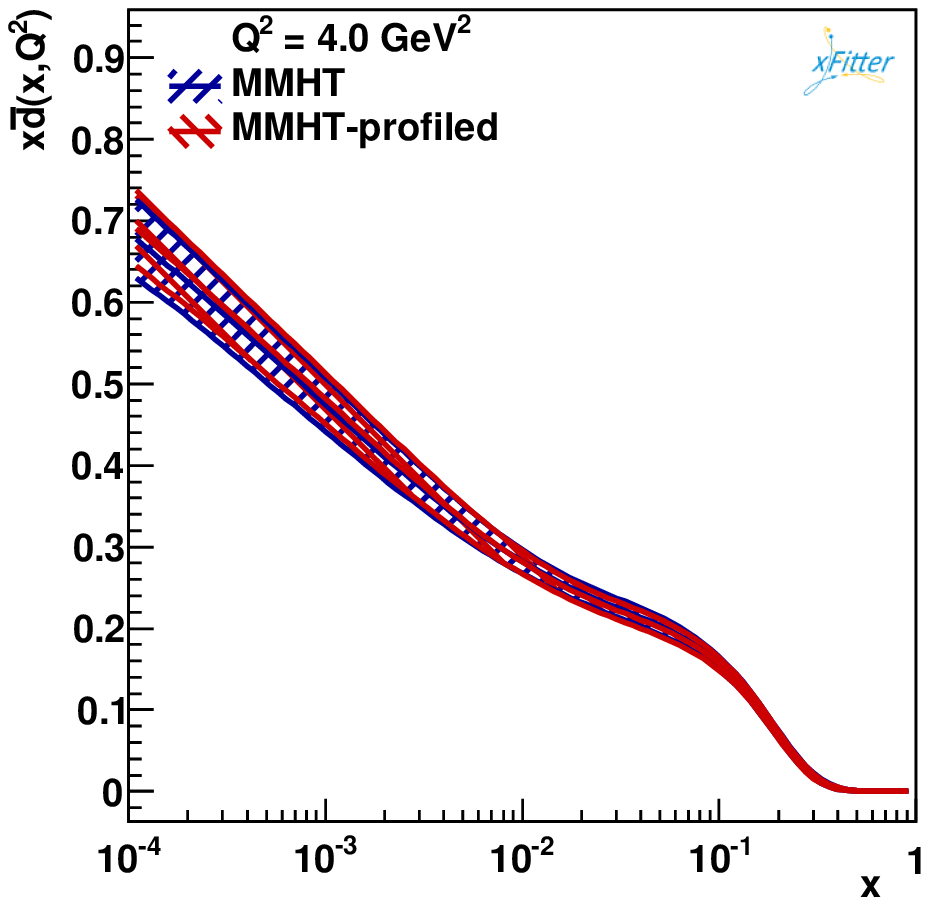}	    	    
	    \includegraphics[scale = 0.35]{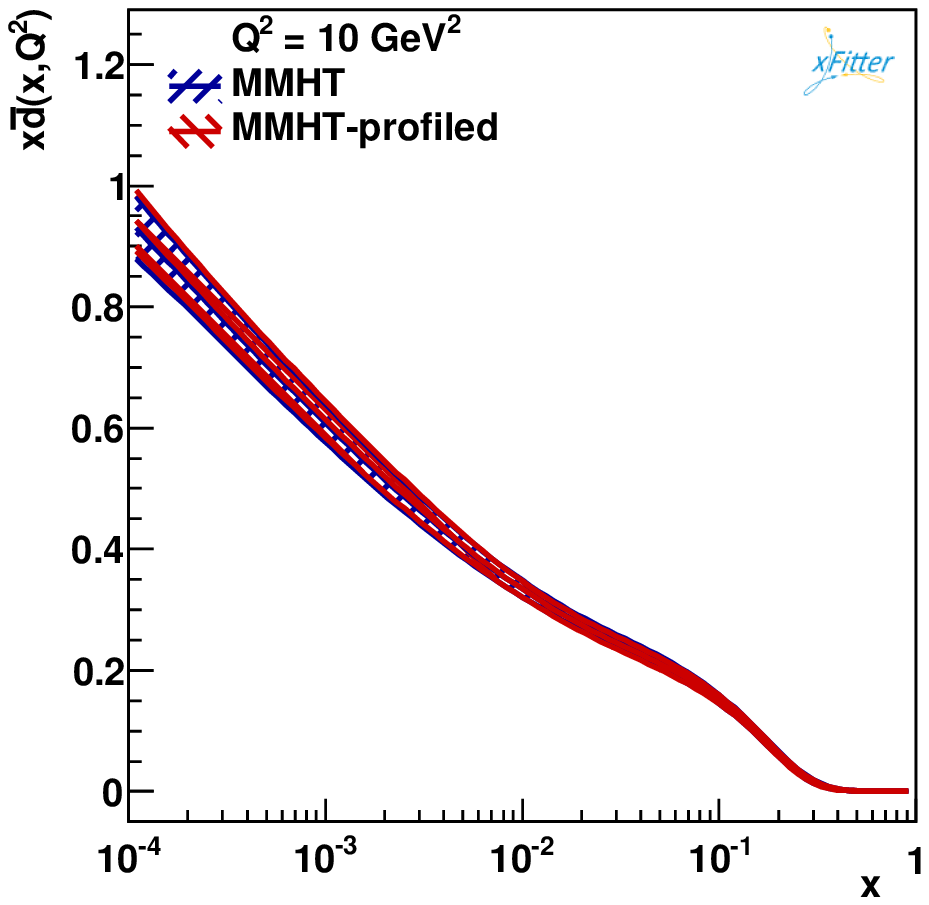}
	    \includegraphics[scale = 0.35]{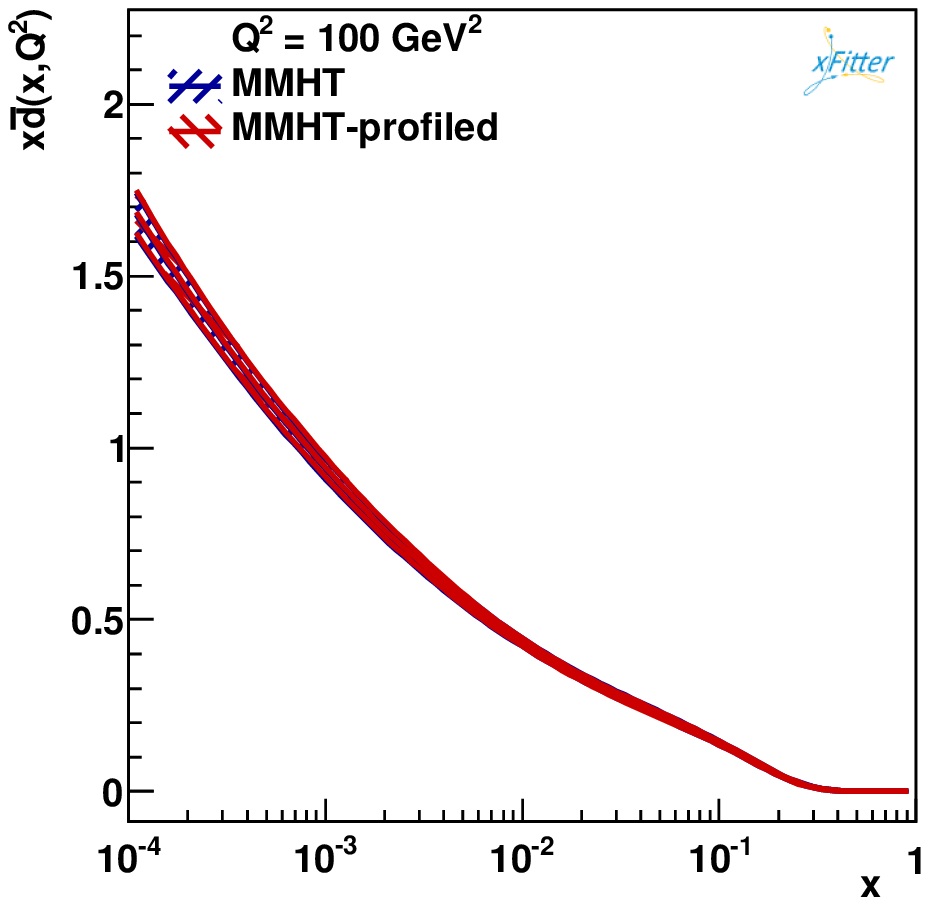}
	    \includegraphics[scale = 0.35]{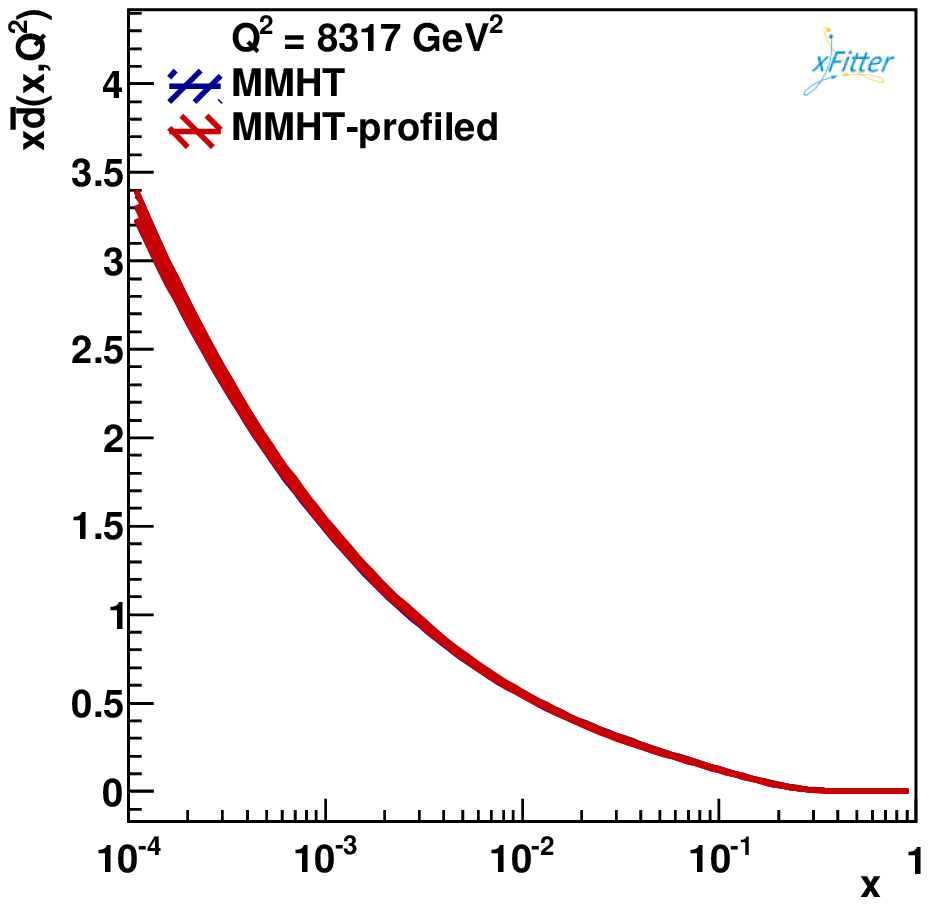}

	    \includegraphics[scale = 0.35]{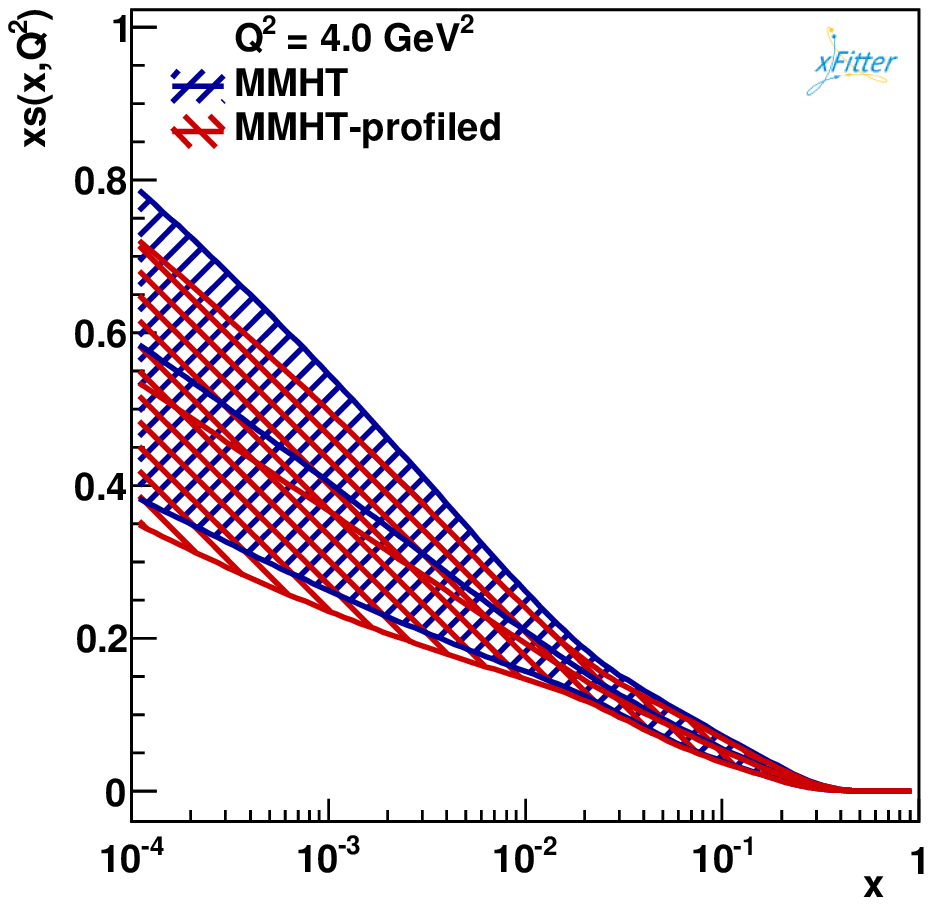}        
        	\includegraphics[scale = 0.35]{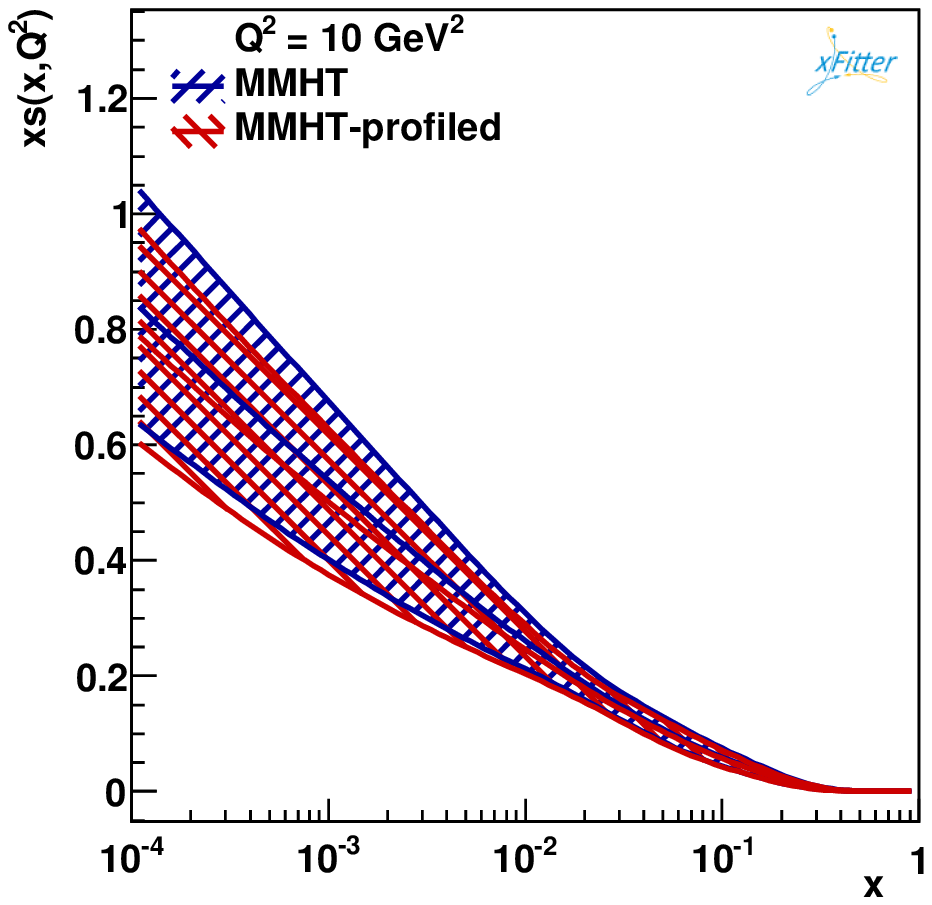}
	    \includegraphics[scale = 0.35]{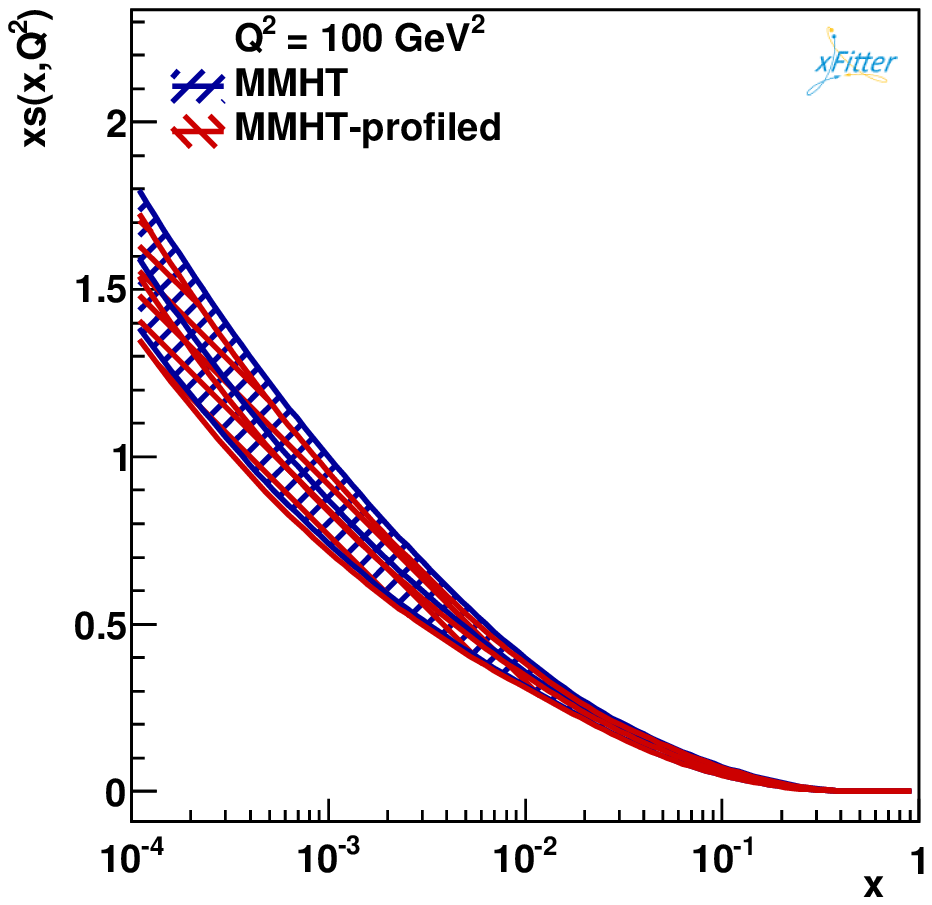}
	    \includegraphics[scale = 0.35]{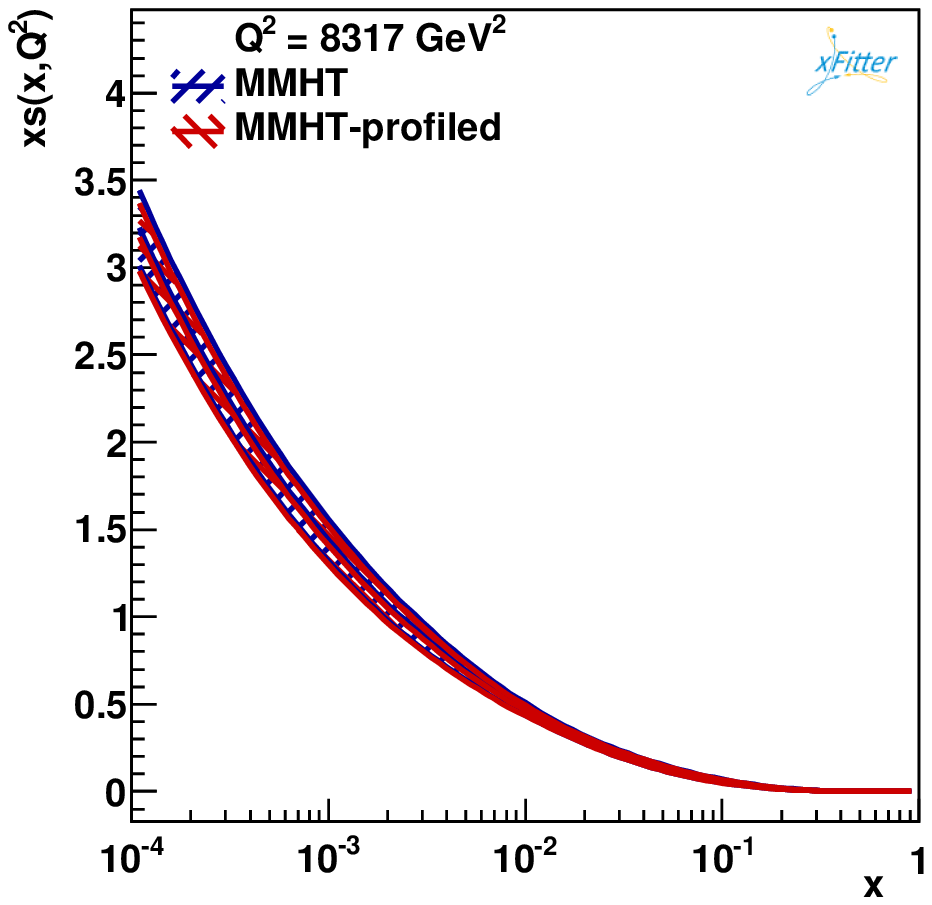}

	    \includegraphics[scale = 0.35]{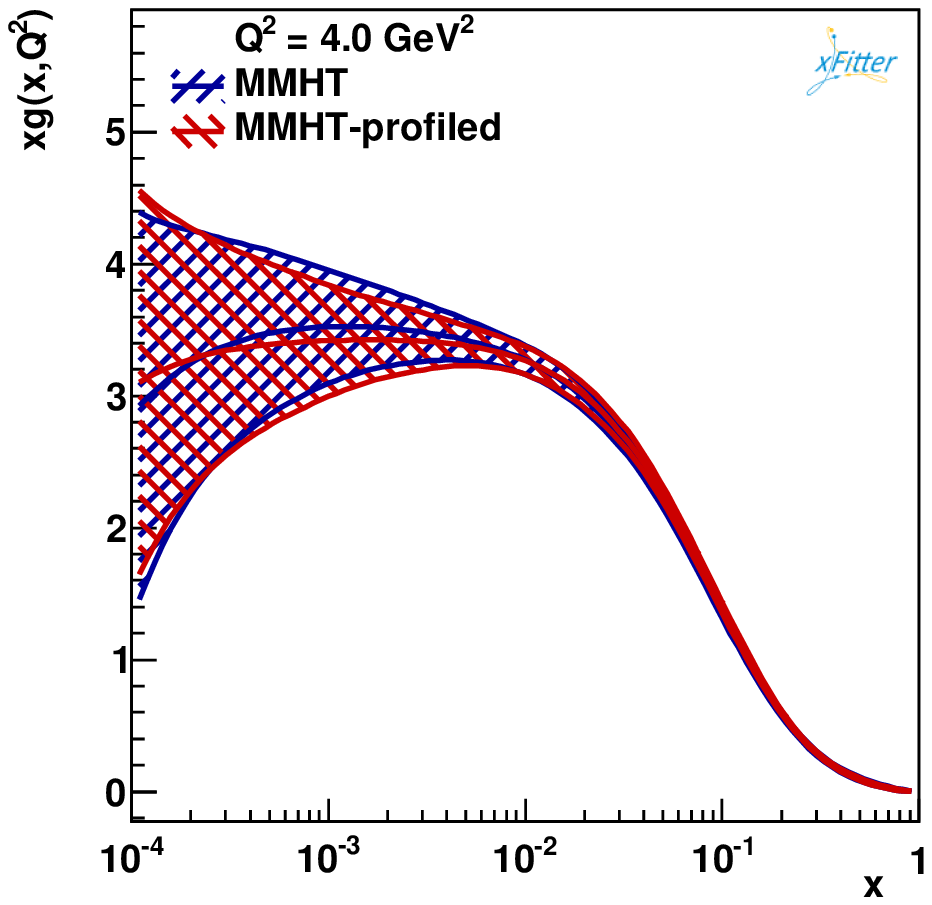}     
	    \includegraphics[scale = 0.35]{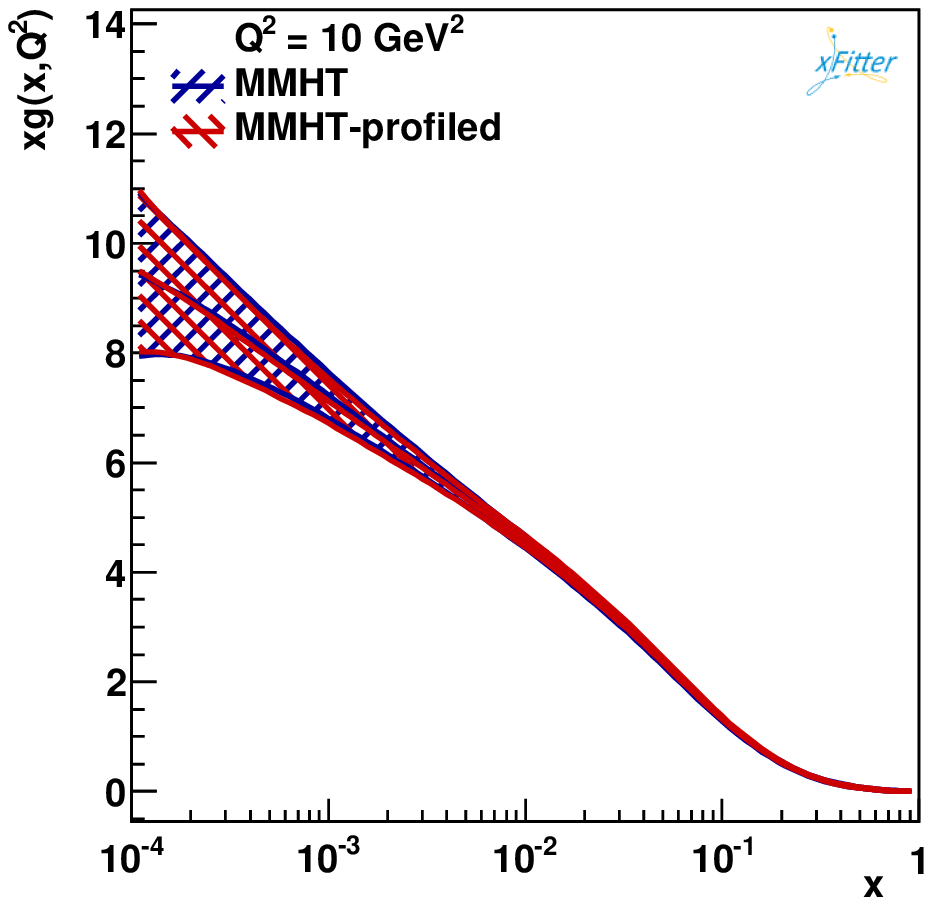}
	    \includegraphics[scale = 0.35]{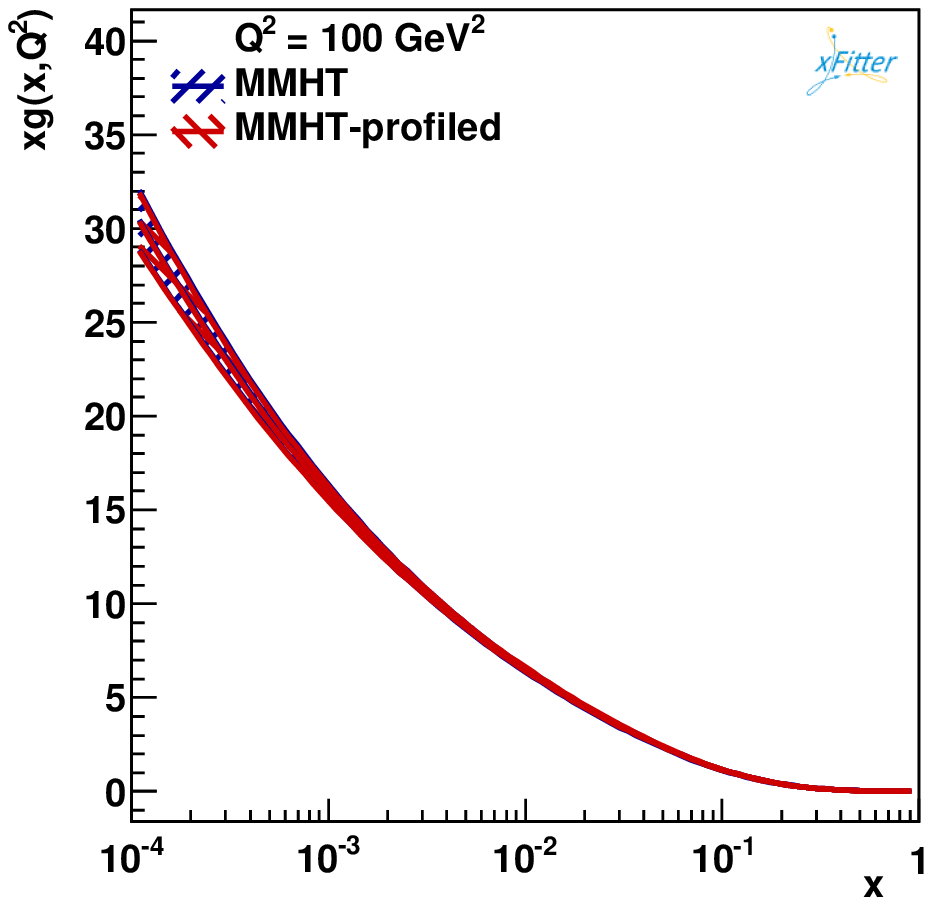}
	    \includegraphics[scale = 0.35]{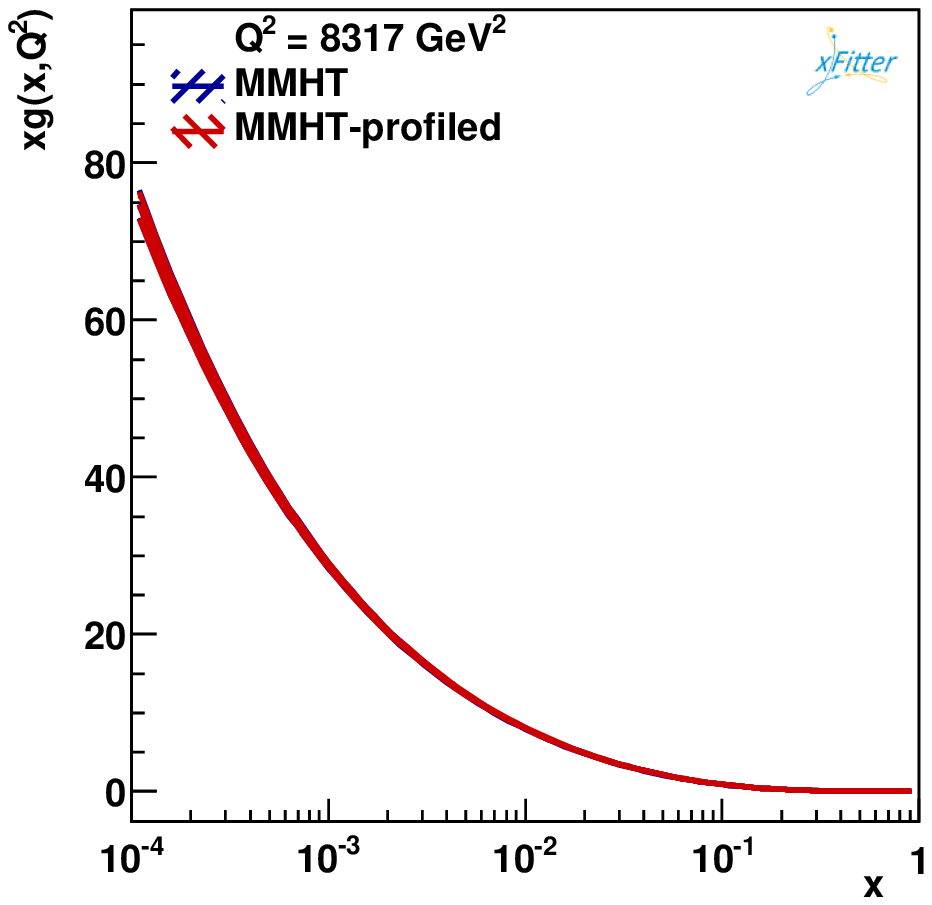}

\caption{The parton distribution of $xu_v, xd_v, x\bar{u}, x\bar{d}, xs$, and $xg$ extracted from MMHT2014 \cite{Harland-Lang:2014zoa} PDFs as a function of $x$ at 4, 10, 100, and 8317 GeV$^2$. The results obtained after the profiling procedure compared with corresponding same features  before profiling.}
		\label{fig:partonDisMMHT}
	\end{center}
\end{figure}

\begin{figure}[!htb]
	\begin{center}

        \includegraphics[scale = 0.35]{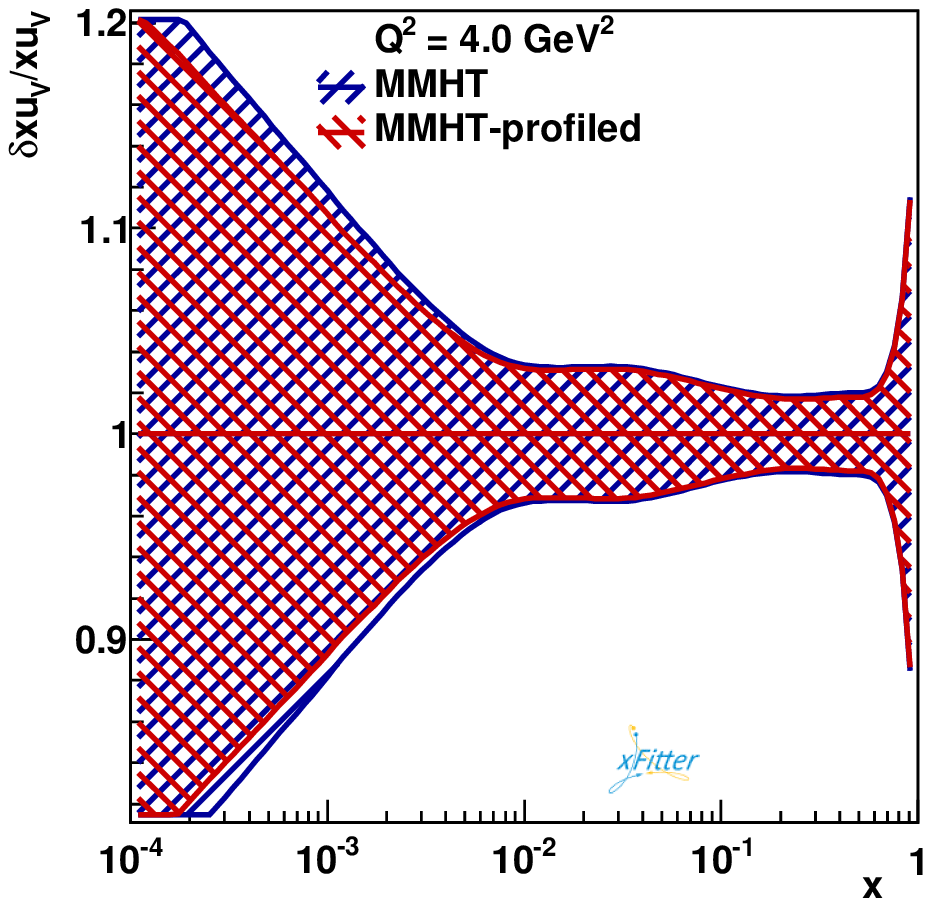}
        \includegraphics[scale = 0.35]{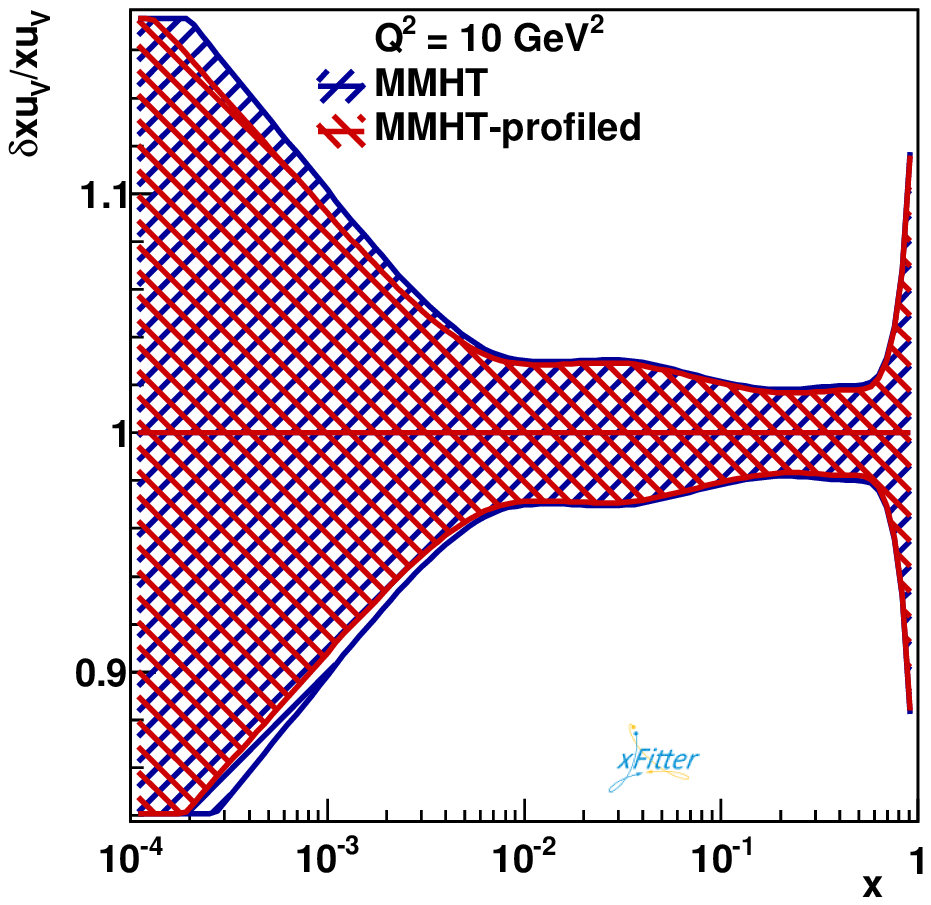}
        \includegraphics[scale = 0.35]{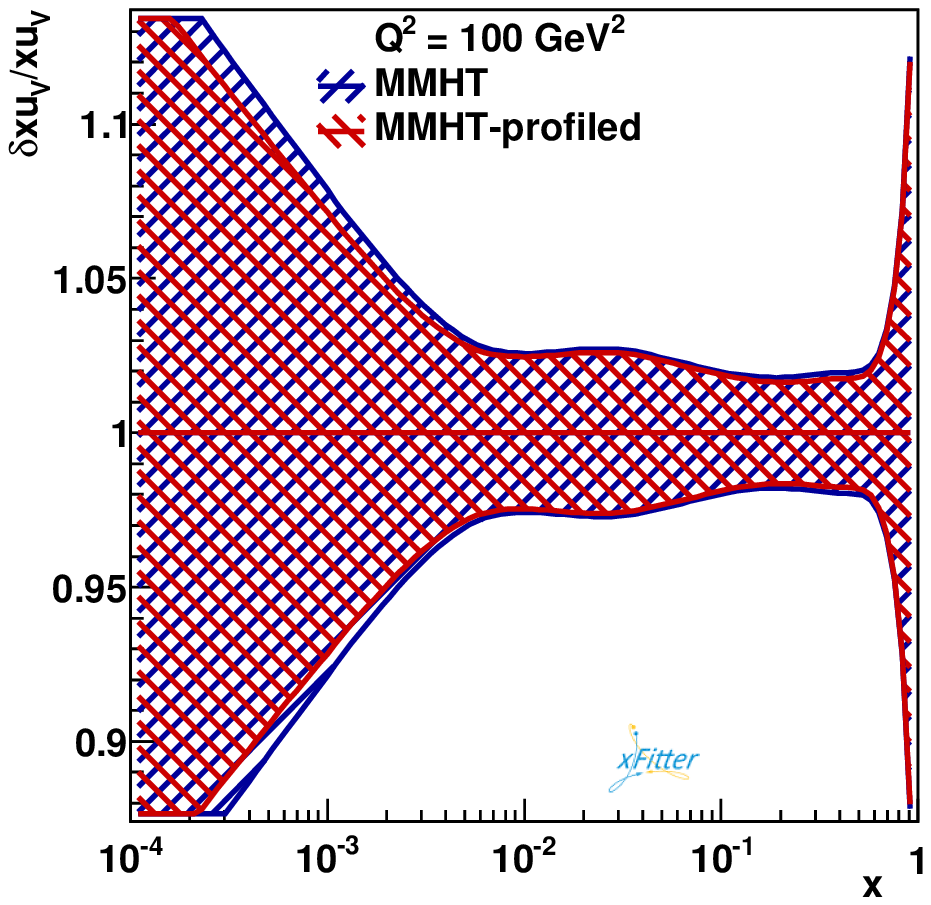}
        \includegraphics[scale = 0.35]{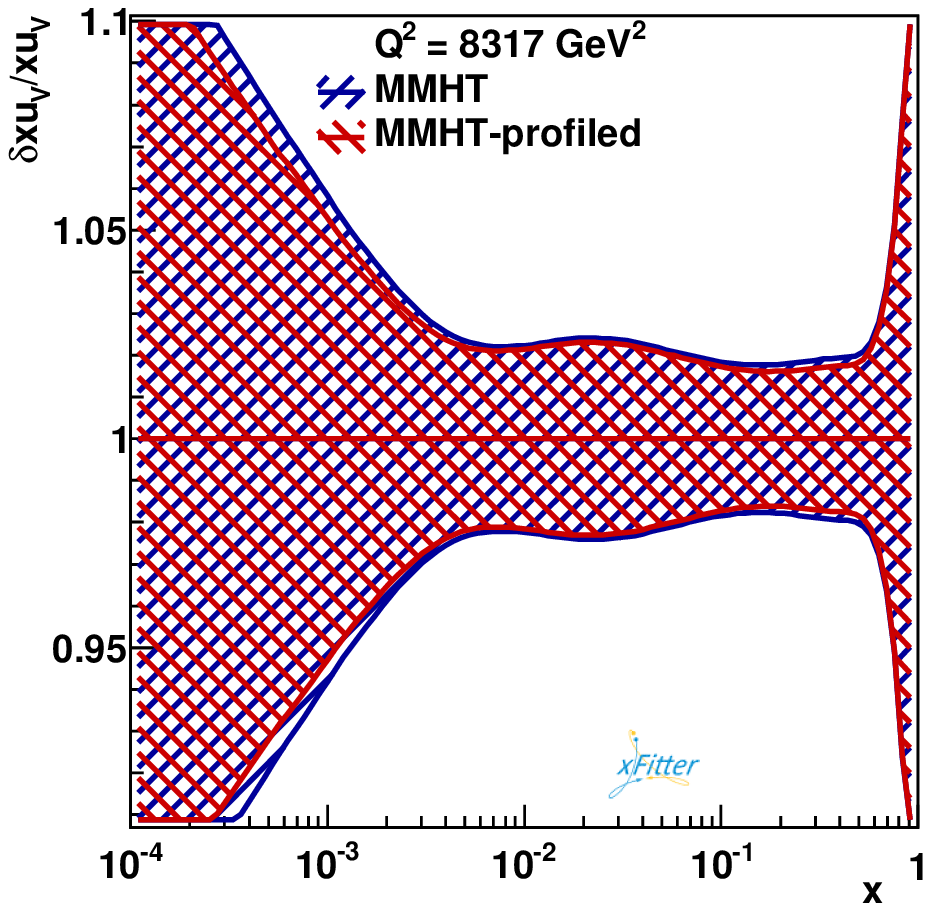}

        \includegraphics[scale = 0.35]{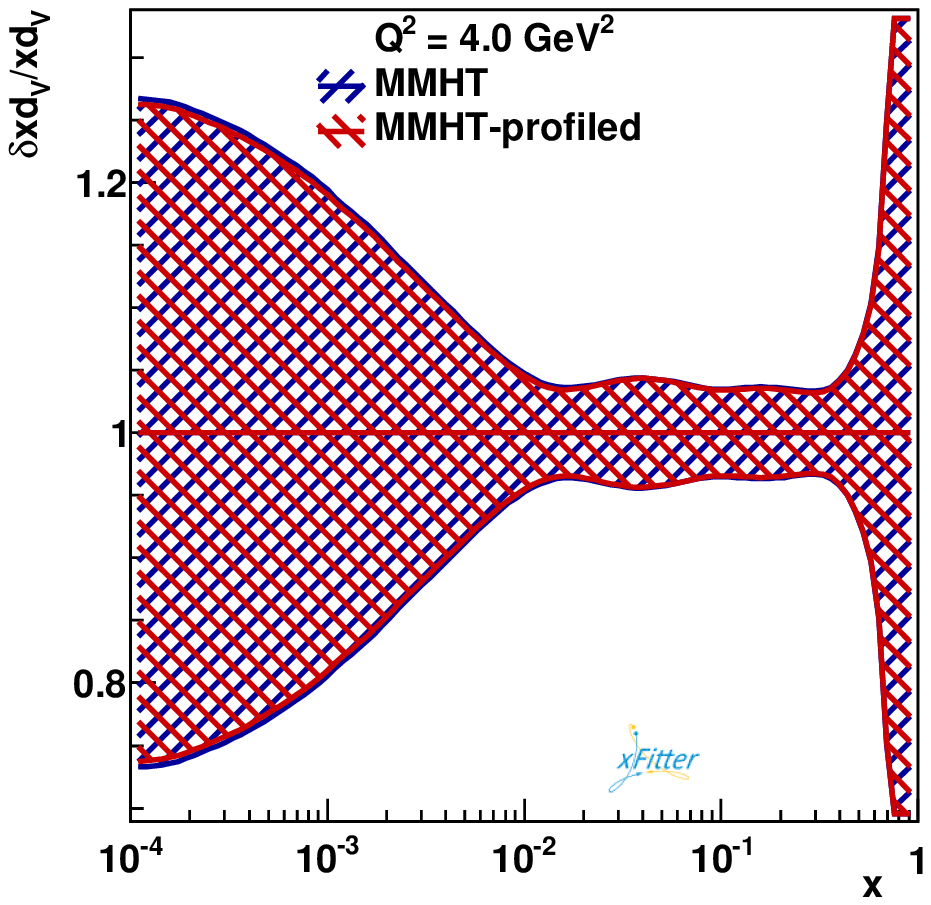}
        \includegraphics[scale = 0.35]{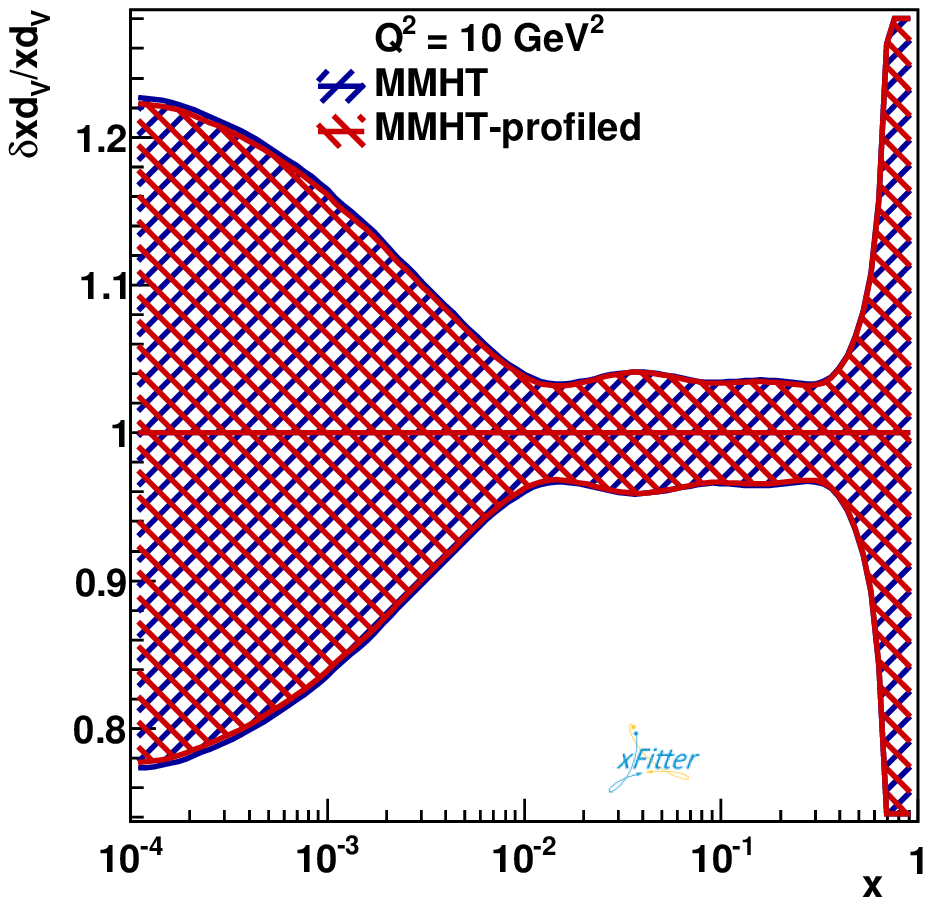}
        \includegraphics[scale = 0.35]{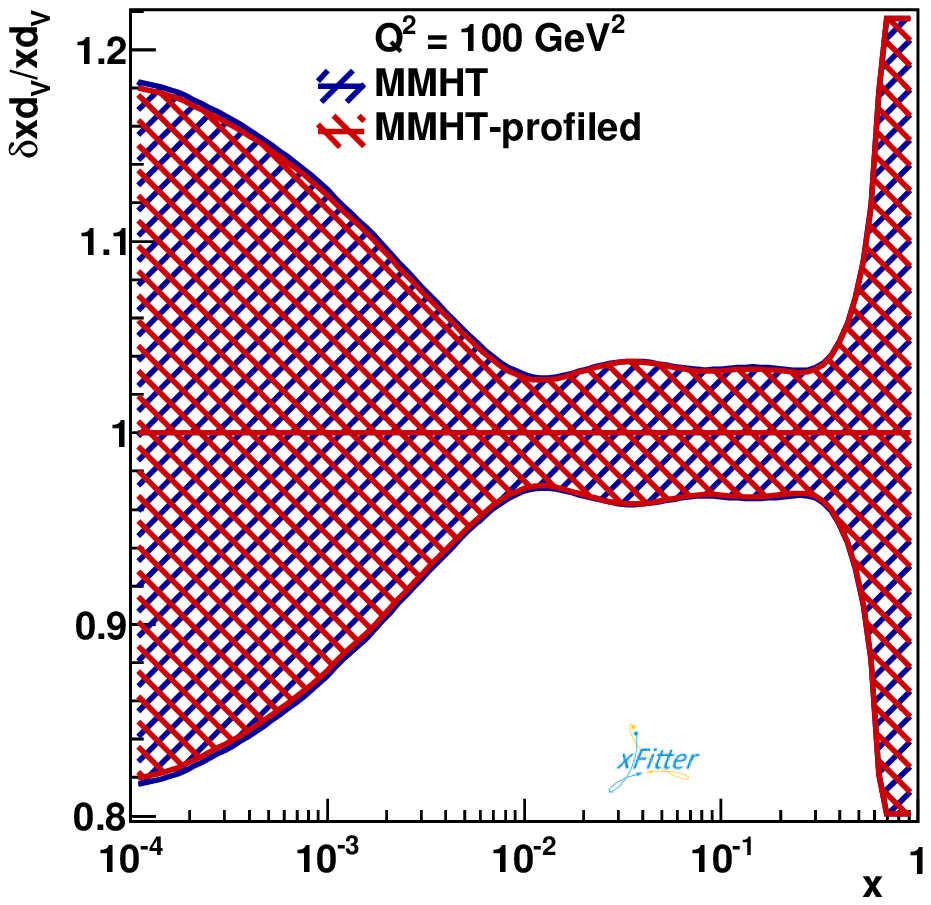}
        \includegraphics[scale = 0.35]{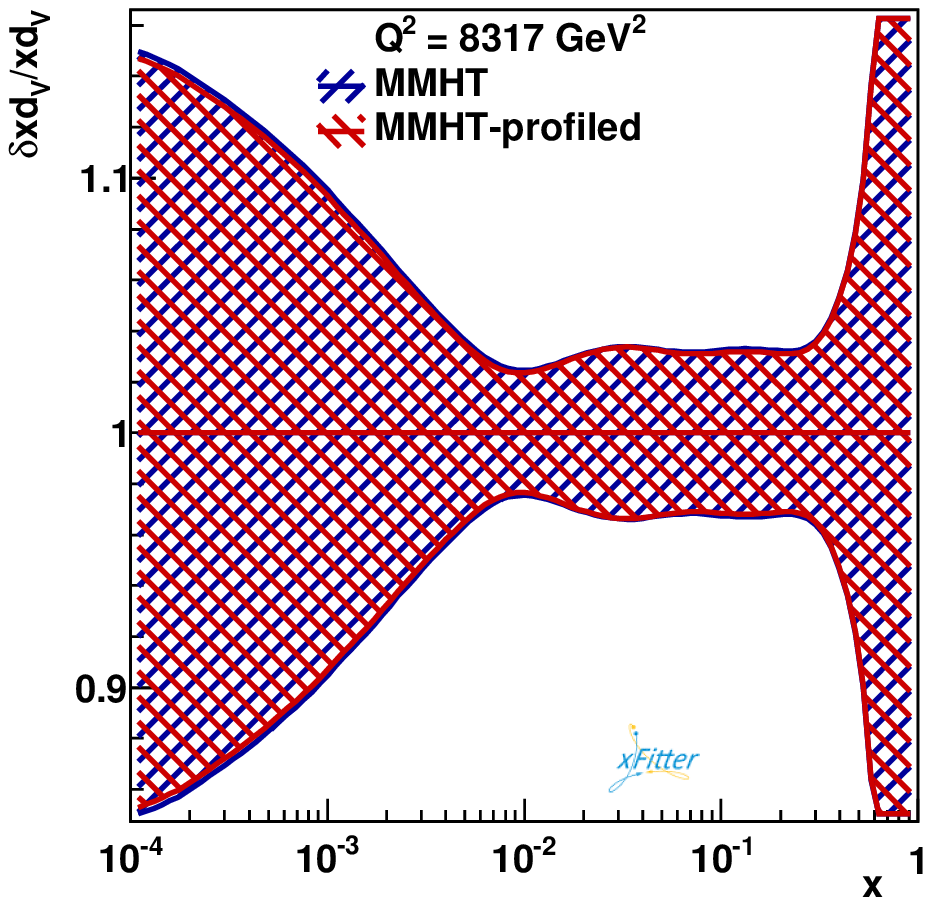}

        \includegraphics[scale = 0.35]{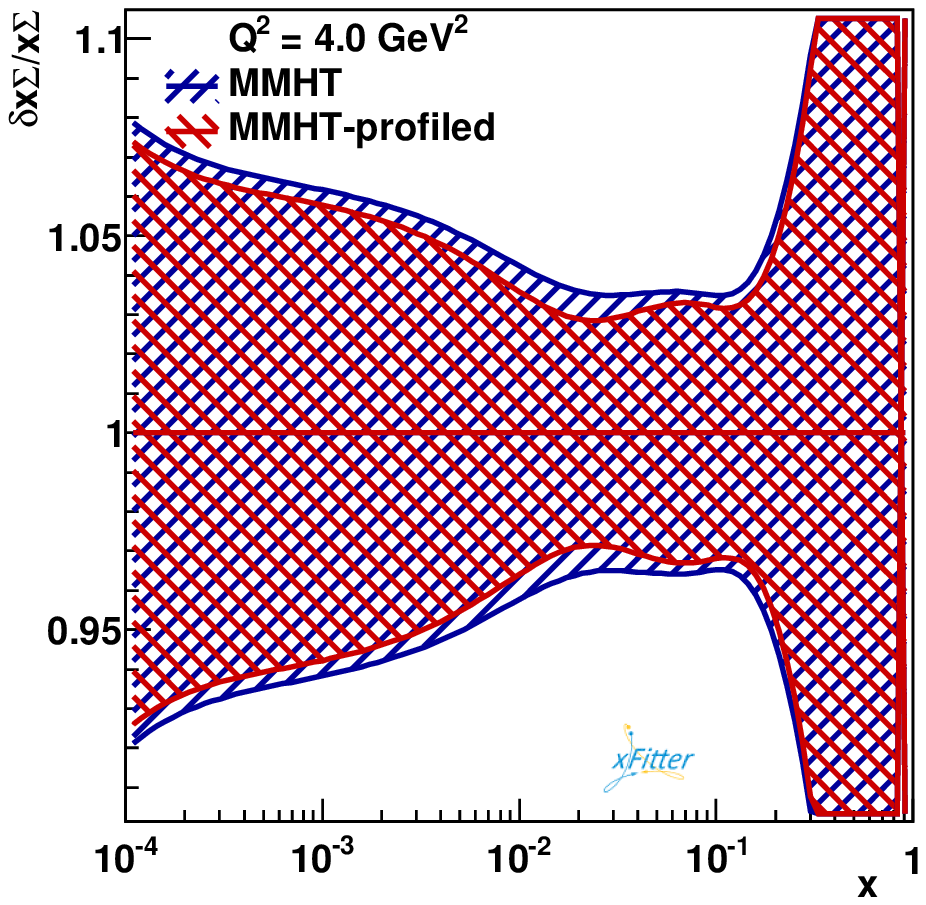}  
        \includegraphics[scale = 0.35]{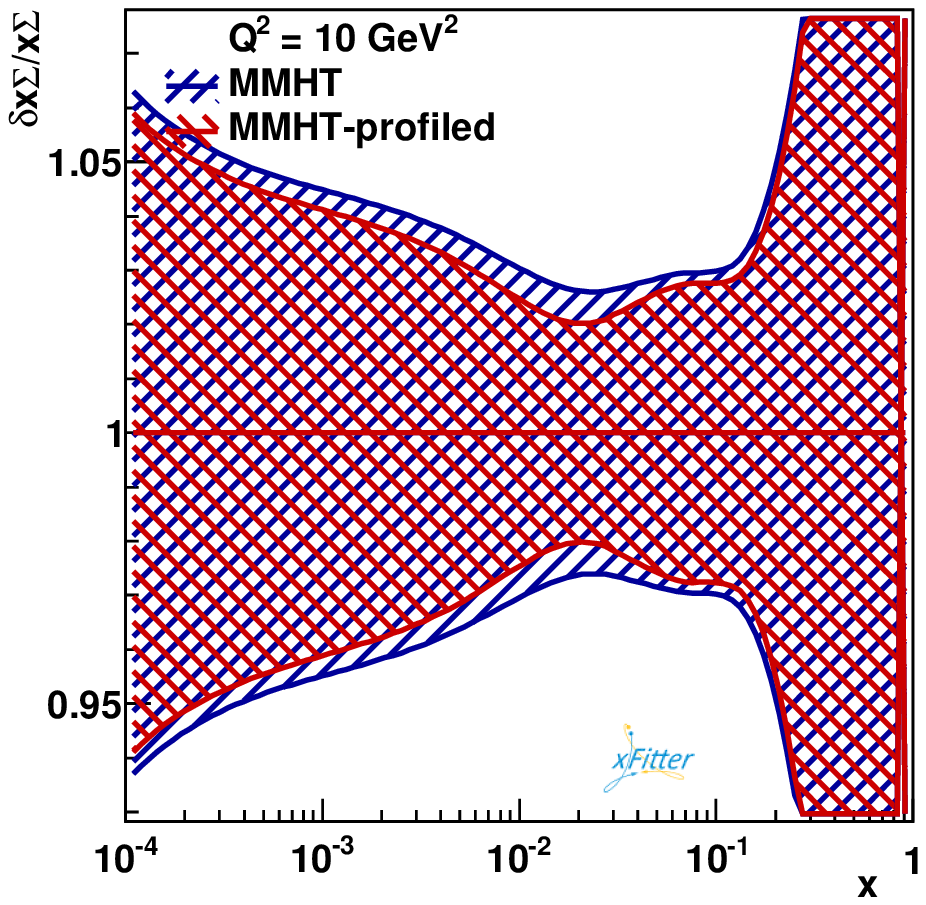}
        \includegraphics[scale = 0.35]{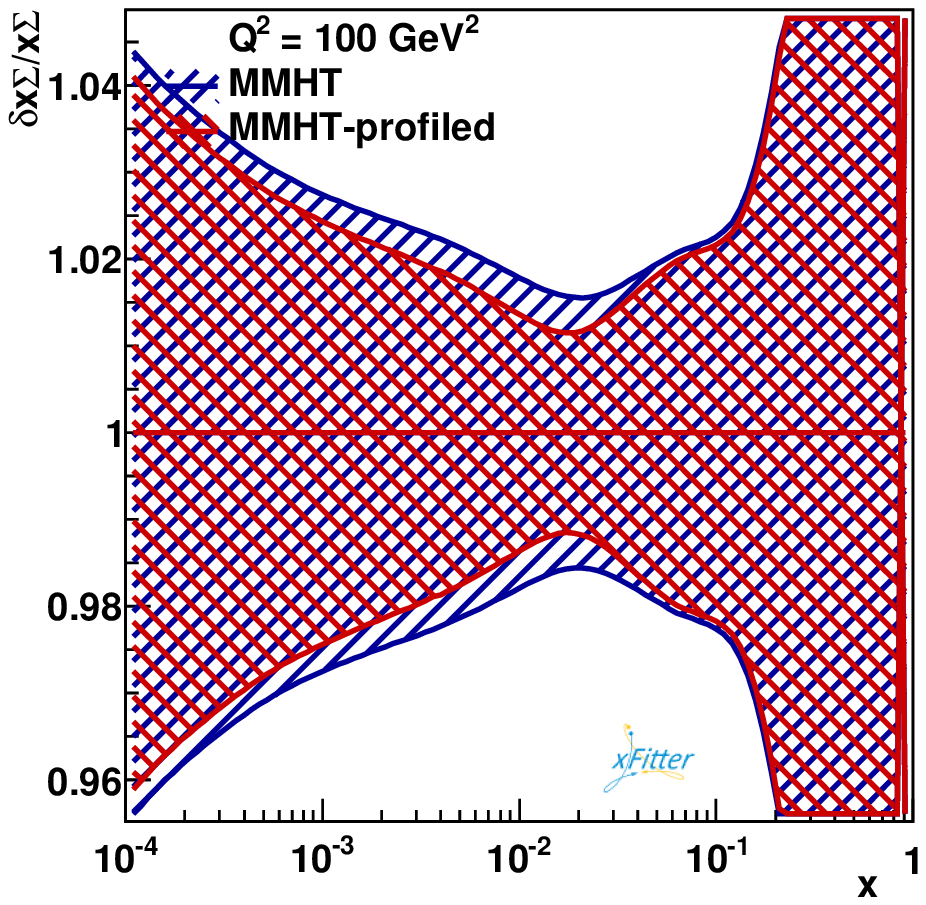}
        \includegraphics[scale = 0.35]{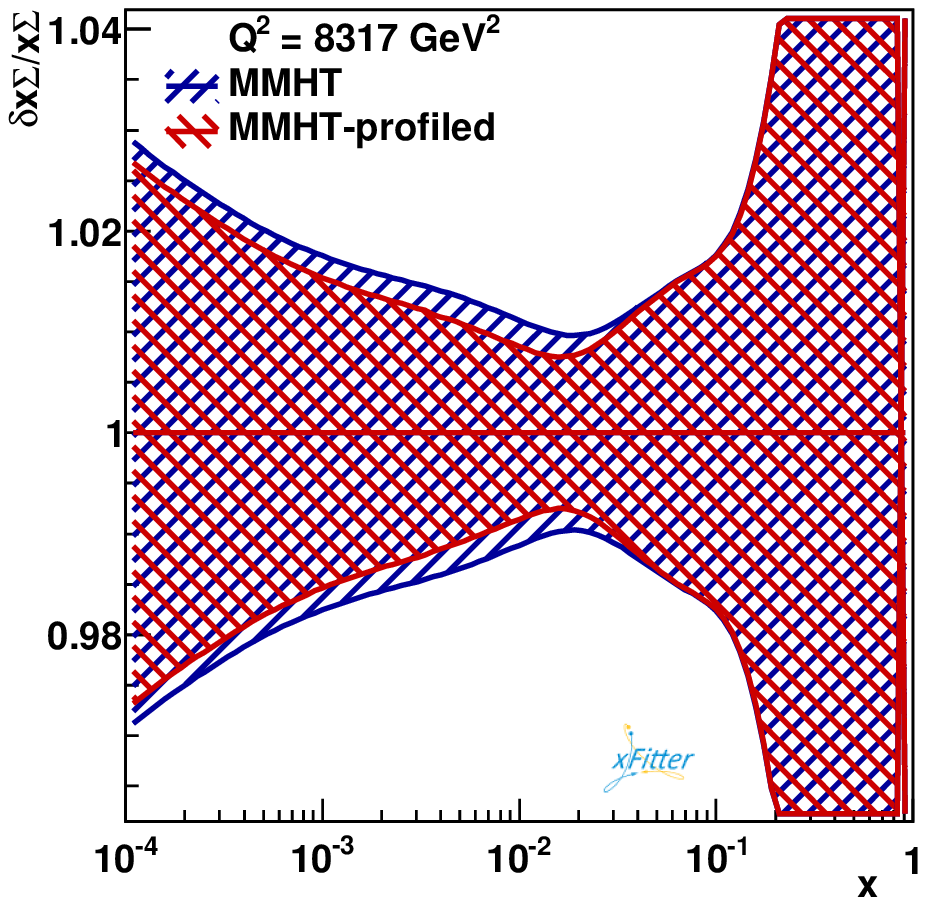}

        \includegraphics[scale = 0.35]{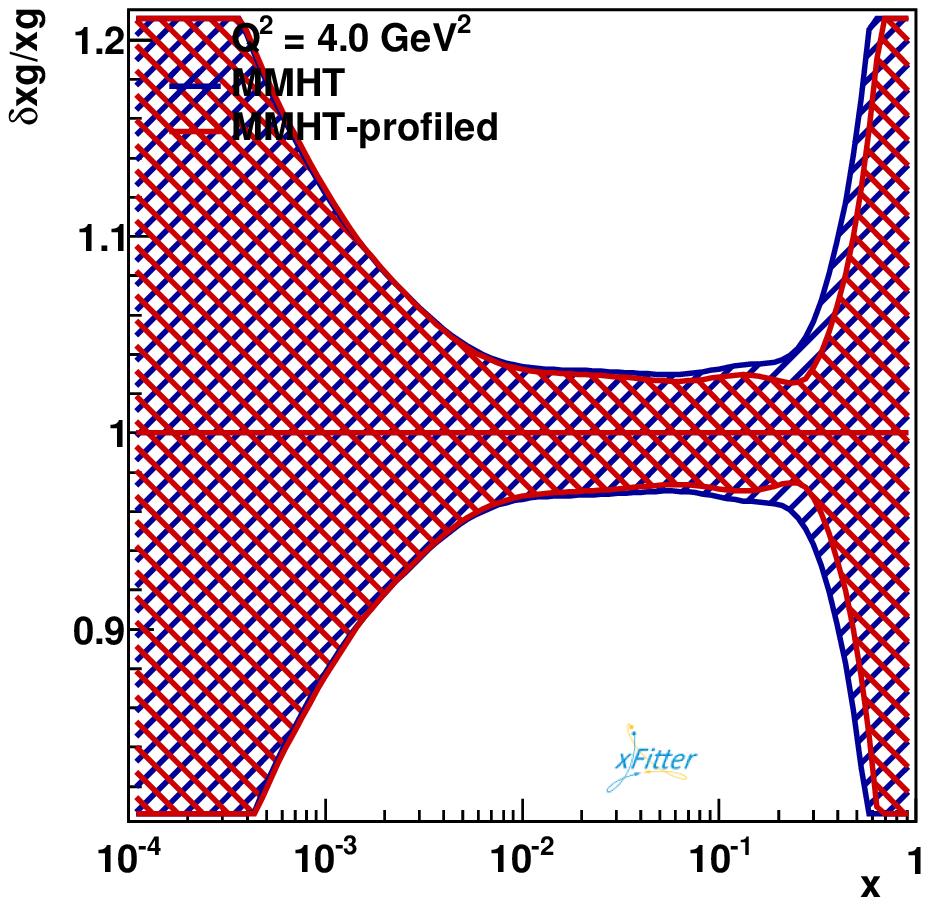} 
        \includegraphics[scale = 0.35]{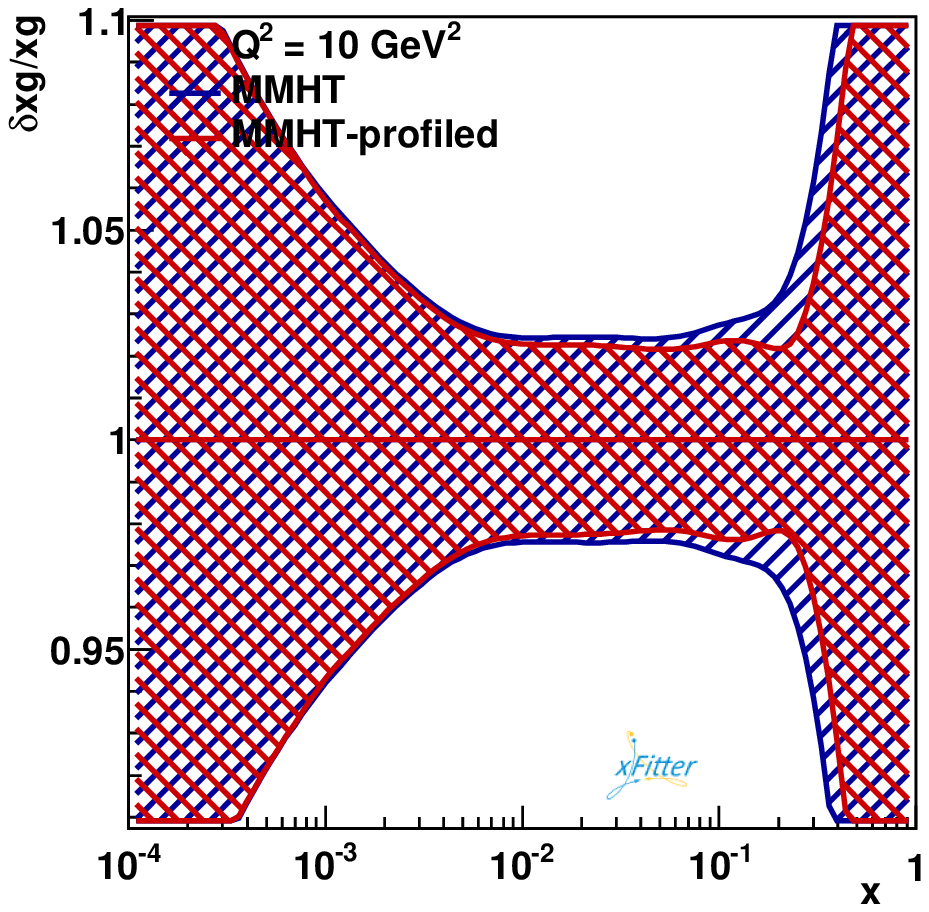}
        \includegraphics[scale = 0.35]{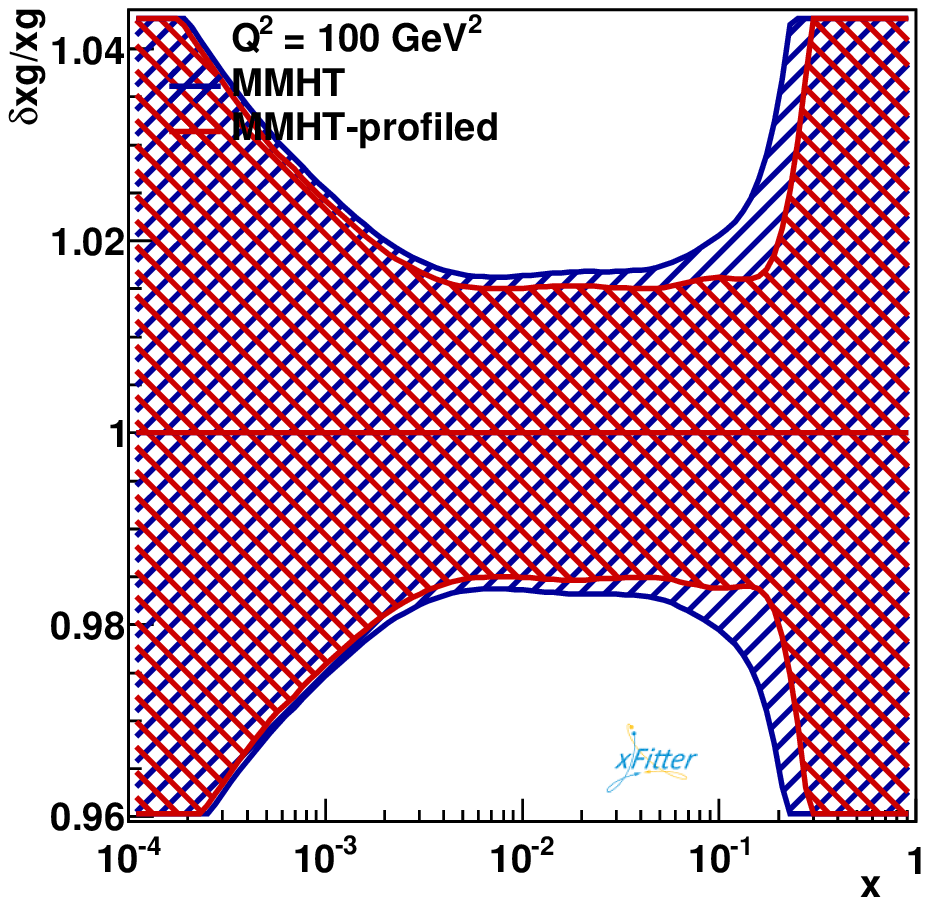}
        \includegraphics[scale = 0.35]{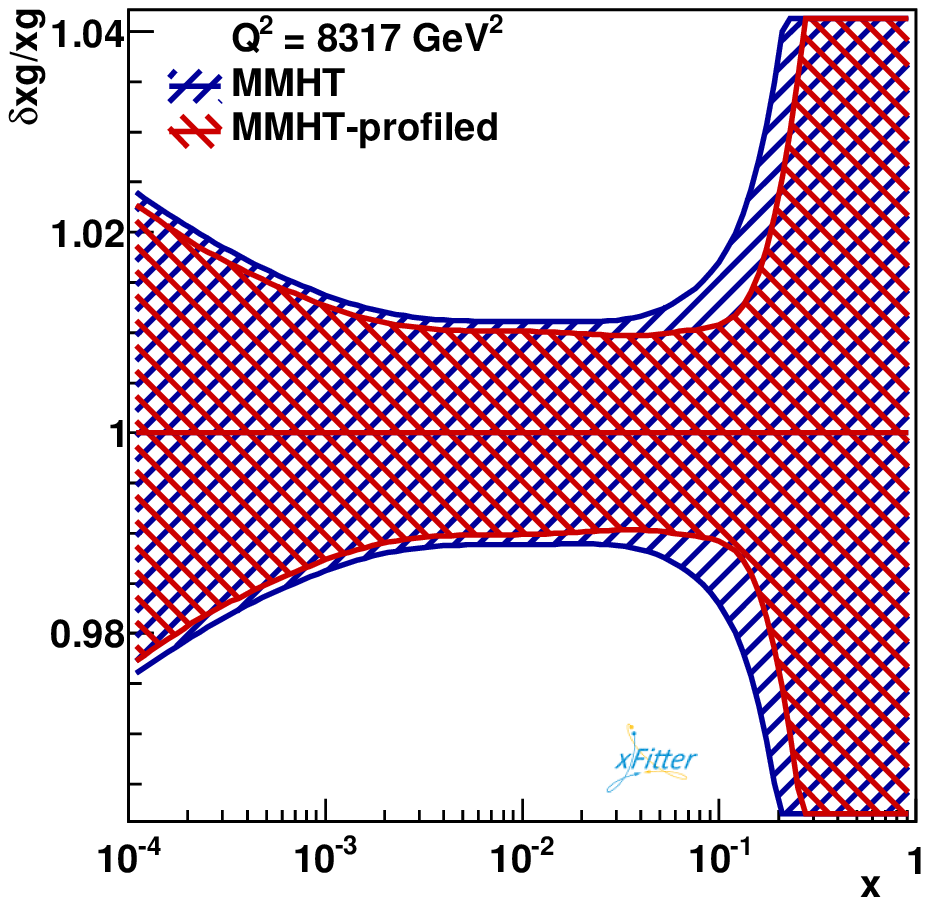}
		
\caption{The relative PDF uncertainties $\delta x u_v/xu_v$, $\delta x d_v/xd_v$, $\delta x \Sigma/x\Sigma$, and $\delta x g/xg$ extracted from MMHT2014 \cite{Harland-Lang:2014zoa} PDFs as a function of $x$ at 4, 10, 100, and 8317 GeV$^2$. The results obtained after the profiling procedure compared with corresponding same features  before profiling. Newly added top quark data obviously constrained distributions of $\delta x \Sigma/x\Sigma$, and $\delta x g/xg$.}
		\label{fig:partonRatioMMHT}
	\end{center}
\end{figure}

Same as figures \ref{fig:partonDisCT14} and \ref{fig:partonRatioCT14}, the results for MMHT2014   are shown in Figs.~\ref{fig:partonDisMMHT} and \ref{fig:partonRatioMMHT}.   It is seen that the new top quark cross section data provide significant changes on the central values and the uncertainties  of $xs$, and $xg$ PDFs.

\begin{figure}[!htb]
	\begin{center}

        \includegraphics[scale = 0.35]{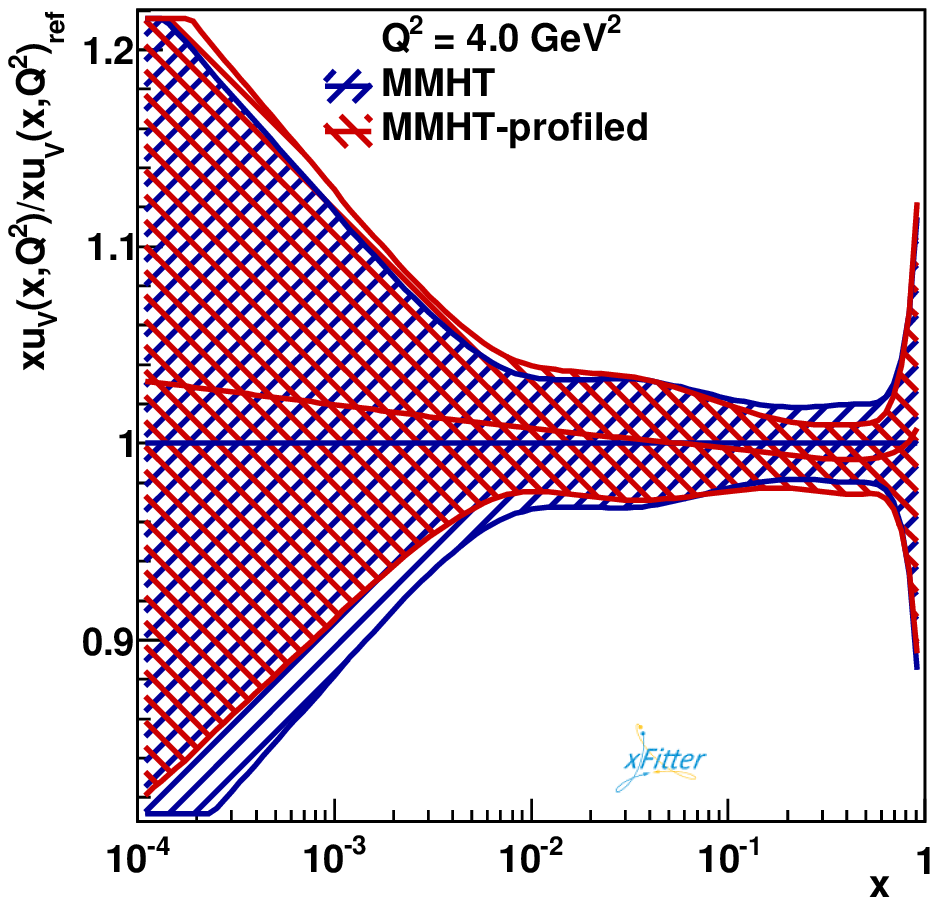}
        \includegraphics[scale = 0.35]{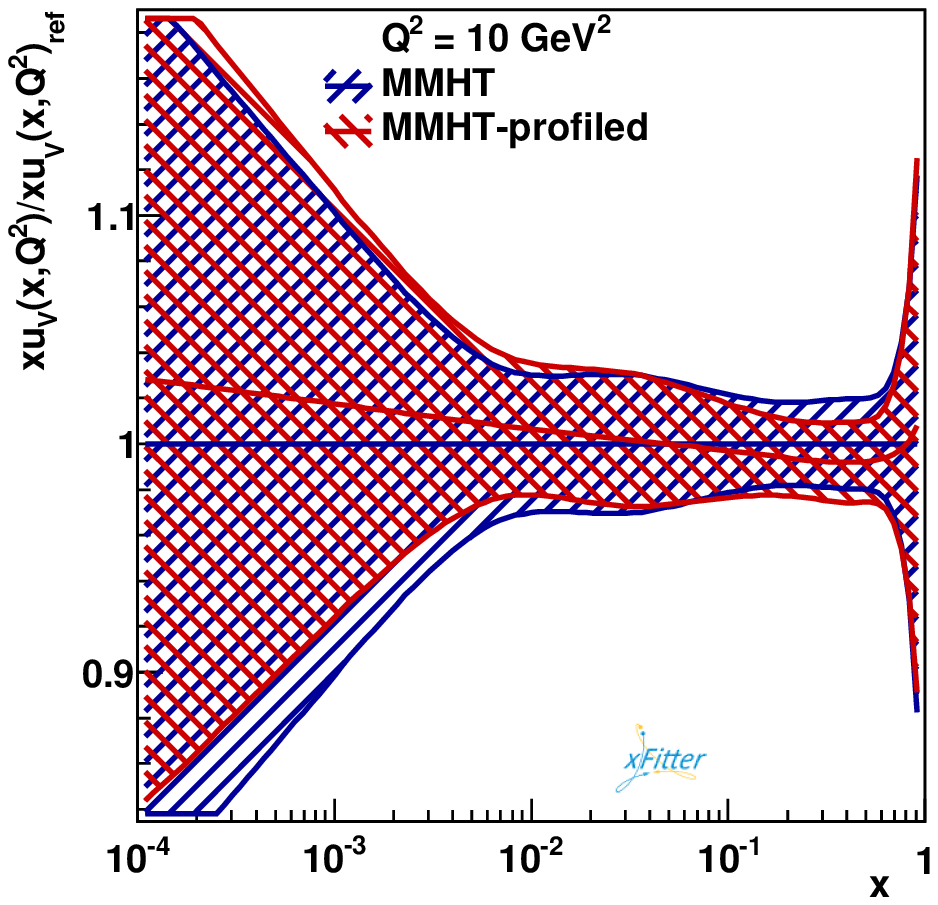}
        \includegraphics[scale = 0.35]{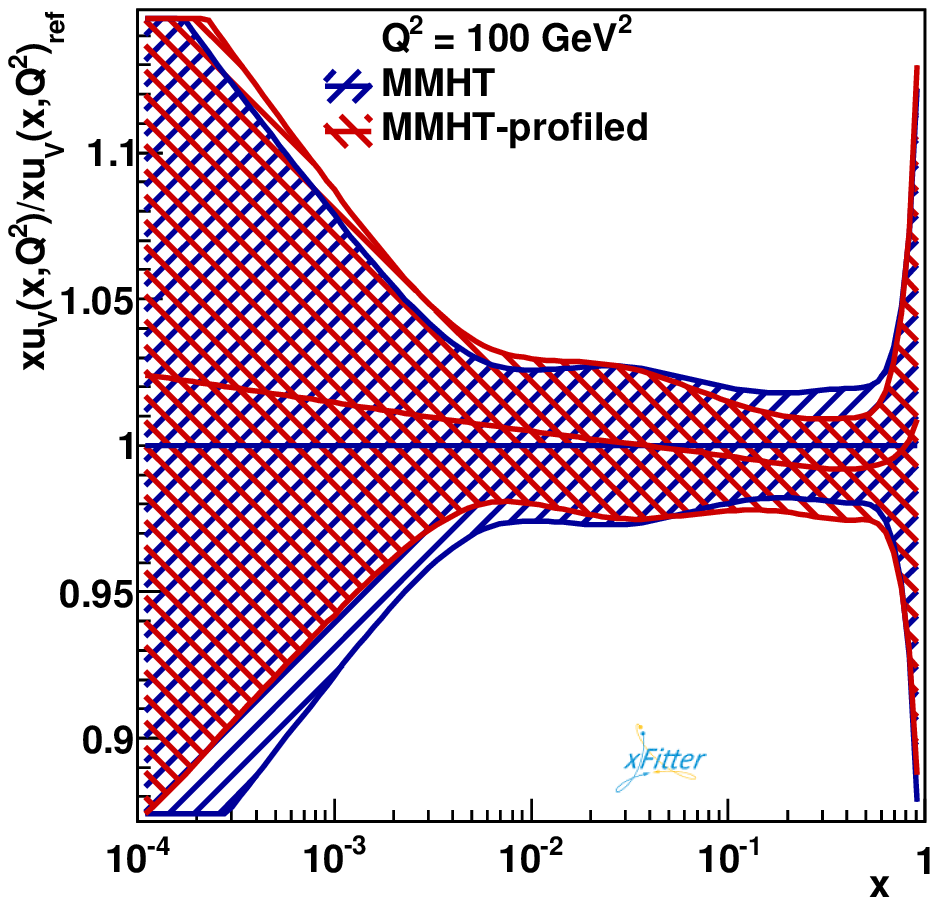}
        \includegraphics[scale = 0.35]{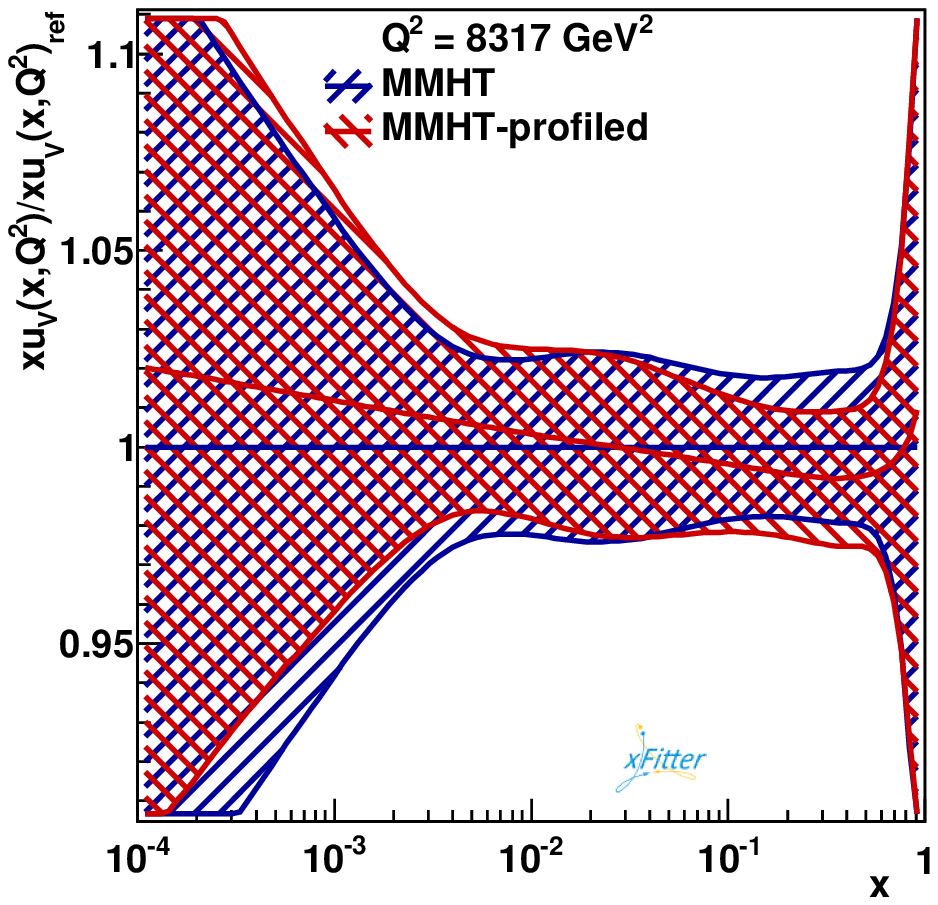}

        \includegraphics[scale = 0.35]{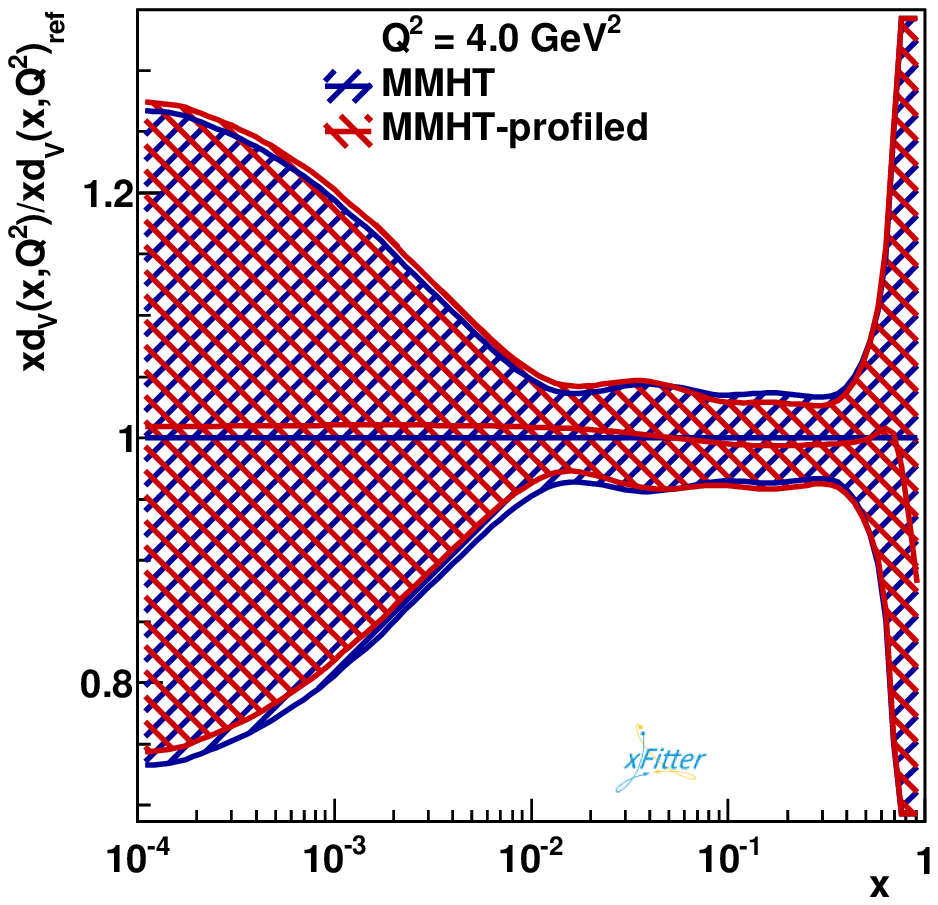}
        \includegraphics[scale = 0.35]{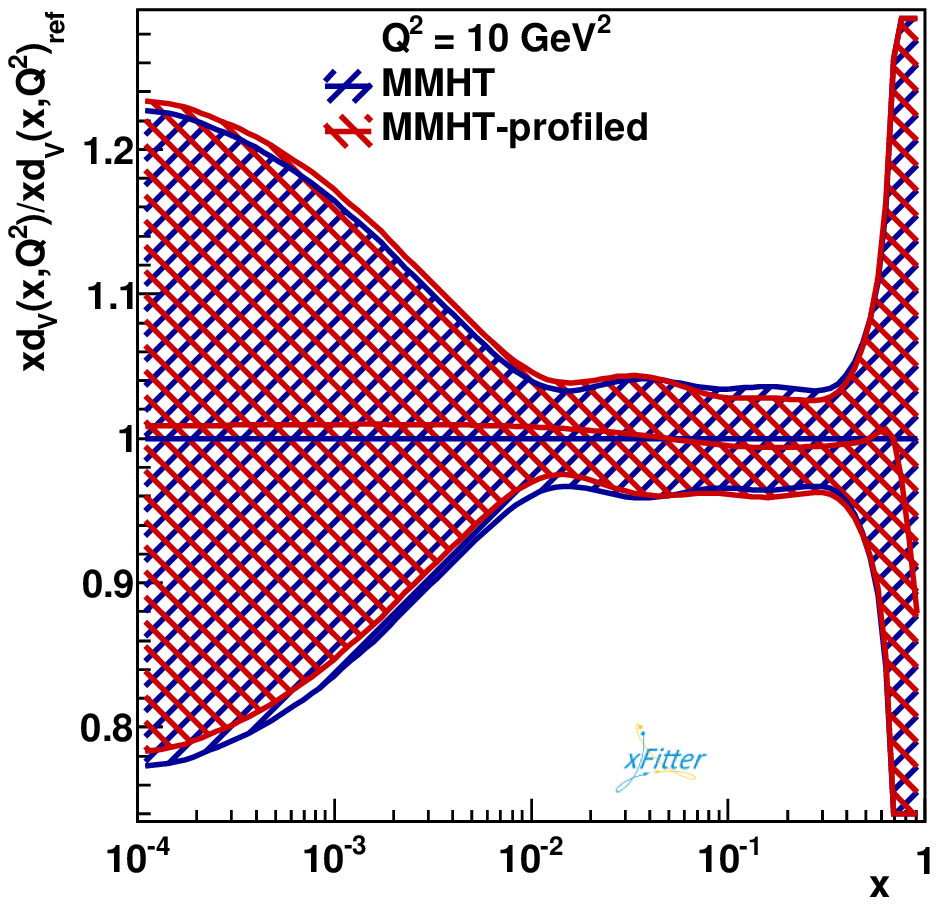}
        \includegraphics[scale = 0.35]{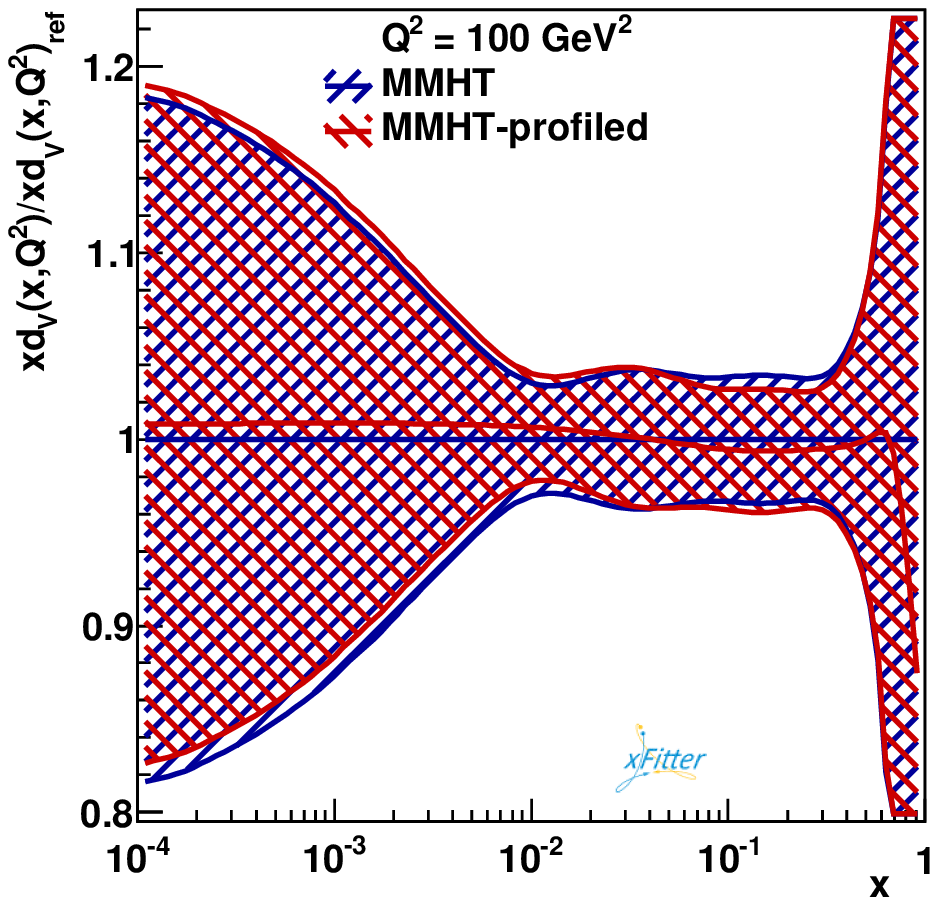}
        \includegraphics[scale = 0.35]{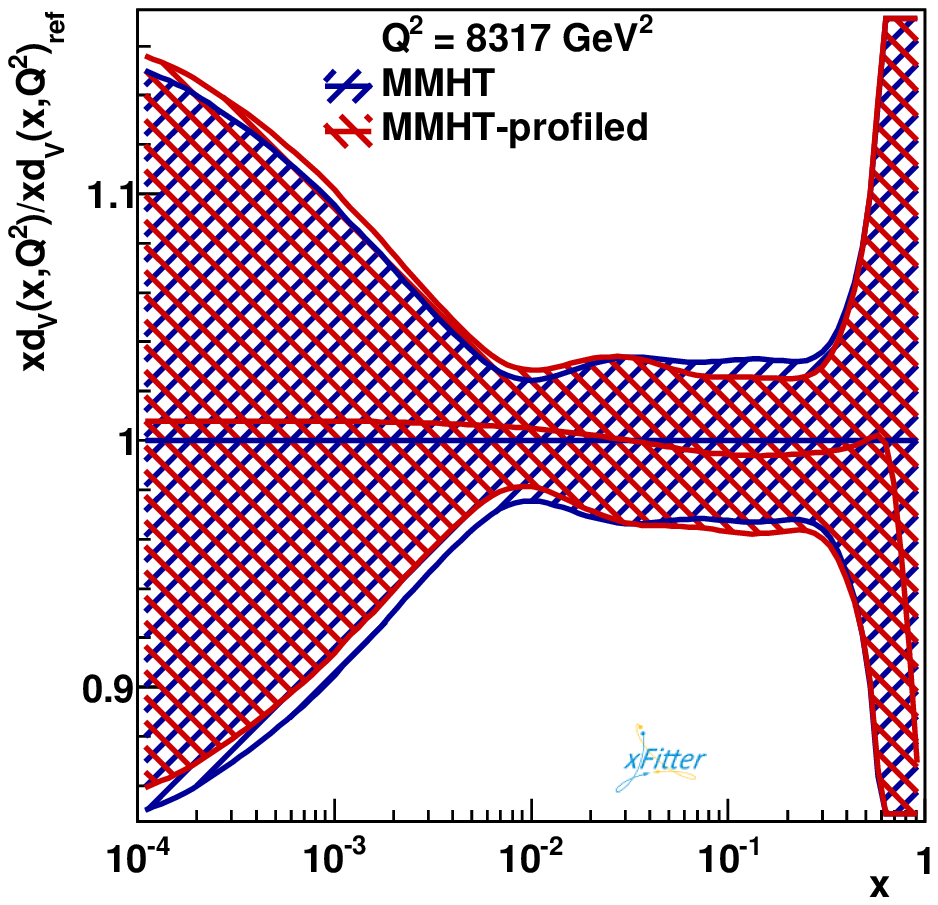}

        \includegraphics[scale = 0.35]{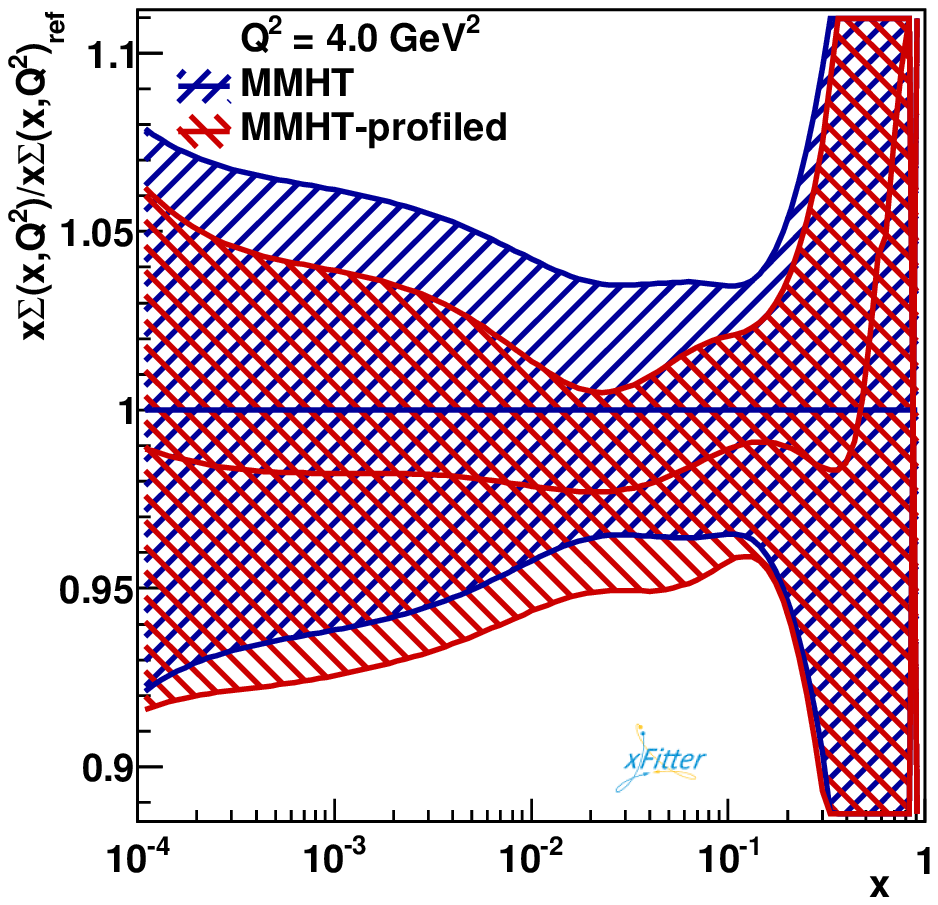}  
        \includegraphics[scale = 0.35]{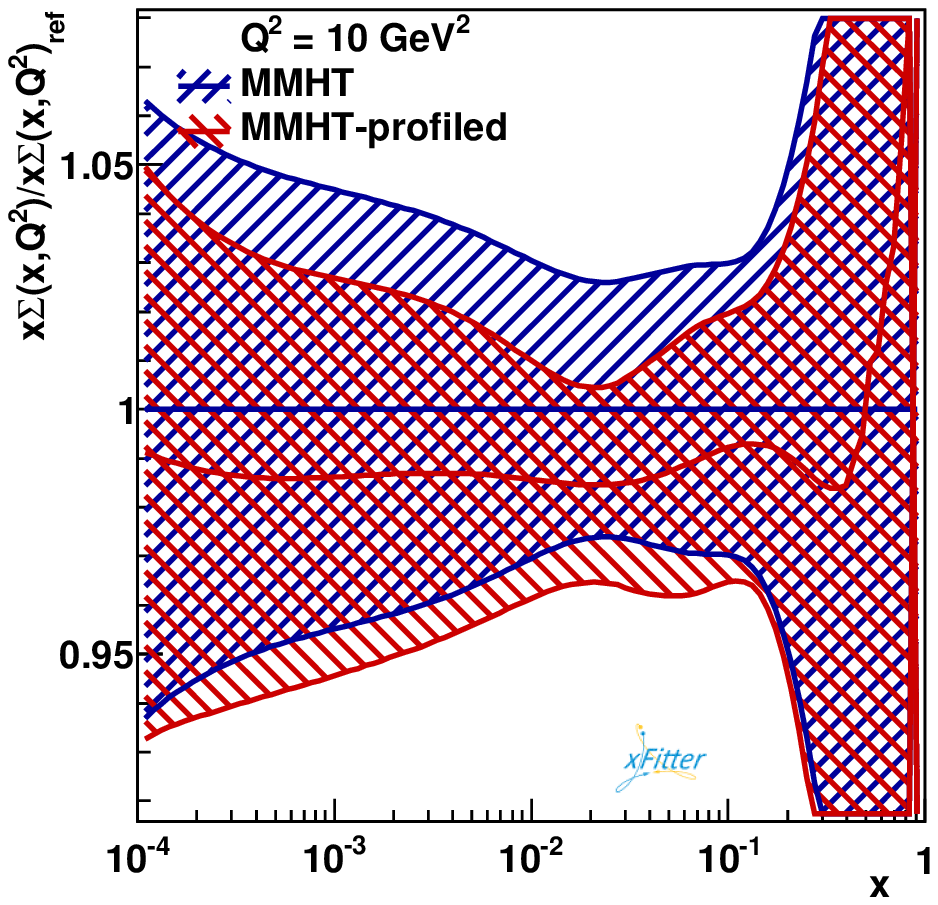}
        \includegraphics[scale = 0.35]{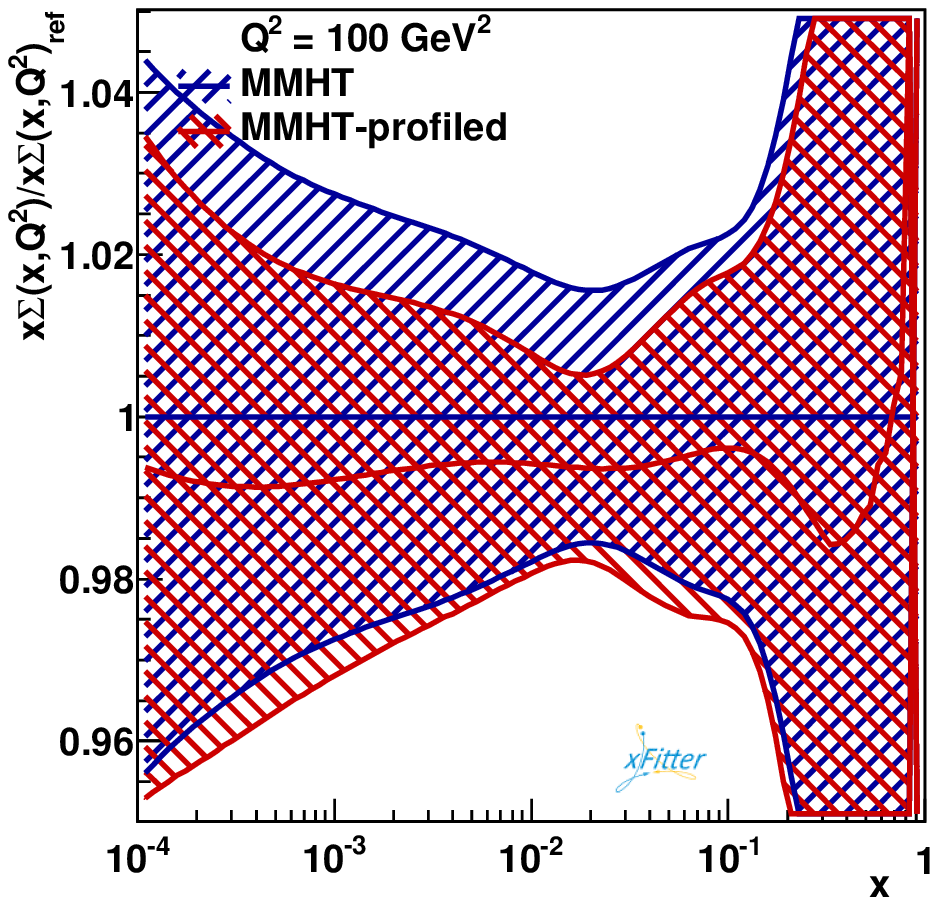}
        \includegraphics[scale = 0.35]{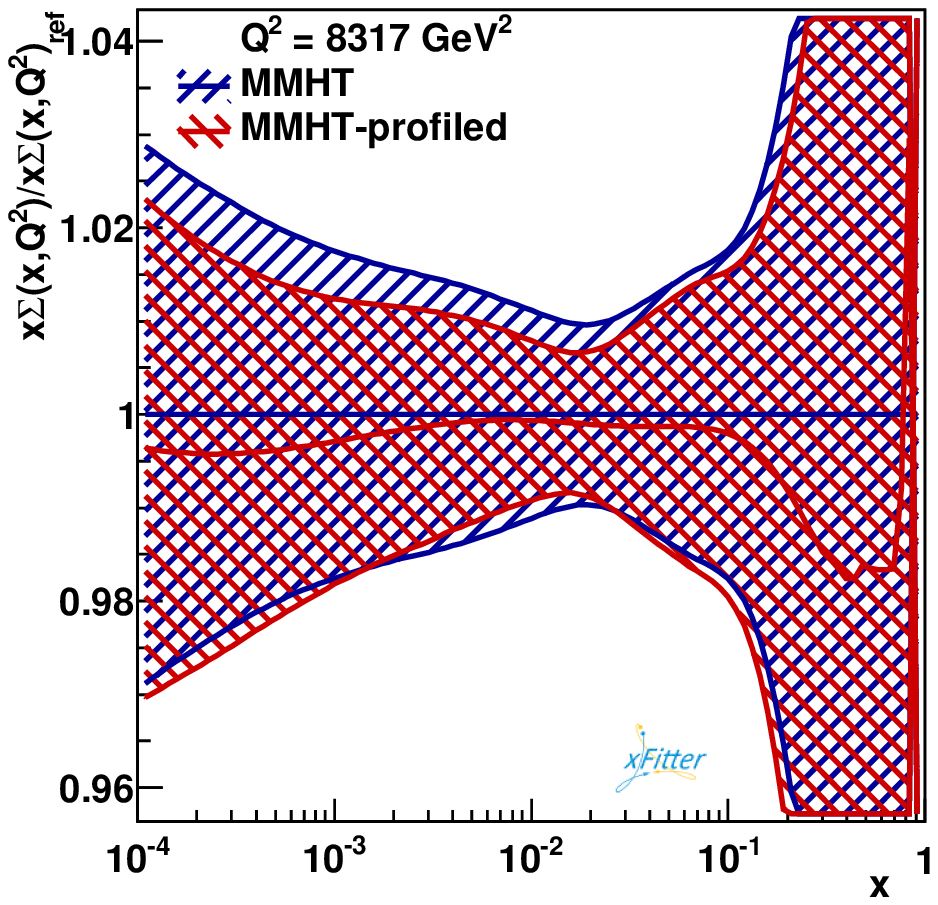}

        \includegraphics[scale = 0.35]{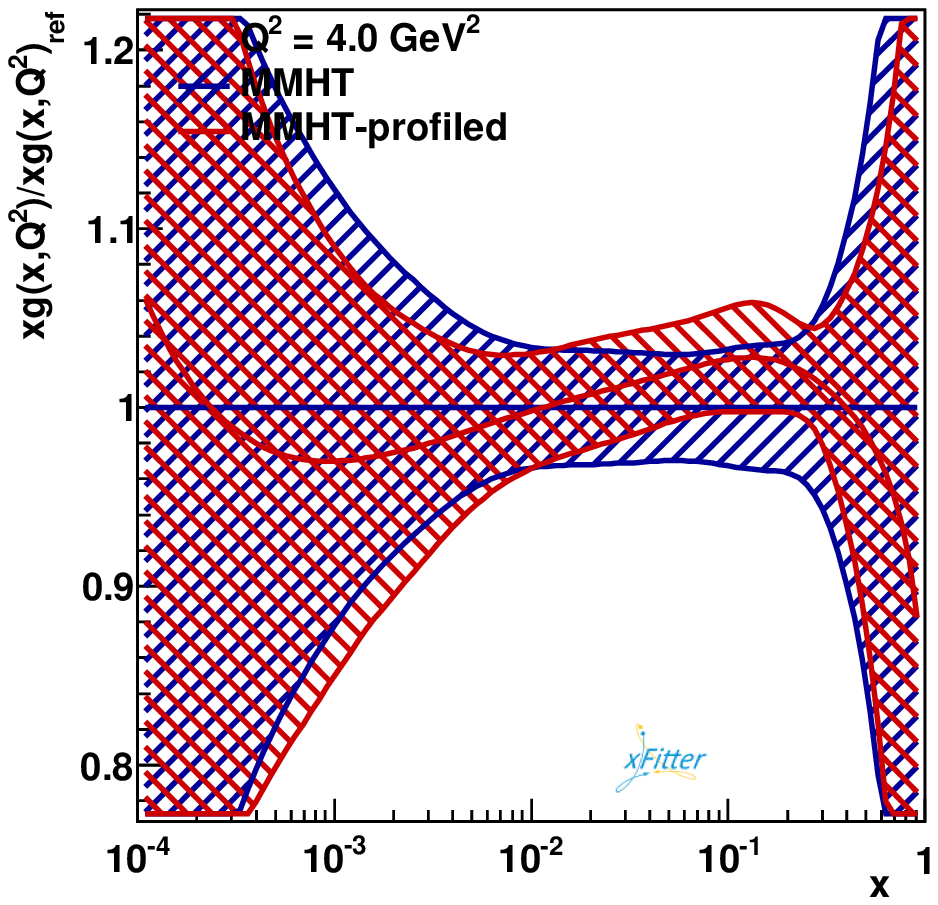} 
        \includegraphics[scale = 0.35]{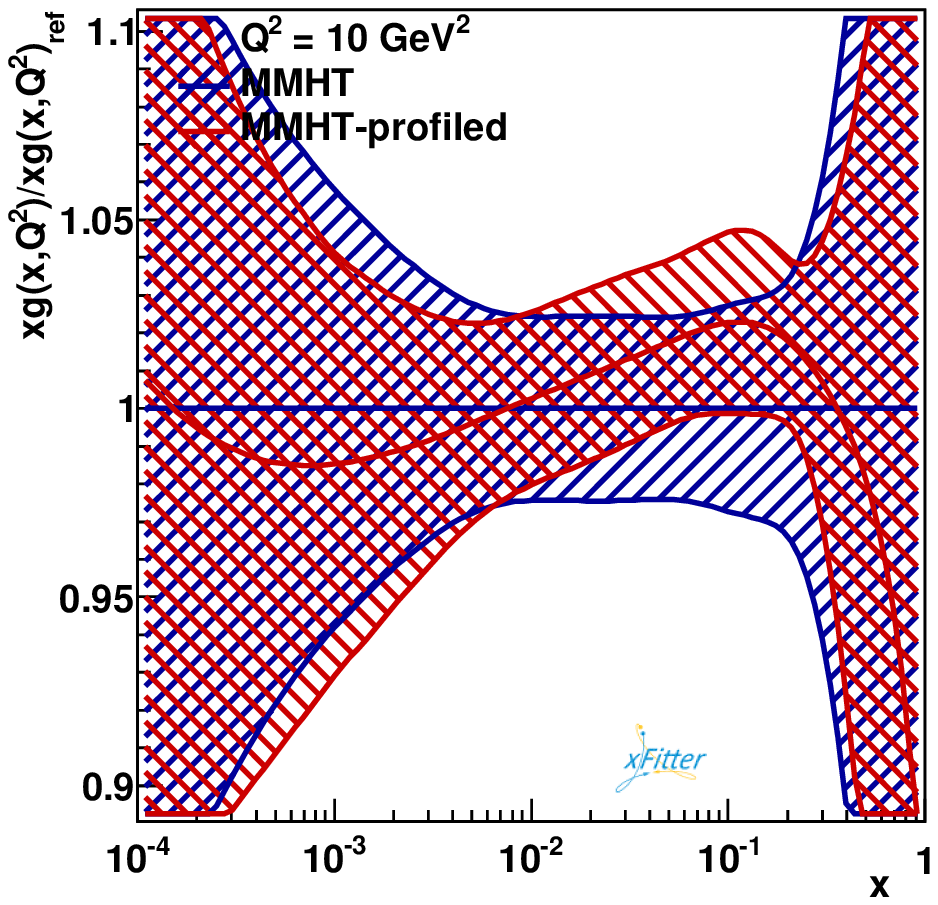}
        \includegraphics[scale = 0.35]{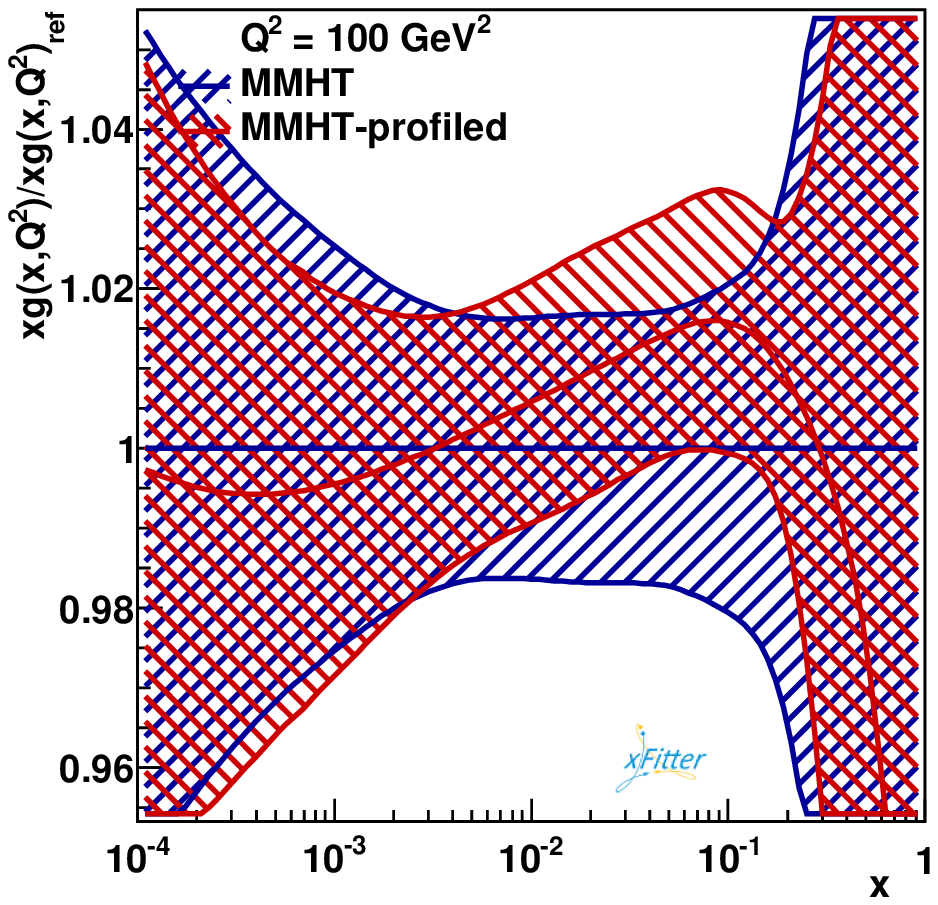}
        \includegraphics[scale = 0.35]{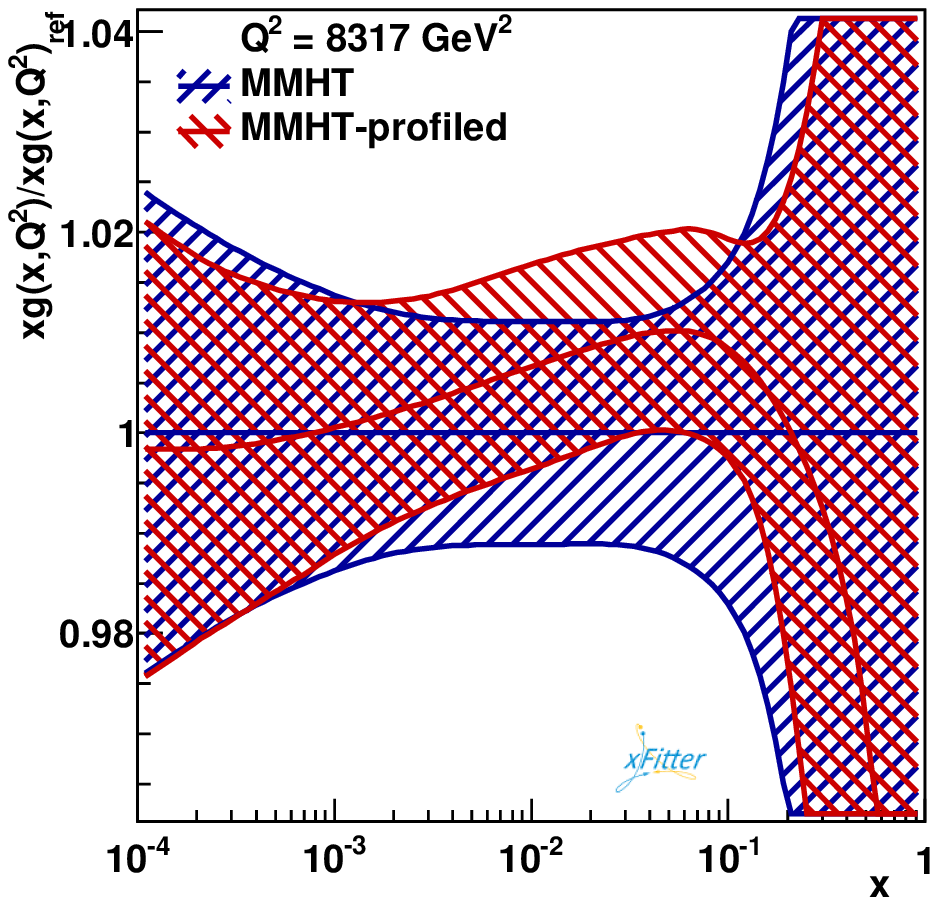}
		
\caption{The parton distribution ratio  $xu_v/xu_{v_{ref}}$, $xd_v/xd_{v_{ref}}$, $x\Sigma/x\Sigma_{ref}$, and $xg/xg_{ref}$ with respect to without profiling procedure, extracted from MMHT2014 \cite{Harland-Lang:2014zoa} PDFs as a function of $x$ at  4, 10, 100, and 8317 GeV$^2$. The results obtained after the profiling procedure compared with corresponding same features before profiling.}
		\label{fig:partonRefMMHT}
	\end{center}
\end{figure}

According to Fig.~\ref{fig:partonRatioMMHT}, the relative PDF uncertainty of $x\delta u_v/xu_v$ is affected at low  and large-$x$. Also, the  relative PDF uncertainty of  $x\delta \Sigma/x\Sigma$ is affected at medium-$x$, but the relative PDF uncertainty of $x\delta g/xg$  decreases significantly at high $x$. Finally, in Fig.~\ref{fig:partonRefMMHT} the impact of the recent top quark cross section data on the parton distribution ratio  $xu_v/xu_{v_{ref}}$, $xd_v/xd_{v_{ref}}$, $x\Sigma/x\Sigma_{ref}$, and $xg/xg_{ref}$  for MMHT2014 PDFs are presented. 

\begin{figure}[!htb]
	\begin{center}

	    \includegraphics[scale = 0.35]{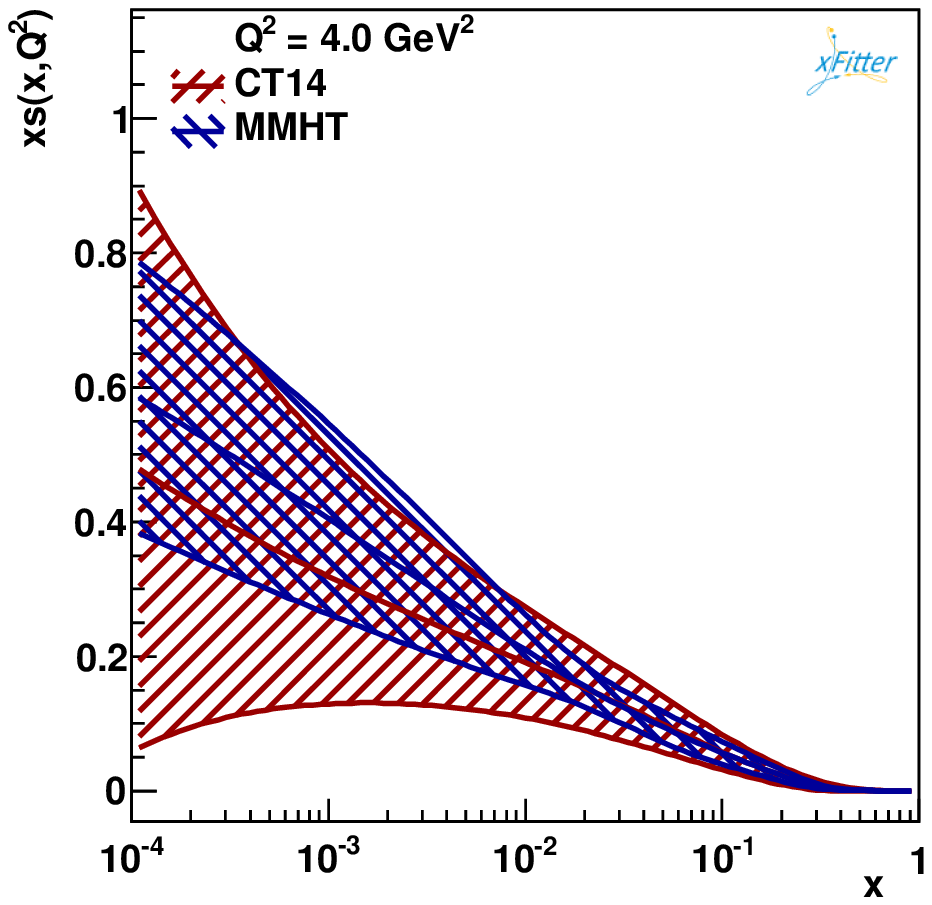}        
        	\includegraphics[scale = 0.35]{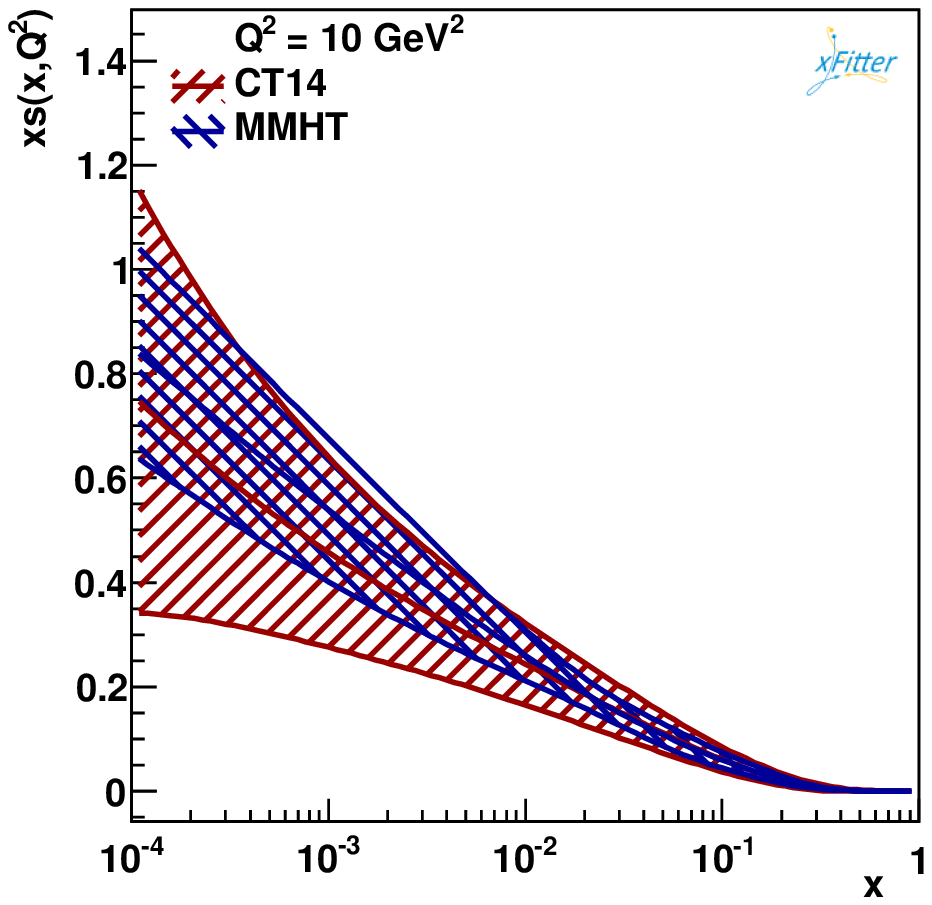}
	    \includegraphics[scale = 0.35]{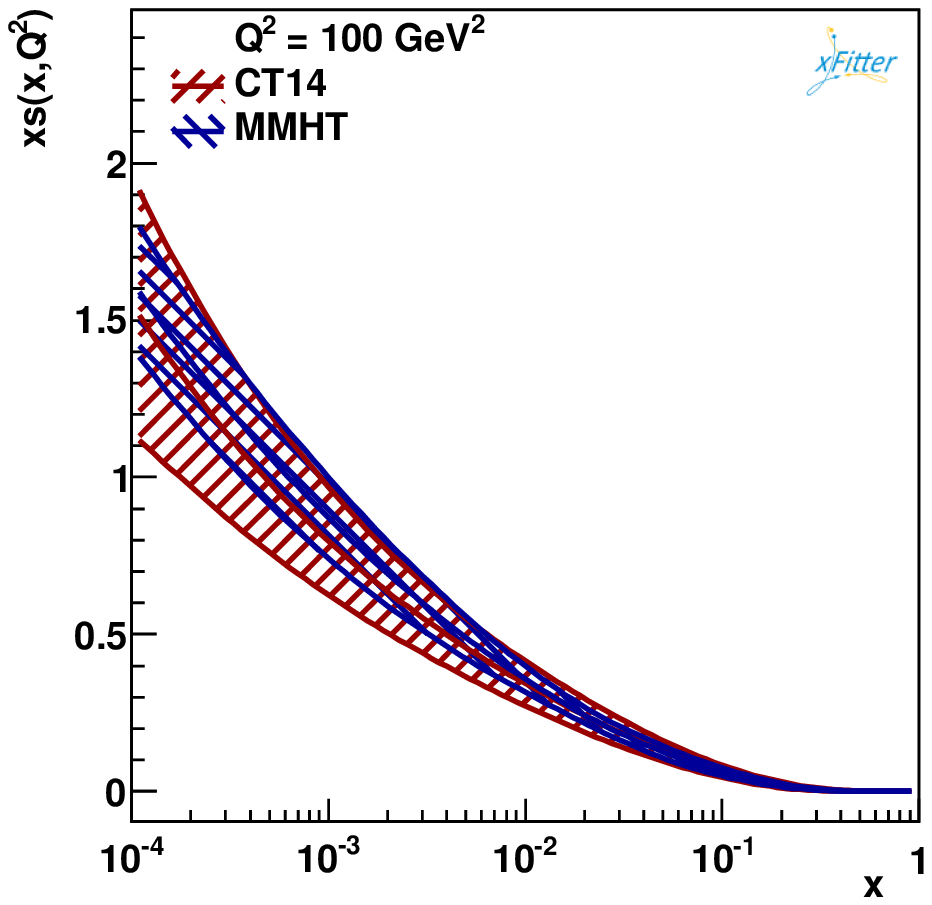}
	    \includegraphics[scale = 0.35]{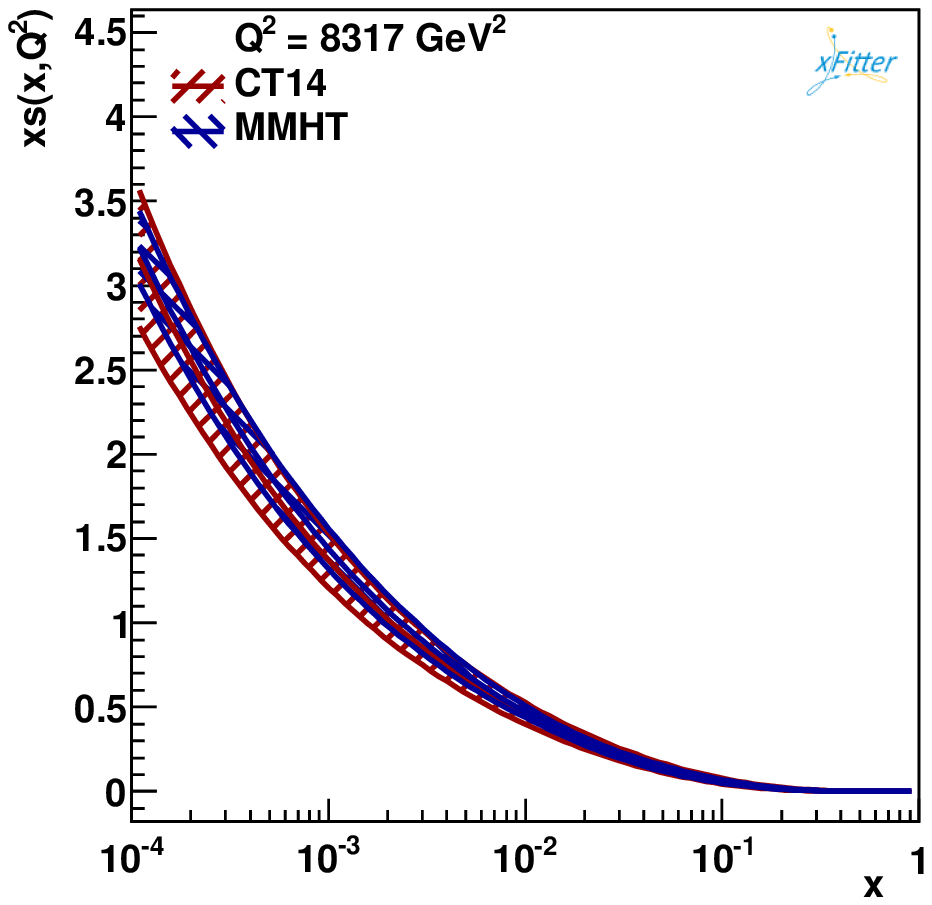}

	    \includegraphics[scale = 0.35]{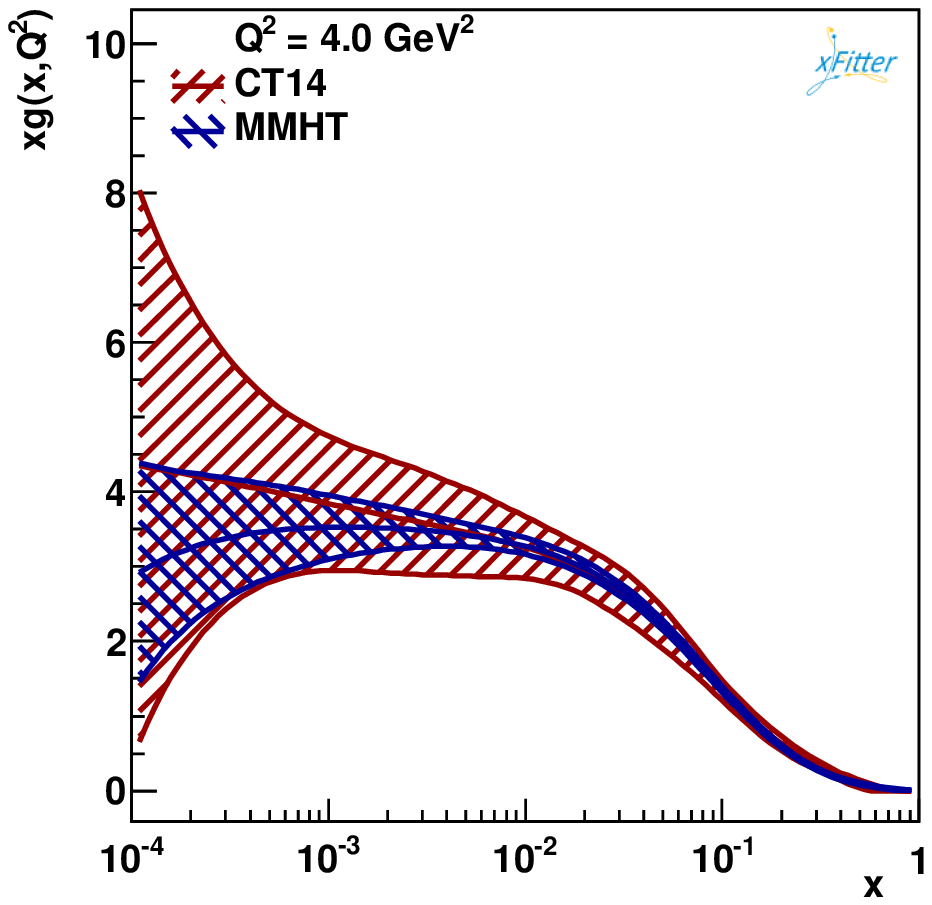}     
	    \includegraphics[scale = 0.35]{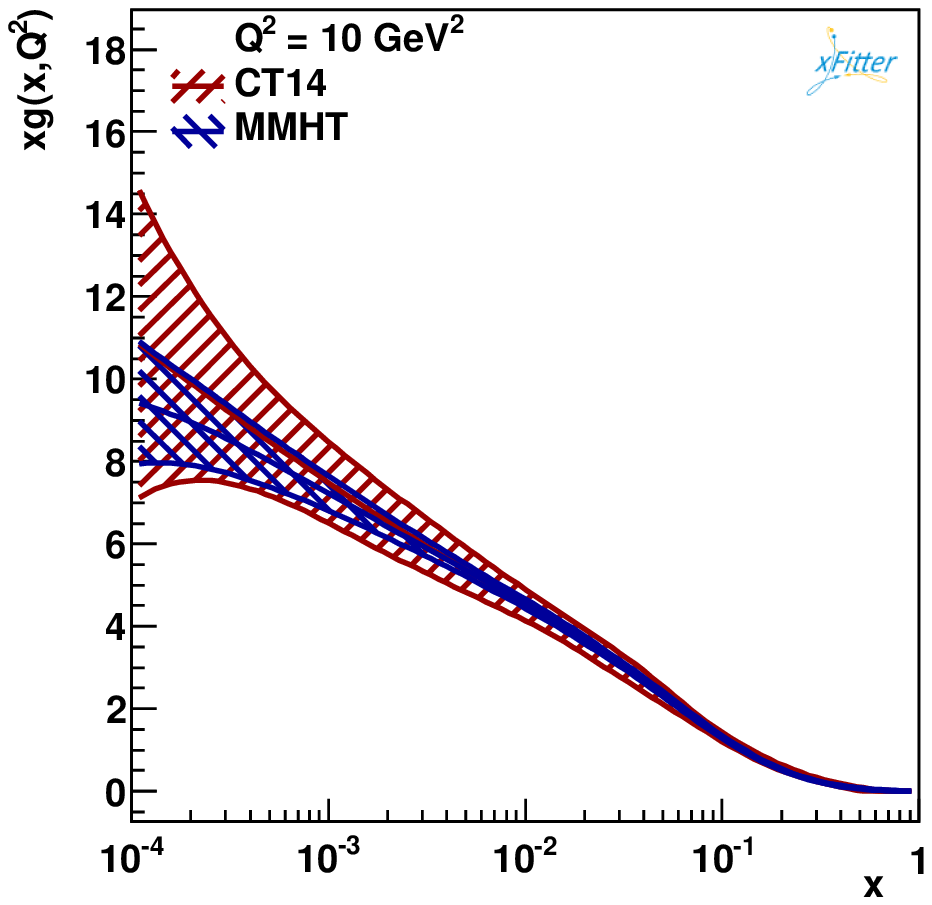}
	    \includegraphics[scale = 0.35]{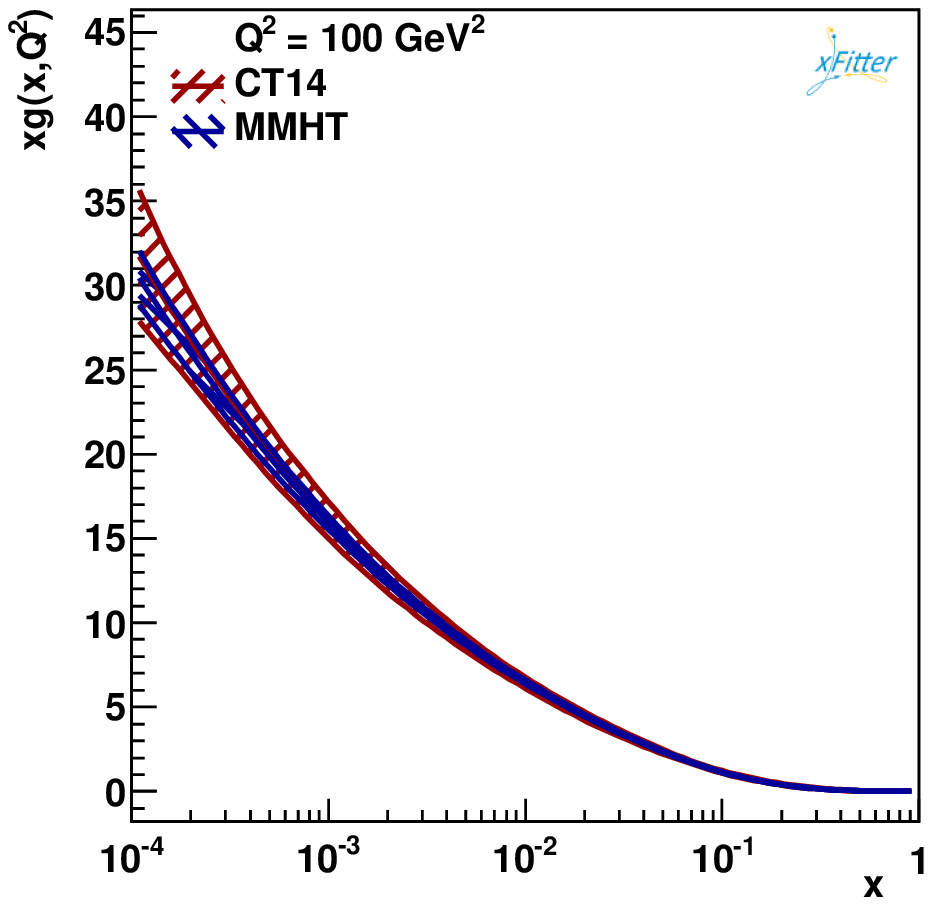}
	    \includegraphics[scale = 0.35]{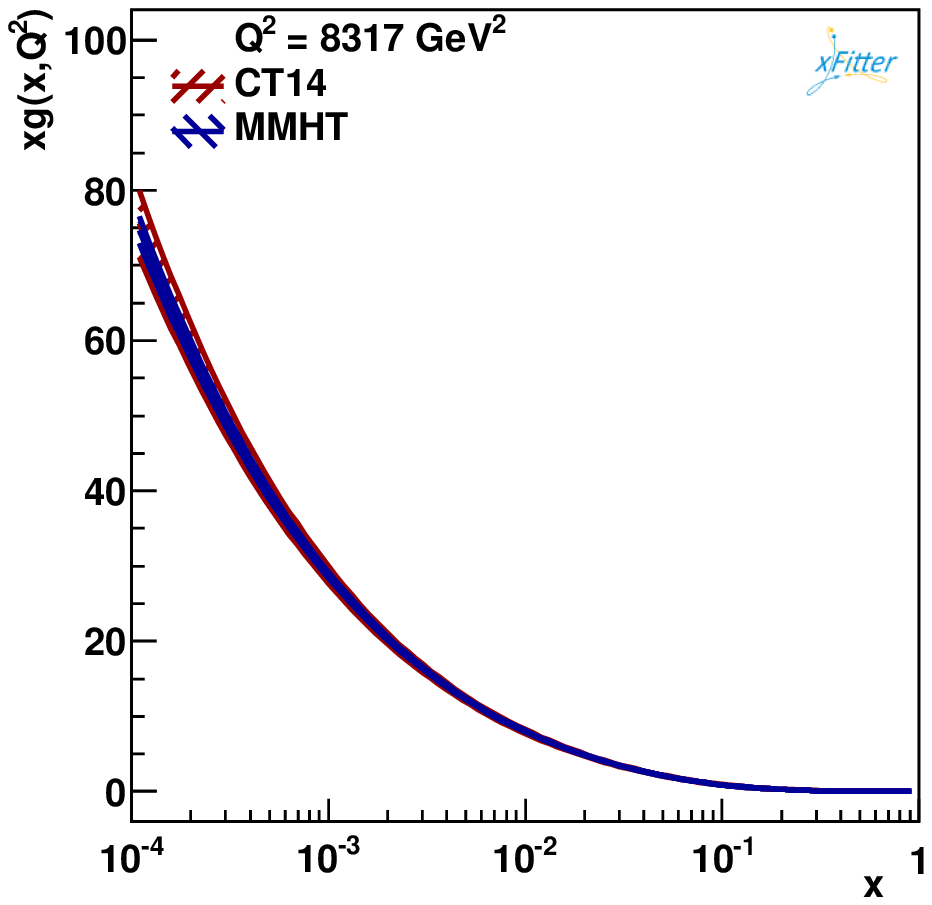}

\caption{The compression of parton distribution of  $xs$, and $xg$ extracted from MMHT2014 \cite{Harland-Lang:2014zoa} and CT14 \cite{Dulat:2015mca} PDFs as a function of $x$ at 4, 10, 100, and 8317 GeV$^2$. The results obtained before profiling.}
		\label{fig:MMHT&CT14}
	\end{center}
\end{figure}

\begin{figure}[!htb]
	\begin{center}

	    \includegraphics[scale = 0.35]{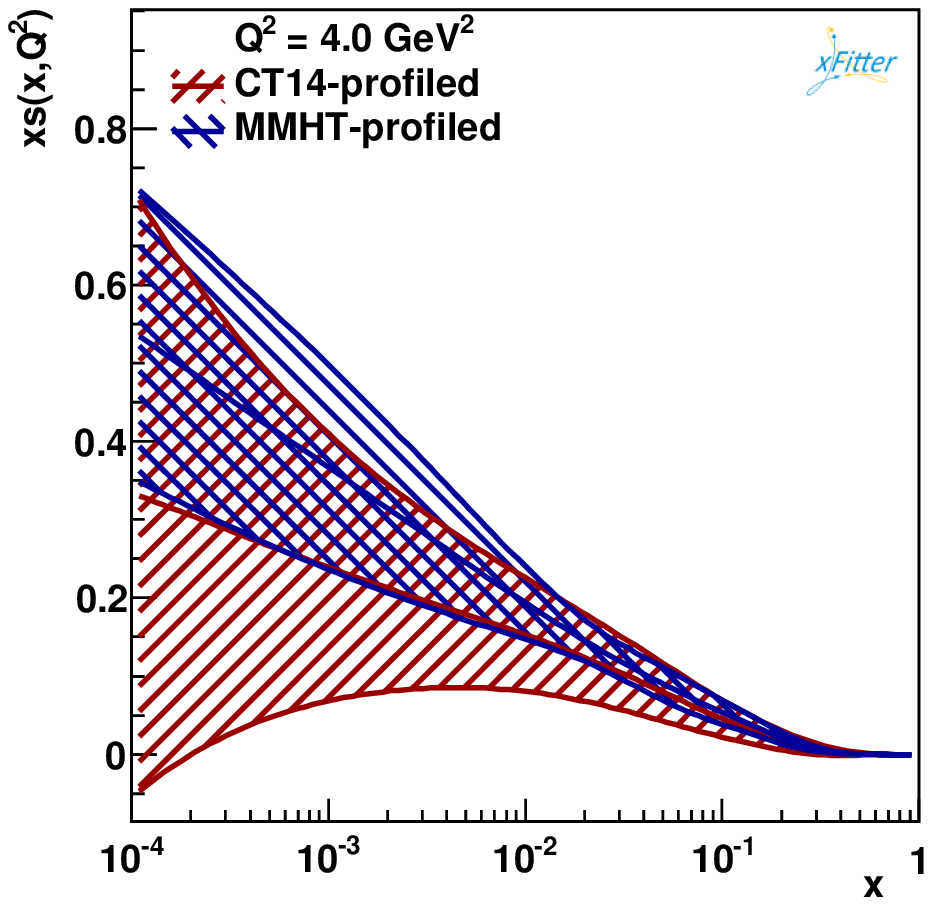}        
        	\includegraphics[scale = 0.35]{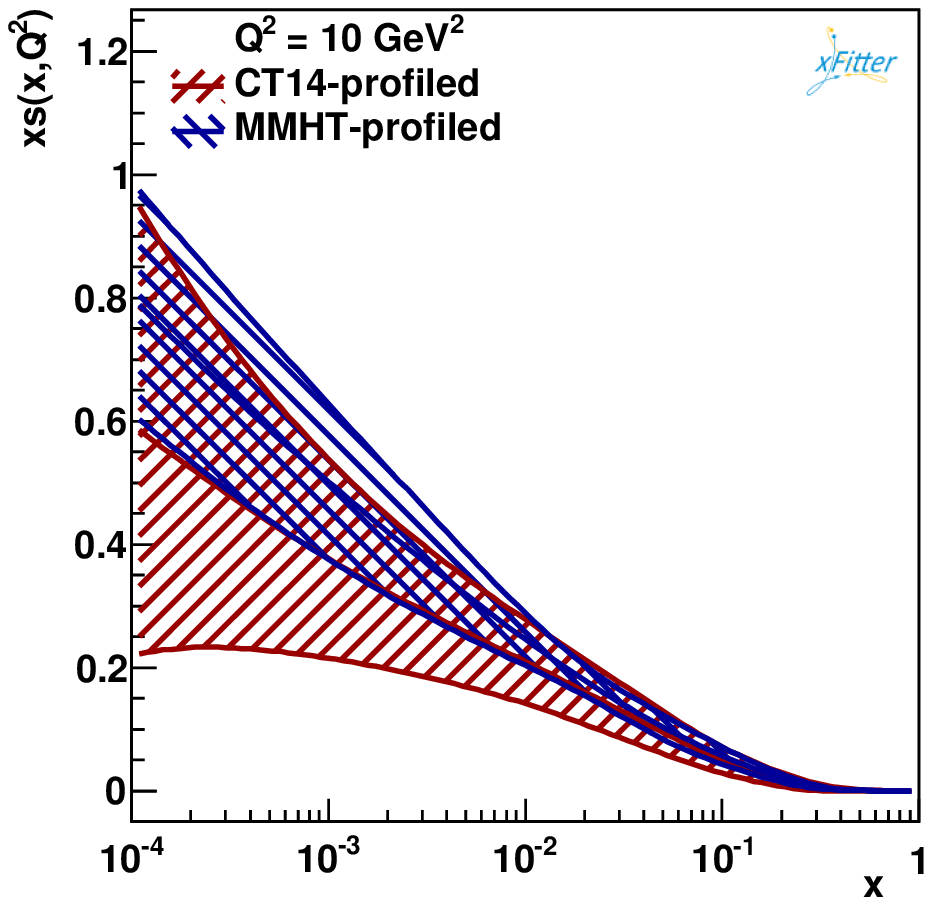}
	    \includegraphics[scale = 0.35]{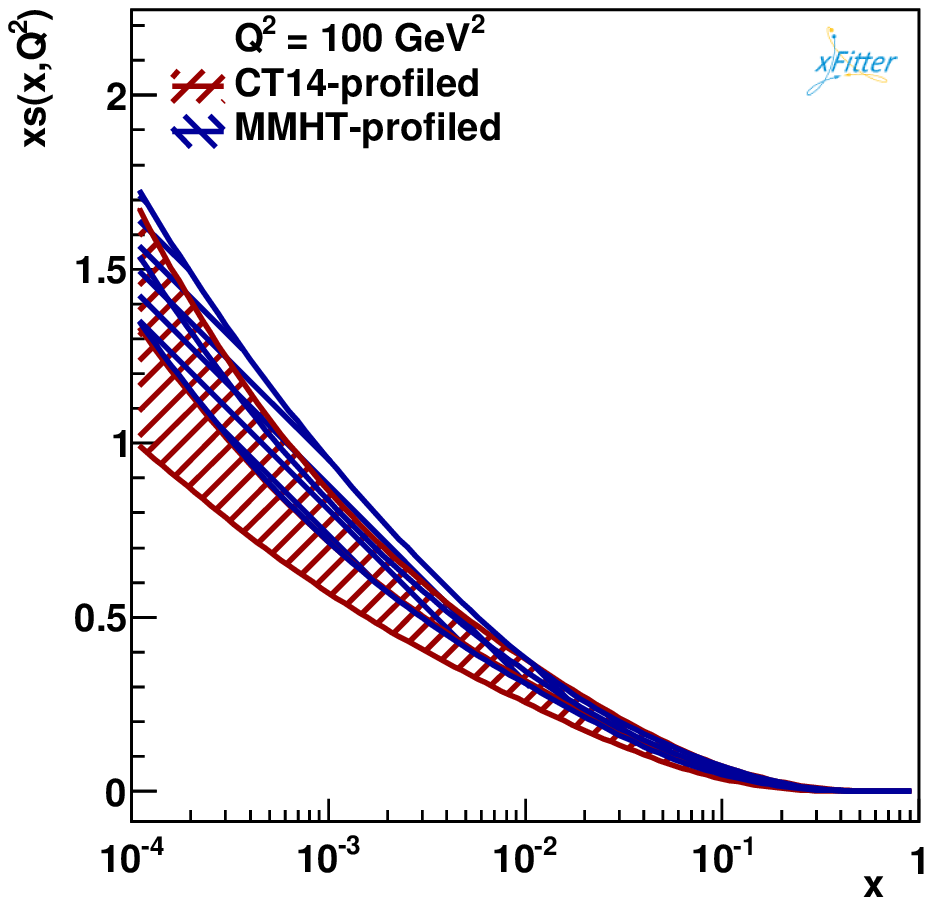}
	    \includegraphics[scale = 0.35]{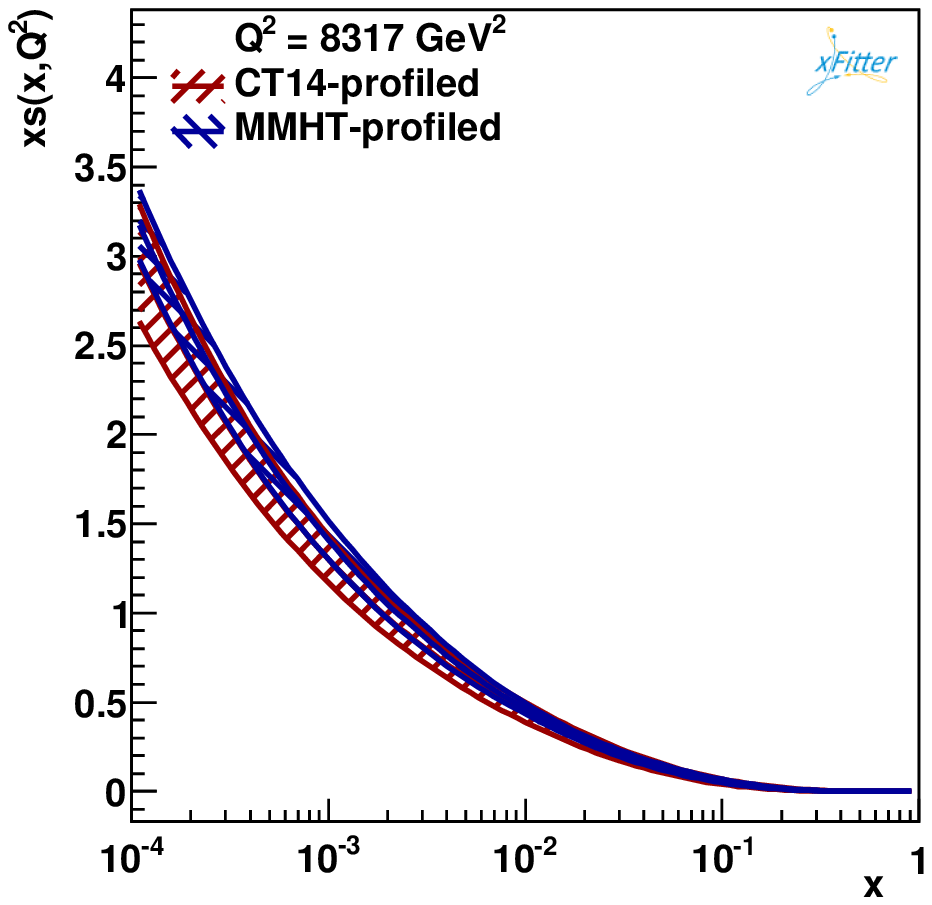}

	    \includegraphics[scale = 0.35]{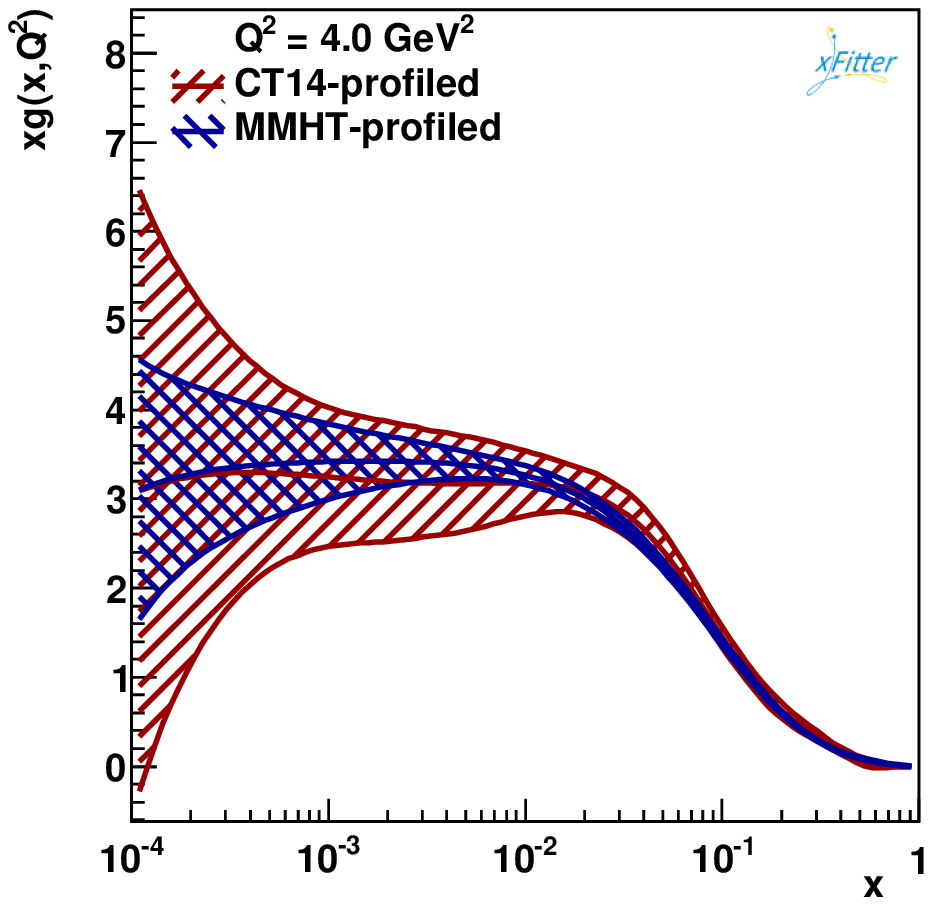}     
	    \includegraphics[scale = 0.35]{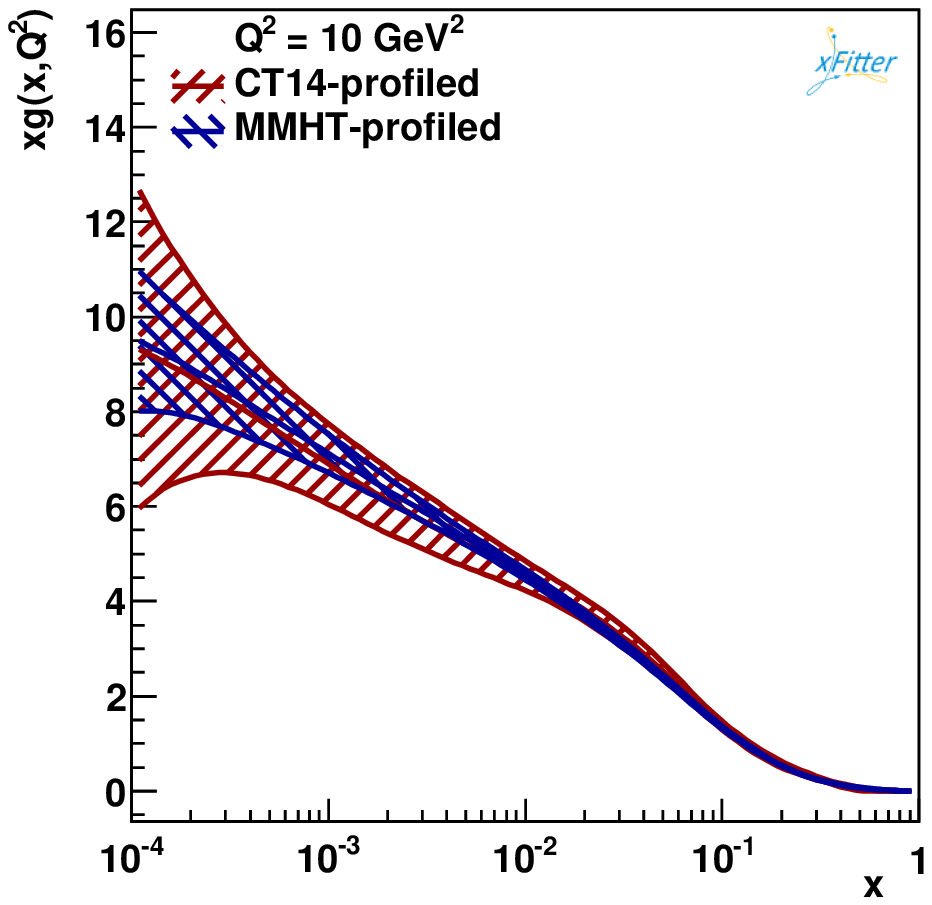}
	    \includegraphics[scale = 0.35]{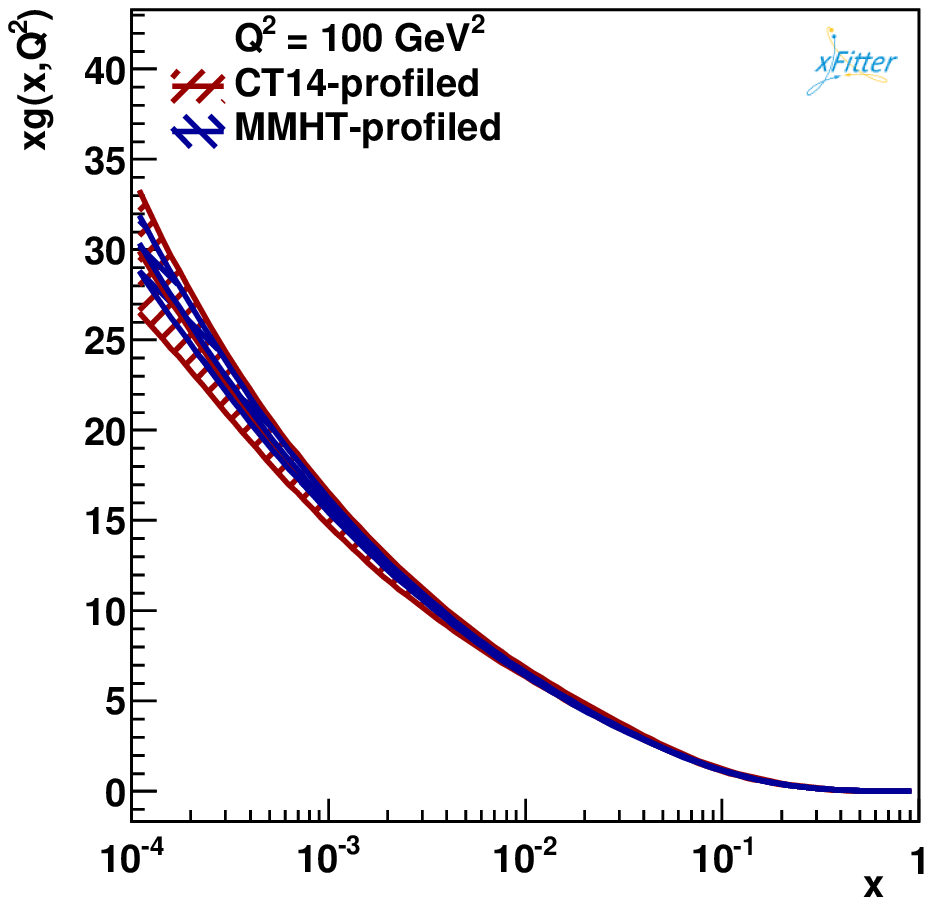}
	    \includegraphics[scale = 0.35]{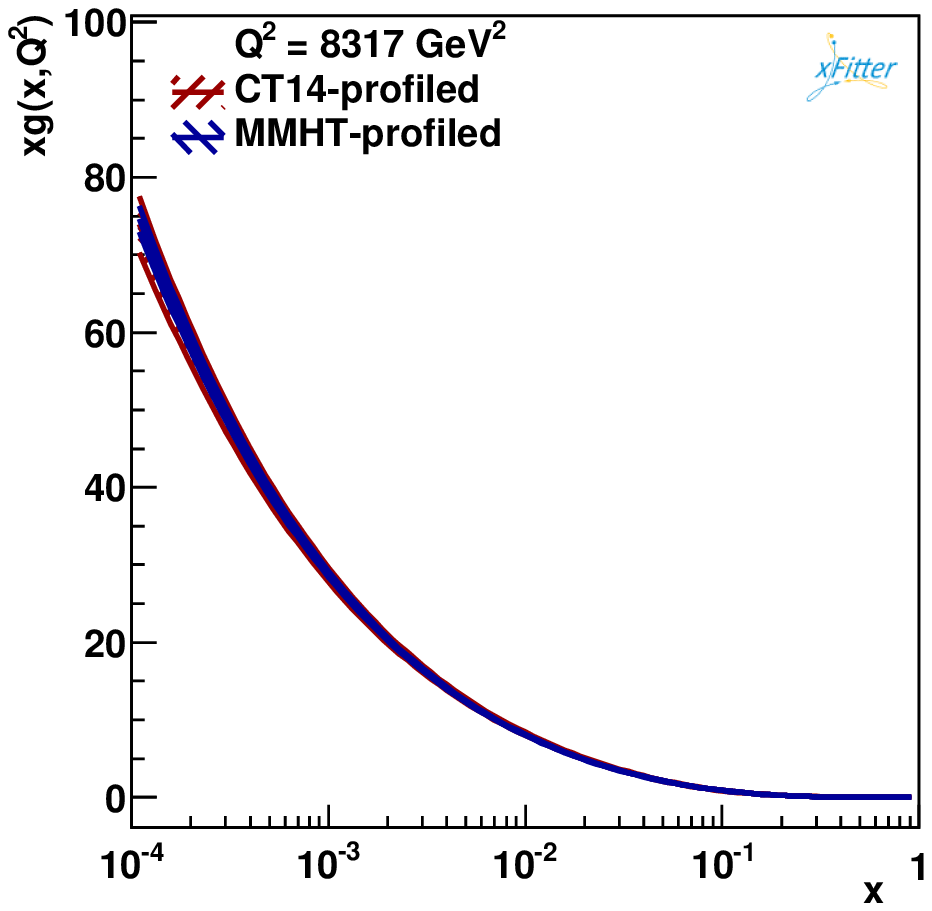}

\caption{The compression of parton distribution of  $xs$, and $xg$ extracted from MMHT2014 \cite{Harland-Lang:2014zoa} and CT14 \cite{Dulat:2015mca} PDFs as a function of $x$ at 4, 10, 100, and 8317 GeV$^2$. The results obtained after profiling.}
		\label{fig:MMHT&CT14-profiled}
	\end{center}
\end{figure}

The profiling procedure using  new set of top quark pair production  data improves agreement
of the strange $xs$ and gluon distributions between the CT14 and MMHT2014 PDF sets. Figures \ref{fig:MMHT&CT14} and \ref{fig:MMHT&CT14-profiled} show a direct comparison between CT14 and MMHT2014 sets.

\section{Conclusion}
\label{sec:conclusion}
In fact, a large part of high energy collider physics depends on the knowledge of PDFs in QCD. The PDFs determination in global analyses is a complex procedure, which needs the parametrization using the fits of experimental data. Although different PDF parameterizations are available for a general user, finding the impact of new measurements of the data on PDFs without doing a global QCD analysis would be worthful. For example, we can find which kind of PDFs can be affected in the presence of a specific new data. In this regard, the Hessian profiling technique is a good choice.  
	
We have discussed how to investigate the effects that a new set of
top quark pair production measurements have within an existing PDF set.  
Using the profiling formalism we have determined the impact of the recent top quark pair production data on the different PDFs.
The gluon PDF is one of the worth known parton distribution functions and deep inelastic scattering data
constrain the gluon only indirectly, and direct information comes only
from the inclusive jet production measurements.

The CT14 and MMHT2014 PDFs are profiled according to Eq.~(1). 
The results of the profiling on $s$ and gluon PDFs, their relative uncertainties, and  on the PDF ratios  with respect to  before profiling procedure are shown. The profiling affects the shape of the PDF more for the CT14 when compared to MMHT2014 PDF set. 
In the present paper, we observed a small  impact on $x\bar u$, a significant on  both central values and uncertainties  of $xs$ 
and gluon PDFs for CT14 and MMHT2014.

A significant reduction of the relative gluon uncertainties is obtained in the large $x$ for MMHT2014 PDF set. However, a significant reduction of the uncertainties is observed in the medium and large $x$ for CT14 PDF set.

The profiling procedure using  new set of top quark affected  the strange $xs$ and gluon distributions at the CT14 and MMHT2014 PDF sets.
These findings are interesting  and show the significance of the top quark production cross section data to constrain gluon and strange PDFs, and suggest that the data should be used in the future global QCD analyses.

\section{Acknowledgments}
We appreciate R. Placakyte and A. Glazov for useful comments about the xFitter framework.  
The authors are grateful to CMS, ATLAS and \dzero collaborations for their fantastic measurements.

\newpage

           


\end{document}